\documentclass[manuscript]{emulateapj}

 \usepackage{amsfonts,amsmath,graphicx,natbib,subfigure,color,gensymb,longtable}

\newcommand{\SNANA}{{\tt SNANA}}
\newcommand{\SMP}{{\tt SMP}}

\def\SAMPLENAME{DES-SN3YR}

\def\statwerr{0.042}
\def\systwerr{0.042}
\def\werr{0.059}

\def\sigmacal{0.021}

\def\ghrstep{0.009}
\def\ghrsteperr{0.018}
\def\chrstep{0.004}
\def\chrsteperr{0.017}

\def\gammasigma{2.4}

\def\desgsigint{0.066}
\def\descsigint{0.088}

\def\desgsigtot{0.129}

\def\totsn{329}

\def\fakestatbias{$-0.0029 \pm 0.0035$}
\def\fakezpbias{$-0.0039 \pm 0.0072$}
\def\fakescatbias{$-0.0046 \pm 0.0053$}

\def\lowzcutsys{0.016}

\newcommand{\werrSystBiasAstro}{0.026}
\newcommand{\werrSystBiasSurvey}{0.023}
\newcommand{\werrSystRedshift}{0.012}
\newcommand{\werrSystNew}{0.024}

\def\LCDM{$\Lambda$CDM}
\def\sigint{\sigma_{\rm int}}

\newcommand{\lowz}{low-$z$}

\def\KEYPAPER{\cite{KEYPAPER}}
\def\KEYPAPERalt{\citealt{KEYPAPER}}

\newcommand{\urldr}{https://des.ncsa.illinois.edu/releases/sn}

\newcommand{\dmuBias}{\delta\mu_{\rm bias}}
\newcommand{\dmuHost}{\delta\mu_{\rm host}}

\received{November 9, 2018}
\accepted{February 20, 2019}

\usepackage{lineno}
\usepackage{color}
\usepackage{soul} 

\def\sc{14}

\newcommand{\wCDM}{$w$CDM}

\shorttitle{\textit{Brout~et~al.} First Cosmology Results From DES-SN: Analysis, Systematic Uncertainties, and Validation}
\shortauthors{\textit{Brout~et~al.} First Cosmology Results From DES-SN: Analysis, Systematic Uncertainties, and Validation}
\newpage
\newpage
\begin{document}

{
\begin{nolinenumbers}
\vspace*{-\headsep}\vspace*{\headheight}
\footnotesize \hfill FERMILAB-PUB-18-541-AE\\
\vspace*{-\headsep}\vspace*{\headheight}
\footnotesize \hfill DES-2018-0356
\end{nolinenumbers}
}

\title{First Cosmology Results Using Type Ia Supernovae From the Dark Energy Survey: Analysis, Systematic Uncertainties, and Validation}

\def\andname{}

\author{
D.~Brout\altaffilmark{1},
D.~Scolnic\altaffilmark{2},
R.~Kessler\altaffilmark{3,2},
C.~B.~D'Andrea\altaffilmark{1},
T.~M.~Davis\altaffilmark{4},
R.~R.~Gupta\altaffilmark{5},
S.~R.~Hinton\altaffilmark{4},
A.~G.~Kim\altaffilmark{5},
J.~Lasker\altaffilmark{3,2},
C.~Lidman\altaffilmark{6}
% ...
\\ \vspace{0.2cm} (DES Collaboration) \\
}

\affil{$^{1}$ Department of Physics and Astronomy, University of Pennsylvania, Philadelphia, PA 19104, USA}
\affil{$^{2}$ Kavli Institute for Cosmological Physics, University of Chicago, Chicago, IL 60637, USA}
\affil{$^{3}$ Department of Astronomy and Astrophysics, University of Chicago, Chicago, IL 60637, USA}
\affil{$^{4}$ School of Mathematics and Physics, University of Queensland,  Brisbane, QLD 4072, Australia}
\affil{$^{5}$ Lawrence Berkeley National Laboratory, 1 Cyclotron Road, Berkeley, CA 94720, USA}
\affil{$^{6}$ The Research School of Astronomy and Astrophysics, Australian National University, ACT 2601, Australia}
% ...

\begin{abstract}
We present the analysis underpinning the measurement of cosmological parameters from 207 spectroscopically classified type Ia supernovae (SNe~Ia) from the first three years of the Dark Energy Survey Supernova Program
(DES-SN), spanning a redshift range of $0.017<z<0.849$. 
We combine the DES-SN sample with an external sample of 122 low-redshift ($z<0.1$) SNe~Ia, resulting in a ``{\SAMPLENAME}'' sample of 329 SNe~Ia.
Our cosmological analyses are blinded: after combining our {\SAMPLENAME}
distances with constraints from the Cosmic Microwave Background \citep[CMB;][]{Planck16_cosmo}, our uncertainties in the measurement of the dark energy equation-of-state parameter, $w$, are \statwerr~(stat) and \werr~(stat+syst) at 68\% confidence. 
We provide a detailed systematic uncertainty budget, which 
has nearly equal contributions from
photometric calibration, astrophysical bias corrections, and instrumental bias corrections. We also include several new sources of systematic uncertainty. While our sample is $<1/3$ the size of the Pantheon sample, our constraints on $w$ are only larger by $1.4\times$, showing the impact of the DES SN~Ia light curve quality.
We find that the traditional stretch and color standardization parameters of the DES SNe~Ia are in agreement with earlier SN~Ia samples such as Pan-STARRS1 and the Supernova Legacy Survey. However, we find smaller intrinsic scatter about the Hubble diagram (0.077~mag). Interestingly, we find no evidence for a Hubble residual step ($0.007 \pm 0.018$ mag) as a function of host galaxy mass for the DES subset, in 2.4$\sigma$ tension with previous measurements.
We also present novel validation methods of our sample using simulated SNe~Ia inserted in DECam images and using large catalog-level simulations to test for biases in our analysis pipelines.

\end{abstract}
\section{Introduction}
\label{Sec:intro}

\setcounter{footnote}{0}

The discovery of the accelerating expansion of the universe (\citealt{riess98}; \citealt{perlmutter99}) has motivated an era of cosmology surveys with the goal of measuring the mysterious properties of dark energy. The use of standardizable type Ia supernovae (SNe~Ia) to measure distances has proven to be a vital tool in constraining the nature of dark energy because they probe the geometry of the universe throughout a large portion of cosmic time. 

The Dark Energy Survey Supernova Program (hereafter DES-SN) has found thousands of photometrically classified SNe~Ia at redshifts from $0.01<z<1.2$ using repeated observations in the southern celestial hemisphere searching over an area of $27~ {\rm deg}^2$ (\citealt{des12}). Over the full five years of the survey, DES-SN is expected to obtain the largest single dataset of photometrically classified SNe~Ia to date. DES-SN has spectroscopically confirmed a subset of $\sim$500 SNe~Ia at redshifts from $0.017<z<0.849$. In this work, we analyze the first three years of spectroscopically confirmed SNe~Ia and combine our dataset with an external low redshift SN~Ia sample. This combined sample is hereafter called \SAMPLENAME. The subset of DES SNe~Ia is hereafter denoted `the DES subset' and the subset of low-$z$ SNe~Ia from CfA3, CfA4, and \mbox{CSP-1} is hereafter denoted `the low-$z$ subset' (CfA3-4; \citealt{Hicken09a,Hicken12}; \mbox{CSP-1}, \citealt{contreras10}).

Over the past two decades, there have been three parallel and overlapping major developments in using SNe~Ia to measure cosmological parameters, upon which the DES-SN has made improvements.  The first development is the order-of-magnitude growth in the number of spectroscopically confirmed SNe~Ia. Original datasets at low-redshift had tens of SNe~Ia (e.g., CfA1-CfA2, \citealt{CFAriess}; \citealt{CFAjha}) and the next generation of low-redshift and high-redshift datasets had hundreds of SNe-Ia (e.g. CfA3-4; \mbox{CSP-1}; ESSENCE: \citealt{essence3}; SDSS-II: \citealt{frieman08},  \citealt{Sako18}); SNLS: \citealt{Guy10}; PS1: \citealt{rest14}, \citealt{Scolnic17a}). Today, with the addition of DES-SN, there are now more than 1500 spectroscopically confirmed SNe~Ia in total.

The second development has been in detector sensitivity, which has resulted in improved light curve quality and distance measurement uncertainties. The 570 megapixel Dark Energy Camera (DECam; \citealt{decam}), with its fully-depleted CCDs and excellent $z$-band response, facilitates well-measured optical light curves at high redshift (\citealt{diehl14}).

The third major development has been the increasingly sophisticated analyses of the samples. As SN~Ia datasets grow in size, analyses are better able to characterize SN~Ia populations and expected biases from observational selection and analysis requirements. Improvements in the analysis over the last decade have included scene modeling photometry (\SMP\; \citealt{Holtzman09}, \citealt{astier13}, B18-SMP: \citealt{Brout18-SMP}-SMP) instead of classical template subtraction, the modeling and correction of expected biases using large simulations (\citealt{perrett10}, \citealt{kessler09cosmo}, \citealt{jla}), and measuring filter transmissions to achieve sub 1\% calibration uncertainty (\citealt{Astier06}, \citealt{Doi10}, \citealt{Tonry12}, \citealt{Marshall13}, \citealt{burke17}). Recent cosmological parameter analyses (B14: \citealt{betoule14}, S18: \citealt{scolnic18}) have found that systematic uncertainties are roughly equal to the statistical uncertainties; this is due to the improving ability to understand and reduce systematic uncertainties with larger samples and reduced statistical uncertainties.  Each new cosmology analysis (\citealt{essenceCosmology}, \citealt{kessler09cosmo}, \citealt{Sullivan11}, B14, S14: \citealt{scolnic14}, S18, \citealt{jones18cosmo}) has built on previous analyses in their treatment of systematic uncertainties.  
Here we continue in this tradition of improvements, and also study several previously uninvestigated sources of uncertainty.

Improvement in understanding of systematic uncertainties is crucial to taking advantage of the order-of-magnitude increases in statistics expected in the coming years. From DES-SN alone, there is the full sample of $\sim$2000 photometrically classified SNe~Ia. Additionally, the next generation of photometric transient surveys (LSST: \citealt{ivzec08}; \citealt{lsst09}; WFIRST: \citealt{hounsellwfirst}) expects tens to hundreds of thousands of photometrically classified SNe~Ia.

\begin{figure}
\begin{centering}
\includegraphics[width=0.34\textwidth]{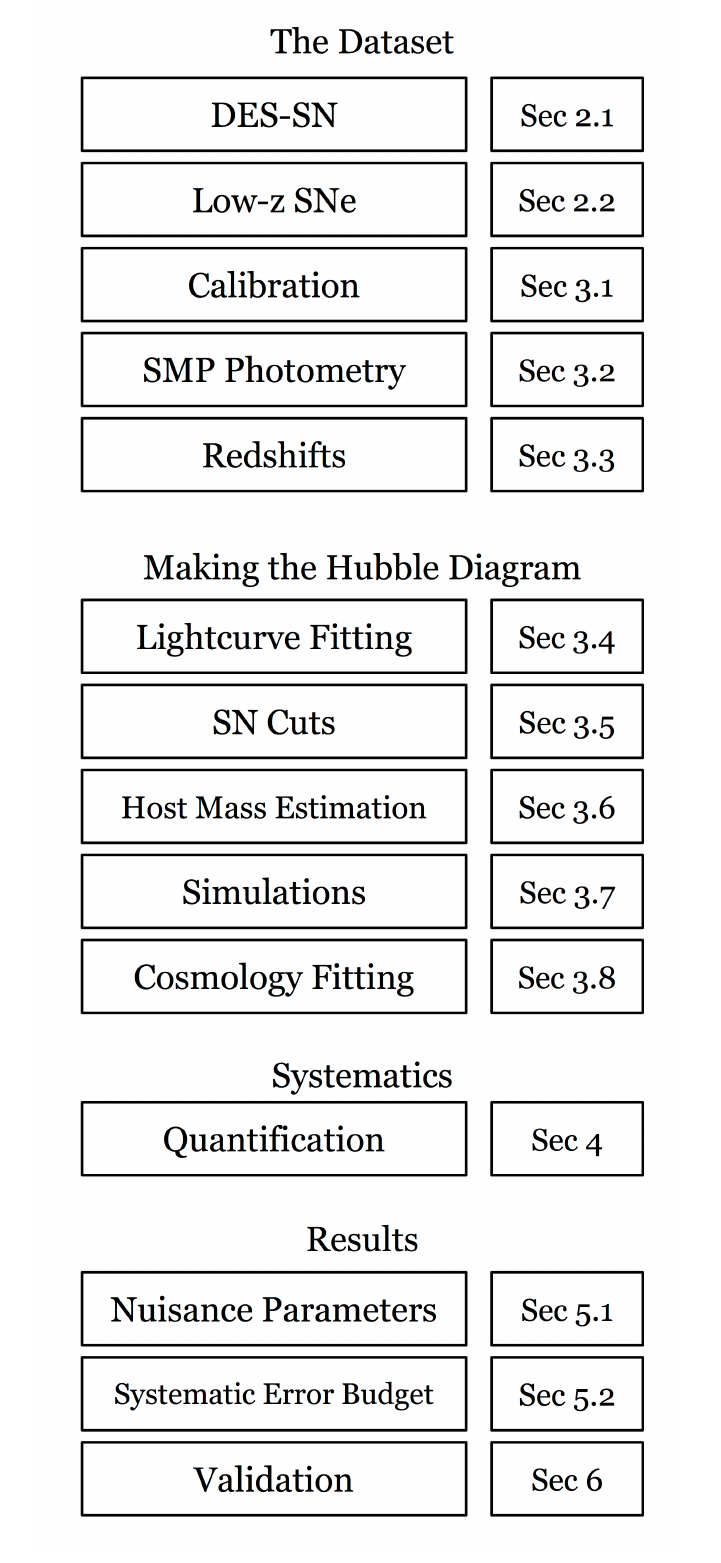}
\label{fig:flowchart}
\caption{Analysis flowchart of this paper. Nuisance parameters, the systematic error budget, and the results of validation are considered the ``Results'' of this work (Section \ref{results}) and the unblinded cosmological parameter best fit values are presented in \KEYPAPER.}
\end{centering}
\end{figure}

The key analysis steps to produce cosmological constraints from our spectroscopically confirmed dataset are 1) absolute calibration of the DES-SN photometric system, 2) precision photometry for light curve fluxes, 3) simulation of large samples to predict biases, 4) light curve fits to standardize the SN brightness and measure luminosity distance,  5) construction of bias-corrected Hubble diagram, 6) construction of full statistical and systematic covariance matrix, and 7) cosmological parameter fits. Step 1 (\citealt{burke17}, \citealt{Lasker18}), step 2 (B18-SMP), and step 3 (\citealt{Kessler18}) are discussed in detail in companion papers and they are discussed within this paper in the context of understanding systematic uncertainties. Steps 4-7 are described here in detail.

There are two main results of this paper. First we present the nuisance parameters involved in the standardization of SNe~Ia. 
Historically $\alpha$ and $\beta$, the correlation coefficients for stretch and color of supernova light curves respectively, have been used to standardize SN~Ia luminosities, and $\sigma_{\rm int}$ has been used to characterize the scatter in SN~Ia luminosities that is not covered by the measurement uncertainties. Additionally, several groups in the last decade have shown that more massive galaxies tend to host overluminous SNe~Ia after color and stretch brightness standardization, suggesting improved standardizability of the SNe~Ia population (\citealt{Kelly10}; \citealt{Sullivan10}; \citealt{Lampeitl10}). This effect has been characterized as a step function in Hubble diagram residuals ($\gamma$) across $10^{10} \mathcal{M}_\sun$. However, the size of this effect has been seen to vary in different samples and the physical interpretation is not understood. In this paper we discuss our own findings for these nuisance parameters using \SAMPLENAME. The second main result is the statistical and systematic uncertainty budget from our $w$CDM cosmological analysis after combining with \cite{Planck16_cosmo} CMB priors. Using the analysis and results derived here, cosmological parameter constraints are shown in \KEYPAPER.

In order to improve upon the treatment and validation of systematic uncertainties from past analyses, we use two types of SN~Ia simulations to examine biases in our pipelines and to provide crosschecks of our analysis. The first set of simulations includes hundreds of catalog-level simulations with input sources of systematic uncertainty. We analyze the catalog level simulations with steps 3-7 above to verify our analysis pipeline
and reported statistical and systematic uncertainties. 
These simulations are generated by the SuperNova ANAlysis software package\footnote{\texttt{https://snana.uchicago.edu}} (\texttt{SNANA}: \citealt{snana}), which has been used extensively by previous analyses to quantify expected biases and offers the capability of parallelization for generating and analyzing large simulations of SNe~Ia.

For the second set of validation simulations, we generate 100,000 artificial supernova light curves which are inserted as point sources onto DECam images (hereafter `fakes'). Previous analyses have used artificial point sources to understand photometric uncertainties (\citealt{Holtzman09}; \citealt{perrett10}). In DES-SN, fake supernovae  light curves are used for several reasons. Fakes are used to check for biases in photometry (B18-SMP) and in the determination of SN~Ia detection efficiency as a function of signal-to-noise (S/N) (Kessler et al. 2018 in prep.), thereby modeling subtle pipeline features that cannot be computed from first principles. Additionally, we present a cosmological analysis of 10,000 fake supernovae that have been recovered by the search pipeline, processed by the photometric pipeline, and processed through our cosmological analysis pipeline in the same manner as the real dataset. This crosscheck is sensitive to potential un-modeled biases in the image-processing pipelines and their propagation to cosmological distance and cosmological parameter biases.

Unfortunately, neither of the methods above address the systematic uncertainty due to calibration. To address calibration uncertainties, we compare our absolute calibration with that of the Pan-STARRS survey (\citealt{Tonry12}) and SuperCal (\citealt{supercal}).

The organization of this paper is depicted in Figure~\ref{fig:flowchart} and is described as follows. In \S\ref{dataset}, we introduce the data samples, a combination of high-redshift SNe~Ia from DES-SN and low-redshift SNe~Ia from CfA and \mbox{CSP-1}. In \S\ref{analysis}, we discuss analysis procedures and characterize systematic uncertainties. In \S\ref{systematics}, we quantify each source of systematic uncertainty. In \S\ref{results} we present results for the nuisance parameters, the systematic uncertainty budget, and the total statistical and systematic uncertainty. In \S\ref{validation} we describe our validation methods. In \S\ref{bhm} we discuss a Bayesian Hierarchical method under development, and its performance on validation and the \SAMPLENAME\ sample. In \S\ref{discussion} we discuss our findings and in \S\ref{conclusion} we conclude.

\begin{figure}
\begin{centering}
\includegraphics[width=0.49\textwidth]{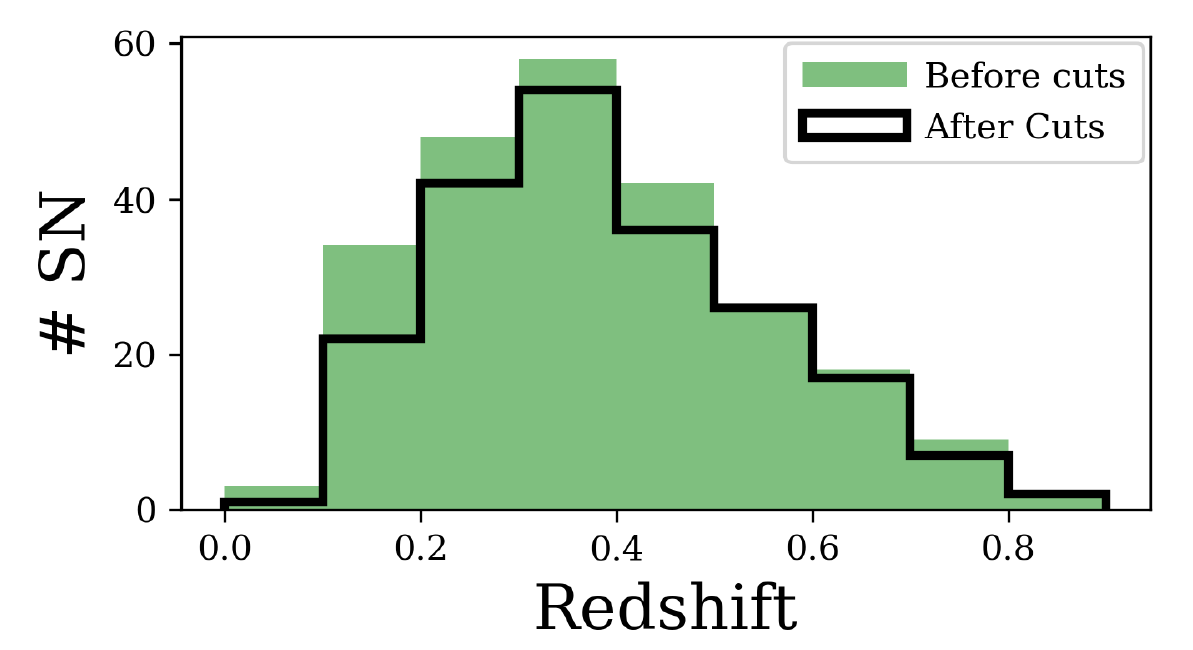}
\caption{Histogram of the 251 spectroscopically confirmed SNe~Ia is shown in green-filled. The sub-sample of SNe~Ia used for cosmological parameter analysis that pass all quality cuts is shown in black.}
\label{fig:zhist}
\end{centering}
\end{figure}

\section{Datasets}
\label{dataset}
\subsection{The Dark Energy Survey Supernova Program}

DES-SN performed a deep, time-domain survey in four optical bands $(g,r,i,z)$ covering $\sim 27$~deg$^2$ over 5 seasons (2013-2018) using the DECam mounted on the \mbox{4-m} Blanco telescope at the Cerro Tololo Inter-American Observatory (CTIO). Exposure processing (\citealt{morganson18}), difference imaging (\texttt{DiffImg}: \citealt{kessler15}), and automated rejection of subtraction artifacts (\citealt{autoscan}) are run on a nightly basis. DES-SN observed in 8 ``shallow" fields (C1,C2,X1,X2,E1,E2,S1,S2) with single-epoch 50\% completeness depth of $\sim$ 23.5 mag; and in 2 ``deep" fields (C3,X3) with depth $\sim$ 24.5~mag in all four bands. The ten DES-SN fields are grouped into 4 separated regions (C,X,E,S), where each group contains adjacent pointings on the sky. For example, C1,C2,C3 are adjacent fields denoted group C, where each field center is separated by 2 degrees. Tables 1\&2 of \cite{kessler15} contain detailed information of the DES-SN observing fields.

For a SN to be considered a `candidate', we require two detections at the same location on two separate nights in any of the four bands. A subset of the candidates are selected for spectroscopic follow-up to obtain a type classification and redshift. A detailed overview of the spectroscopic follow-up program as well as a general overview of the DES-SN program and observing strategy can be found in \cite{D'Andrea18}.

Over the first three years of DES-SN, from Sept. 2013 to Feb. 2016, we discovered roughly $\sim$12,000 candidates of which $\sim$2000 are likely SNe~Ia. In this first analysis we analyze only the spectroscopically confirmed SN~Ia subset of the data. As described in \cite{D'Andrea18}, 307 transients of the likely SNe~Ia were targeted for spectroscopic classification using a variety of spectroscopic resources, and 251 were confirmed as Type Ia over a redshift range of $0.017 < z < 0.849$. The majority of spectra come from the
Anglo-Australian Telescope (AAT) as part of the OzDES program (\citealt{ozdes1}, \citealt{ozdes2}). The distribution of redshifts for the spectroscopically confirmed SNe~Ia from the first 3 years of DES-SN observations is shown in Figure~\ref{fig:zhist}. DES-SN SNe at lower-redshift are preferentially cut from the sample used for cosmological analysis due to poor light curve coverage and light curve fit quality. Quality cuts and selection requirements are discussed in detail in Section \ref{sec:selection}. The [min,mean,max] redshifts after performing the data selection cuts are $[0.08,0.39,0.85]$ respectively.

Additional data are acquired using an in-situ calibration process called ``DECal'' (\citealt{Marshall13}). The Blanco/DECam optical system and filter transmission functions are measured under multi-wavelength illumination. DES-SN also acquires real-time meteorological data using the SUOMINET system\footnote{\texttt{http://www.suominent.ucar.edu}} to track precipitable water vapor levels and auxiliary ``aTmCAM'' instrumentation (\citealt{li14}) to measure atmospheric conditions.

\subsection{External Low-Redshift Samples}
Cosmological constraints from SNe~Ia are best obtained with samples at both low-redshift and high-redshift. We utilize four publicly available low-redshift surveys: CfA3S, CfA3K, CfA4, and \mbox{CSP-1} (\citealt{CFAjha}; \citealt{Hicken09a,Hicken12}; \citealt{contreras10}) consisting of 303 spectroscopically confirmed SNe~Ia in the redshift range 0.01 $< z <$ 0.1. These low-redshift surveys are chosen because of their well-defined calibrations.
B14 and S18 included 22 SNe~Ia from CfA1 \& CfA2 as part of their analyses.  However, we chose not to include them in our analysis because filter transmission functions were not provided for those samples.

\section{Analysis}
\label{analysis}
Here we describe the analysis procedures used to measure cosmological parameters.  The majority of this section describes the analysis of the DES subset itself, though we also include our analysis of the low-redshift sample.  The description of systematic uncertainties associated with each step in the analysis is laid out in this section and each source of systematic uncertainty is quantified in Section \ref{systematics}. We rely on complementary work in \cite{Kessler18}, hereafter K18, which details the simulations of DES-SN3YR. These simulations are used for computing bias corrections in Section \ref{simulations}.

\subsection{Calibration}
\label{calibration}

SN~Ia cosmological constraints rely on the ability to internally transform each SN flux measurement in ADU (Analog/Digital Units) into a `top-of-the-galaxy' brightness. This is done in two steps, first via measurements of Hubble Space Telescope (HST) CalSpec\footnote{\texttt{http://www.stsci.edu/hst/observatory/crds/calspec.html}} standard stars to obtain a top-of-the-atmosphere brightness, which is discussed here. Second, we obtain top-of-the-galaxy brightness by accounting for the Milky Way extinction along the line of sight, values for which are obtained from \cite{schlegel98} \& \cite{Schlafly11}.
Measurements of cosmological parameters using SNe~Ia are sensitive to filter calibration uncertainties (internal) due to the fact that at higher redshift, constraints of the SN light curve models rely on observed fluxes in a different set of filters than at lower redshift. A dependence in SN cosmological distances as a function of redshift could arise from differences in the calibration between the \lowz~and DES subsets (external). Below we discuss the steps taken to both internally and externally calibrate the DES-SN measurements.

\subsubsection{Star Catalog}
\label{starcat}
\begin{figure}
\begin{centering}
\includegraphics[width=0.48\textwidth]{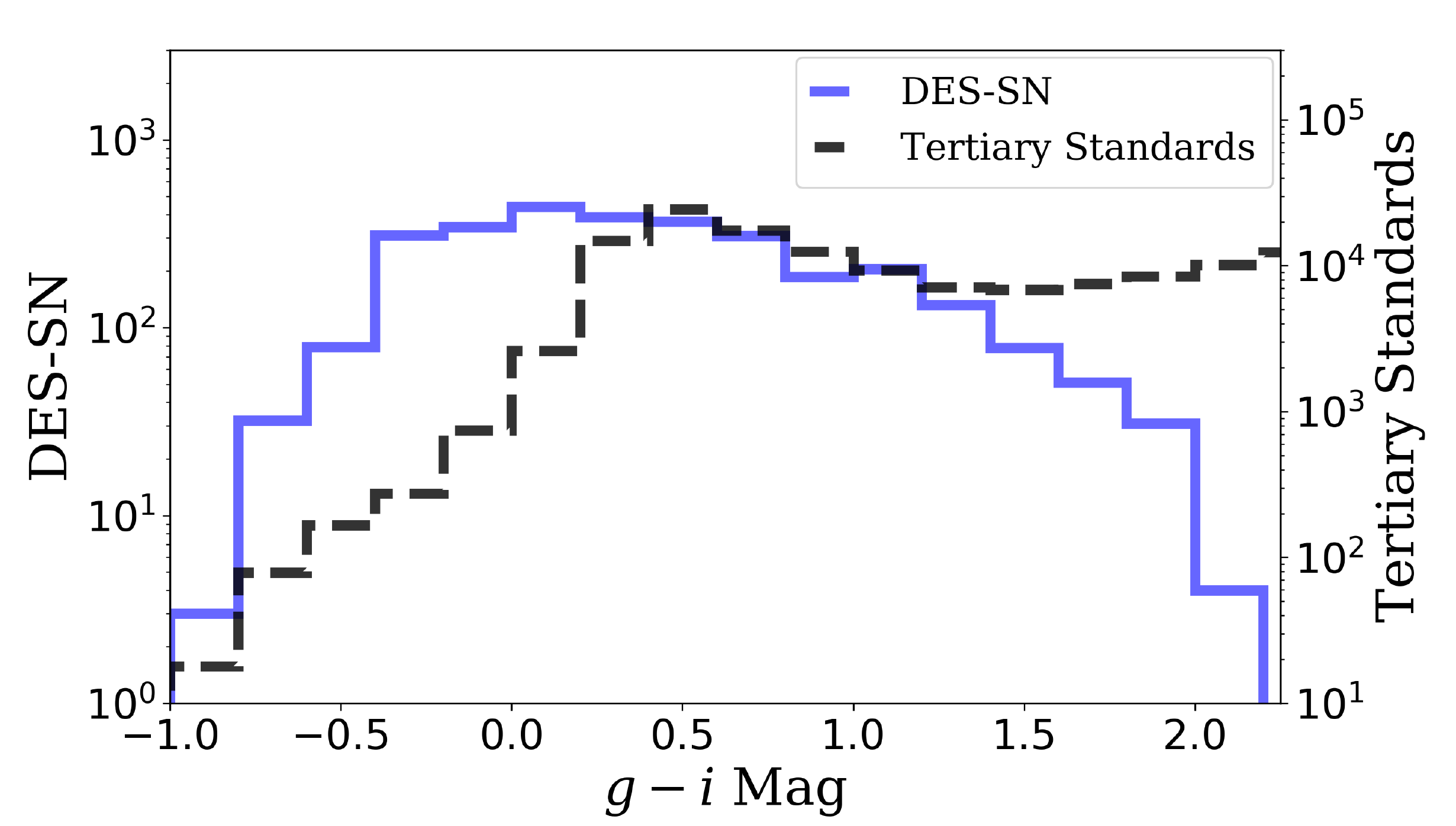}
\caption{Blue: Distribution of observed $g-i$ colors for the DES-SN sample observations. Epochs with S/N $>$ 10 are shown. Black: Distribution of $g-i$ colors for the tertiary standard stars used for internal calibration. The validity of chromatic corrections is evaluated over the stellar color range (black) but the corrections are applied to the DES-SN fluxes (blue).}
\label{fig:colorhist}
\vspace{.1in}

\end{centering}
\end{figure}

Here we describe the process of calibrating each of the DES-SN images. Photometry of approximately 50 tertiary standard stars are used to determine a zero point for each DECam CCD image. The catalog of tens of thousands of tertiary star magnitudes is described in \cite{burke17}. These stars are internally calibrated using a `Forward Global Calibration Method' (FGCM) to an RMS of 6~mmag. FGCM models the rate of photons detected by the camera by utilizing measurements of instrument transmission, atmospheric properties, a model of the atmosphere, and a model of the source. Spectral Energy Distribution (SED)-dependent chromatic corrections are applied to the standard stars which extend the 6~mmag calibration uncertainty to be valid over a very wide color range ($-1$ $<$ $g-i$ $<$ 3). The $g-i$ color distribution of the tertiary standard stars is shown in Figure~\ref{fig:colorhist}. The color distribution of the DES subset light curves is different from that of the standard stars and is discussed in Section \ref{photometry}.

\subsubsection{AB offsets}
\label{aboffsets}

\begin{figure*}
\begin{centering}
\includegraphics[width=0.95\textwidth]{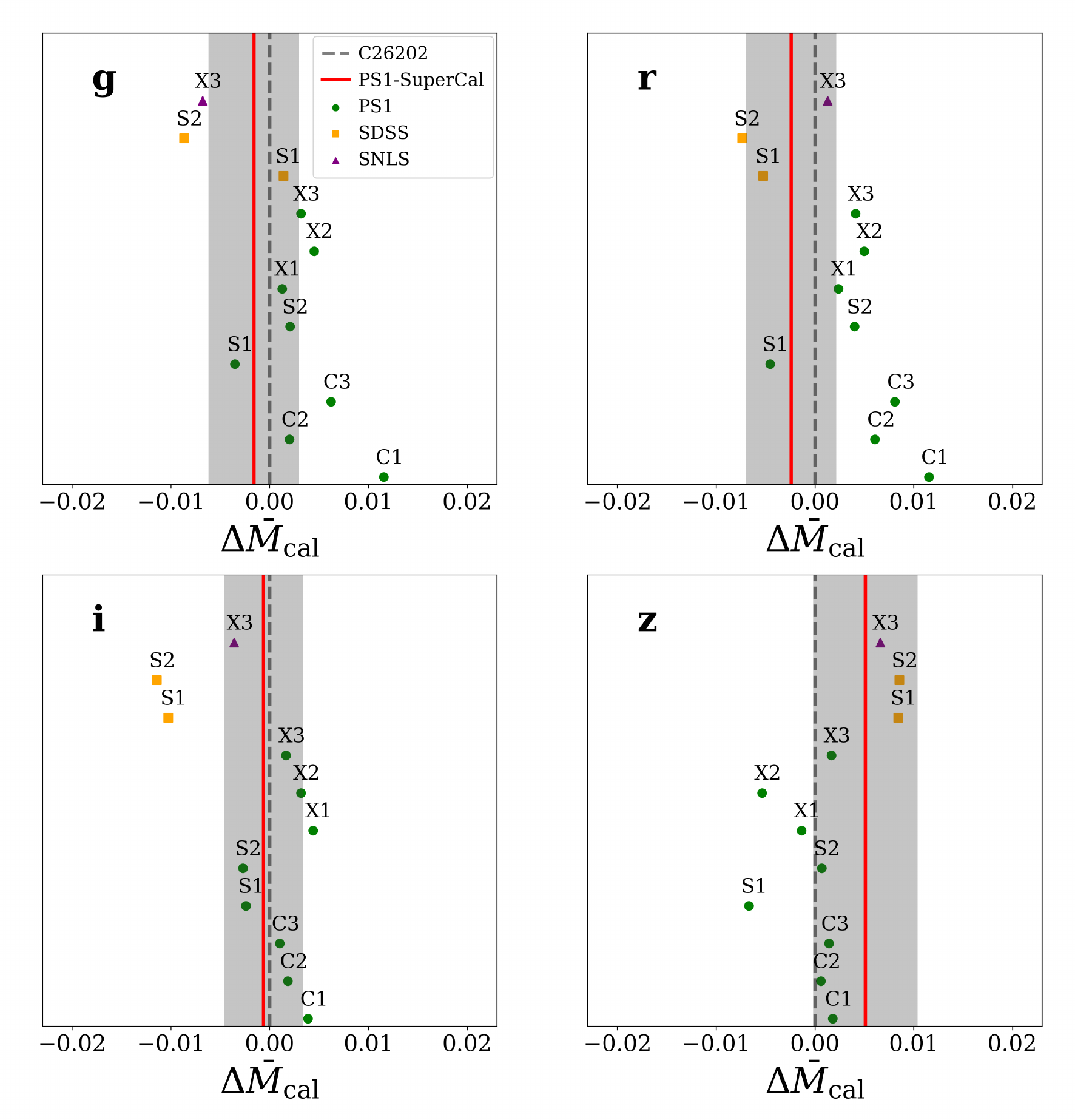}
\caption{The relative offsets in stellar magnitudes when comparing PS1, SDSS and SNLS to overlapping DES fields ($\bar{\Delta M_{{\rm cal}}}$). Offsets are further broken down by field. In each panel, $\bar{\Delta M_{{\rm cal}}}$ for the HST Calspec magnitude of C26202 is defined to be zero. Each of the points are determined from a comparison of DECam and external survey photometry accounting for difference in filter transmission functions. SNLS and SDSS are shown for reference, however it is only PS1 that is used to determine the goodness of the calibration. The vertical red line is the mean of the PS1-DES overlap (green points) shifted by the PS1 offset to SuperCal. The grey area represents the quadrature sum of the uniformity uncertainty and the SuperCal uncertainty in absolute calibration (\S~\ref{sys:cal})}.
\label{fig:calibration}

\end{centering}
\end{figure*}

The FGCM catalog is calibrated to the AB system (\citealt{oke}) using measurements of the HST CalSpec standard C26202. As detailed in \cite{burke17}, we compute synthetic magnitudes of C26202 by multiplying the CalSpec spectrum with the standard instrumental and atmospheric passbands used in the FGCM calibration\footnote{Y3A1 passbands from \cite{burke17}.} DECam filter transmission functions. The synthetic magnitudes are compared to the FGCM catalog magnitudes of C26202 for each passband, and the magnitude difference is applied to the FGCM catalog so that the observed and synthetic magnitudes of the standard are in perfect agreement. C26202 was chosen because it is located in `C3', which is one of the deep fields and has been observed over 100 times during the course of the survey. C26202 is sufficiently faint to avoid saturation and is observed in a similar range of seeing conditions to that of the DES-SN dataset. Other CapSpec standards in the DES footprint are either
saturated, or were observed with short exposures under
twilight conditions.
We do not find any dependence in the corrected, top-of-the-atmosphere, fluxes of C26202 on airmass, sky brightness, CCD number of the observation, or exposure time. 

A secondary method of calibrating the FGCM catalog is to cross-calibrate with catalogs from other surveys that are also tied to the AB system. Using tertiary standard stars in $8$ of the $10$ of the DES-SN fields (DES Fields: C1,C2,C3,S1,S2,X1,X2,X3) that overlap with the footprint of other surveys, we measure the calibrated brightness differences for stars observed by both surveys, and compare these differences to predictions using a spectral library
and known filter transmissions. We define $\Delta M_{{\rm cal}}$ as the offset between the predicted and observed brightness differences for stars with the same color as the Calspec standard C26202. 
In Figure~\ref{fig:calibration}, we examine the mean difference ($\bar{\Delta M_{\rm cal}}$) for several groupings of overlapping calibration stars. 
For comparison, we examine the agreement between DES and PS1 (green), DES and SDSS (orange), and DES and SNLS (violet). We also define PS1-SuperCal (red) as the agreement between DES and PS1, if the absolute calibration of PS1 were shifted by the weighted average of differences between the PS1, SDSS and SNLS calibration (see \citealt{supercal} for explanation).  

\cite{burke17} apply FGCM to the DECam images and achieve a calibration uniformity across the sky of $\sim6$ mmag. As a crosscheck for our SN fields, we quantify the relative consistency the DES-SN fields from the standard deviation of PS1-DES $\Delta M_{{\rm cal}}$, which is $4.1,4.3,2.5,3.1$ mmags in the $g, r, i, z$ bands respectively. The observed consistency between PS1 and DES is 2-4~mmag, which shows that $\sim6$ mmag is a conservative estimate of the relative calibration uncertainties due to non-uniformity.
Lastly, the observed offsets of stellar magnitudes between PS1, SDSS, and SNLS shown in Figure~\ref{fig:calibration} are consistent with the scatter seen in \cite{supercal}; these differences are shown for reference and are not used in this analysis.

\subsection{SN Photometry}
\label{photometry}

The light curves used in this analysis are provided by B18-SMP, which measures SN brightnesses by adopting a scene modeling approach. In \SMP, a variable transient flux and temporally constant host-galaxy are forward modeled simultaneously. B18-SMP test the accuracy of the \SMP\ pipeline by processing a sample of 10,000 realistic SNe~Ia light curves that were injected as point sources onto DECam images (`fake SNe'). Upon comparison of input and measured fake SNe fluxes, B18-SMP find that biases in the photometric pipeline are limited to 3~mmag (see Figure~3 of B18-SMP).

Analyzing fakes near bright galaxies, B18-SMP also find that the photometric scatter increases with the local surface brightness (denoted ``the host SB dependence''). This increase is similar to what was observed in DiffImg (\citealt{kessler15}). The host SB dependence is accounted for by scaling our photometric uncertainties of fake SNe near bright host-galaxies to match the observed scatter in \SMP\ flux residuals.  This scaling is determined as a function of host-galaxy surface brightness ($m_{\rm SB}$):
\begin{equation}
\label{eq:ssmp}
\hat{S}_{\rm SMP}(m_{\rm SB}) = \frac{{\rm RMS}[~(F_{\rm true}~-~F_{\rm SMP})~/~ \sigma_{\rm Ref}~]_{\rm fake}}{\langle \sigma_{\rm SMP}~/~\sigma_{\rm Ref}\rangle_{\rm fake}}
\end{equation}
where RMS is the root-mean-square in a bin of $m_{\rm SB}$, $\sigma_{\rm SMP}$ is the \SMP\ flux uncertainty, $\langle\rangle$ indicates an average in the $m_{\rm SB}$ bin, $\sigma_{\rm Ref}$ is the calculated uncertainty based on observing conditions (zero point, sky noise, PSF), $F_{\rm SMP}$ is the fit flux from \SMP, and $F_{\rm true}$ is the input flux of the fake SN. The size of $\hat{S}_{\rm SMP}(m_{\rm SB})$ can be seen in Figure~5 of B18-SMP and can be as large as 4 at $m_{\rm SB}=21$. These corrections are applied directly to the DES-SN sample.

After \SMP, there is an additional set of SED-dependent chromatic corrections made to the DES SN~Ia fluxes, similar to the corrections made to the stellar fluxes discussed in Section \ref{starcat}. The impact of these corrections is presented in \cite{Lasker18}, and is discussed here in Section \ref{sys:cal}. One potential issue is the validity of the chromatic corrections applied to the SN fluxes whose color range ($-1.0<g-i<2.2$) is redder than that of the majority of tertiary calibration stars ($0.2< g-i < 3$), and is shown in Figure~\ref{fig:colorhist}. For $g-i < 0.2$, there is a drop-off in tertiary standard star counts as the star distribution enters the realm of blue horizontal branch stars and white dwarfs. While we do not have the statistics to validate the 6~mmag calibration uncertainty for the bluest stars ($-1.0<g-i < 0.2$), we assume that chromatic corrections are valid for SN~Ia fluxes in this color range.  The chromatic corrections applied to the tertiary standards in the color range of $g-i \in [0,3]$ show no significant trends at the bluest colors and thus we have confidence in applying the corrections to the fraction of bluest SN~Ia epochs in the color range $g-i \in [-1,0]$.

\subsection{Redshifts}
\label{redshifts}

Redshifts for the DES subset are presented in \cite{D'Andrea18}. Redshifts of the low-redshift sample are obtained from their respective surveys to which we make peculiar velocity corrections. The corrections due to coherent flows of SN host galaxies has been performed in the same manner as S18. Peculiar velocities are calculated using the matter density field calibrated by the 2M$++$ catalog (\citealt{2mass}) out to $z \sim$ 0.05, with a light-to-matter bias of $\beta$ = 0.43 and a dipole as described in \cite{Carrick15}. We adopt the error in peculiar velocity correction of $250 ~{\rm km/s/Mpc}$ motivated by dark matter simulations of \cite{Carrick15} as well as from the comparison of low-redshift and intermediate redshift SNe scatter described in S18.

The redshifts of host galaxies used in this analysis are typically reported with an accuracy of $\sim 10^{-4}$ for \lowz~and to $\sim 5 \times 10^{-4}$ for intermediate-redshift. For 71 SNe in the DES subset, a host-galaxy redshift was not obtained and redshifts were determined from the SN spectrum, resulting in redshift uncertainty $\sim 5 \times 10^{-3}$. These redshift uncertainties propagate to SN scatter in distance. However, more important than the statistical uncertainty is the possibility of a systematic shift in redshift due to cosmological effects. A systematic shift could be caused, for example, by a gravitational redshift due to the density of our local environment (\citealt{Calcino17}).  \cite{Wotjak15} show the expected distributions for typical environments in \LCDM~can be described by a one sigma fluctuation from the mean potential with a shift of $\Delta z \approx 2 \times 10^{-5}$.

\subsection{Light Curve Fits}
\label{sec:light curves}

In order to obtain distance moduli ($\mu$) from SN~Ia light curves, we fit the light curves with the SALT2 model (\citealt{Guy10}) using the trained model parameters from B14 over an SED wavelength range of $200-900$nm. We select passbands whose central wavelength, $\bar\lambda$, 
satisfies $280 < \bar\lambda/(1+z) < 700$nm, and we select epochs satisfying -15 to +45 days with respect to the epoch
of peak brightness in the rest frame. We use the \texttt{SNANA} implementation (\citealt{Kessler09SNANA}) based on \texttt{MINUIT} (\citealt{MINUIT}), and we use the \texttt{MINOS} option for the fitted parameter uncertainties. A discussion about techniques used to avoid pathological fits is described in Appendix \ref{minuit}.

\begin{figure}
\begin{centering}

\includegraphics[page=1,trim={0 18cm 0 2.2cm},clip,width=0.525\textwidth]{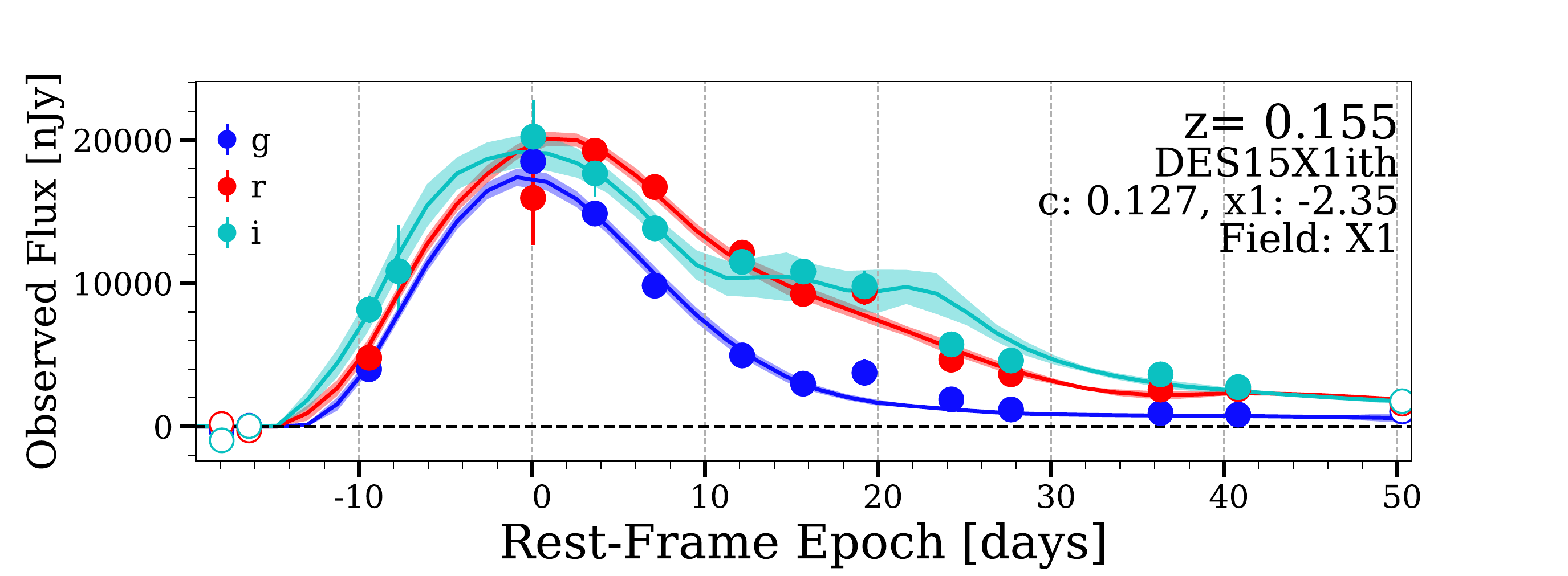}
\includegraphics[page=2,trim={0 18cm 0 2.2cm},clip,width=0.525\textwidth]{representativelc6_withc_and_x1.pdf}
\includegraphics[page=3,trim={0 18cm 0 2.2cm},clip,width=0.525\textwidth]{representativelc6_withc_and_x1.pdf}
\par\medskip\medskip
\caption{Representative light curves of the \SAMPLENAME~Spectroscopic sample with photometric data determined with \SMP\ (points). SALT2 fits to the light curve, are overlaid (curves) and fitted color and stretch values are shown. There is no $g$-band in the bottom panel because z=0.829 is beyond the range of the B14 $g$-band model. Supernovae with C3 (or X3) in the name are found in deep fields, the remaining SNe are found in the shallow fields. Open points are excluded from the SALT2 fits.}
\label{representativelc}

\vspace*{+0.2in} 
\end{centering}
\end{figure}

\begin{figure}
\begin{centering}


\includegraphics[width=0.51\textwidth]{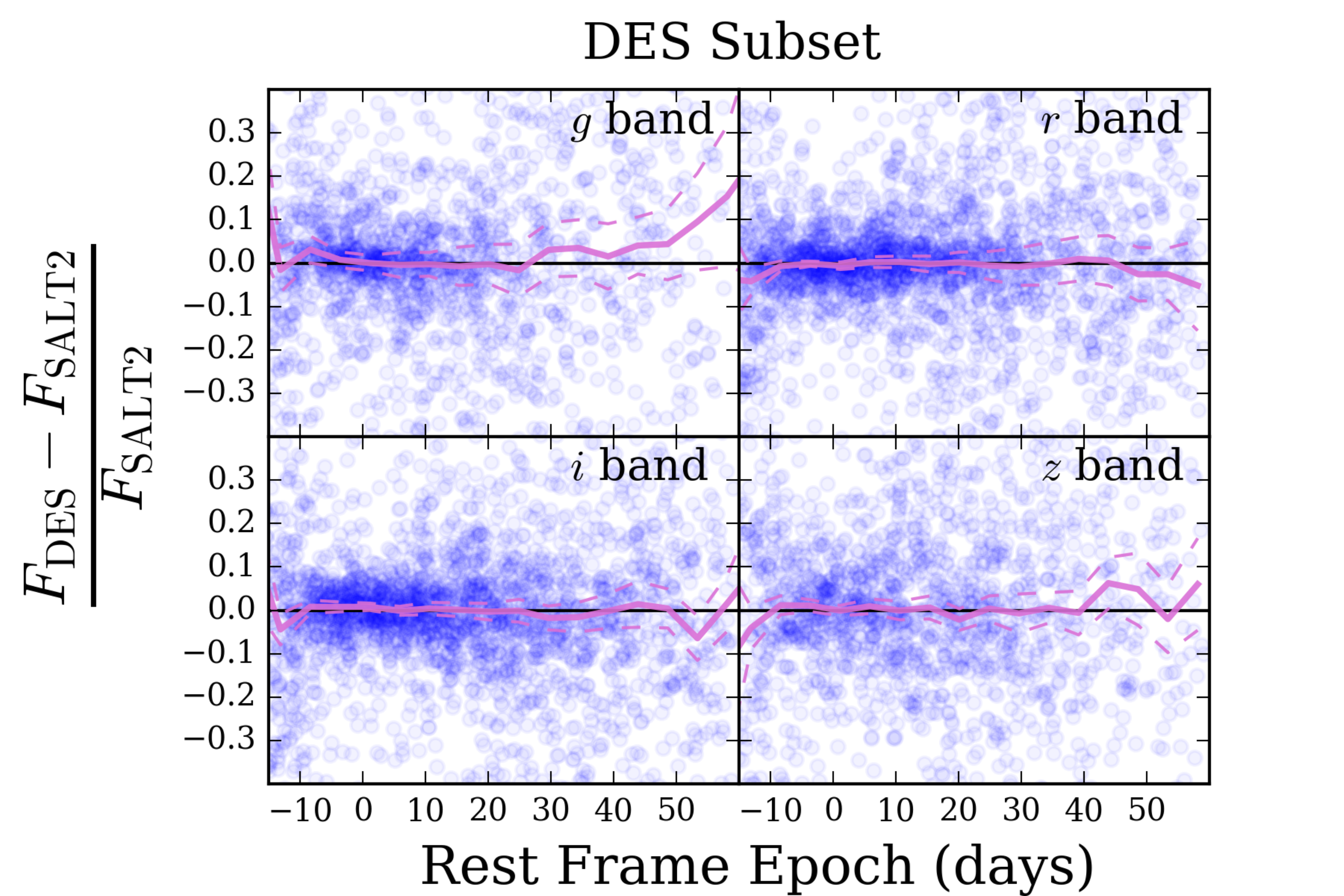}
\medskip

\includegraphics[width=0.51\textwidth]{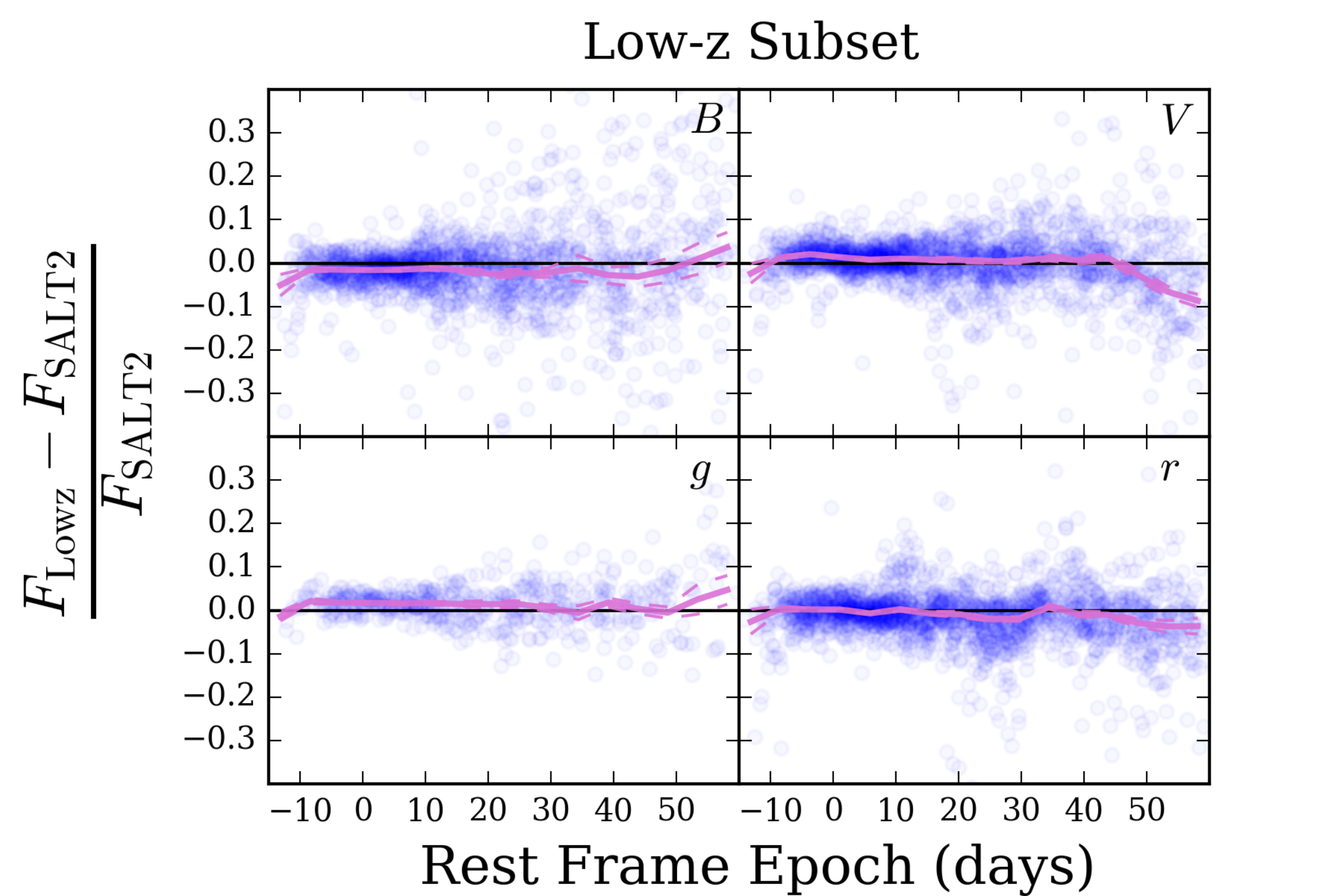}
\par\medskip\medskip
\caption{Fractional flux residuals to the best fit SALT2 light curve model. Top: the \SAMPLENAME~Spectroscopic sample in the four DES filter bands [$griz$]. Bottom: the \lowz\ subset where photometric observations have been grouped by filters with similar wavelength coverage [$BVgr$]. $F_{\rm DES}$ and $F_{Lowz}$ are the SN flux from the data, $F_{SALT2}$ is the flux of the best-fit SALT2 model. The mean of each distribution is shown in solid curve and the uncertainty on the mean is shown as dashed curves.}
\label{saltmodel}

\vspace*{+0.01in} 
\end{centering}
\end{figure}

Each light curve fit determines parameters color $c$, stretch $x_1$, the overall amplitude $x_0$, with $m_B \equiv -2.5\log_{10}(x_0)$, and time of peak brightness $t_0$ in the rest-frame B-band wavelength range. In addition, we compute light curve fit probability $P_{\rm fit}$, which is the probability of finding a light curve data$-$model $\chi^2$ as large or larger assuming Gaussian-distributed flux uncertainties. In Figure~\ref{representativelc}, three representative DES-SN light curves are shown with overlaid light curve fits using the SALT2 model. Normalized flux residuals to the SALT2 light curve model for the \SAMPLENAME~sample are shown in Figure \ref{saltmodel}. Both the DES subset and \lowz\ subset SALT2 model fluxes for all rest-frame passbands are consistent to within $<2$mmag. Calibration offsets to the SALT2 model are adopted as systematic uncertainty; this is described in Section \ref{sys:cal}.   All light curve fit parameters for the \SAMPLENAME~sample are publicly available in machine-readable format as described in Appendix \ref{publiccodes} and in Table \ref{lcfittable}. 

\subsection{Selection Requirements}
\label{sec:selection}

For this analysis, we require all SNe~Ia to have adequate light curve coverage in order to reliably constrain light curve fit parameters and we limit ourselves to a model-training range of SN~Ia properties that limit systematic biases in the recovered distance modulus measurement. The sequential loss of SNe~Ia from the sample due to cuts is shown in Table \ref{Tab:cuts}. We start by requiring $z>0.01$ and our light curve fits to converge. We define $T_{\rm rest}$ as the number of days since $t_0$ in the rest frame of the SN. \cite{DaiWang16} showed that poorly sampled light curves can result in large Hubble residual outliers even though the fit $\chi^2$ shows no indication of a problem. Thus, we require an observation before peak brightness ($T_{\rm rest}<0$), an observation at least 10 days after peak brightness ($T_{\rm rest}>10$), and an observation with S/N $>5$ in at least two bands.  We require $-3 < x_1 < 3$ and $-0.3 < c < 0.3$ over which the light curve model has been trained (\citealt{Guy10}). For the low redshift samples we require limited Milky Way extinction following B14 and S18, $E(B-V)_{\rm MW}< 0.25$.  The DES-SN Fields have low MW extinction and thus the $E(B-V)_{\rm MW}$ cut has no effect.

\begin{figure*}
\begin{centering}
	\centering
    \textbf{DES-SN Sample Data and Simulations}\par\medskip
  \includegraphics[width=0.98\textwidth]{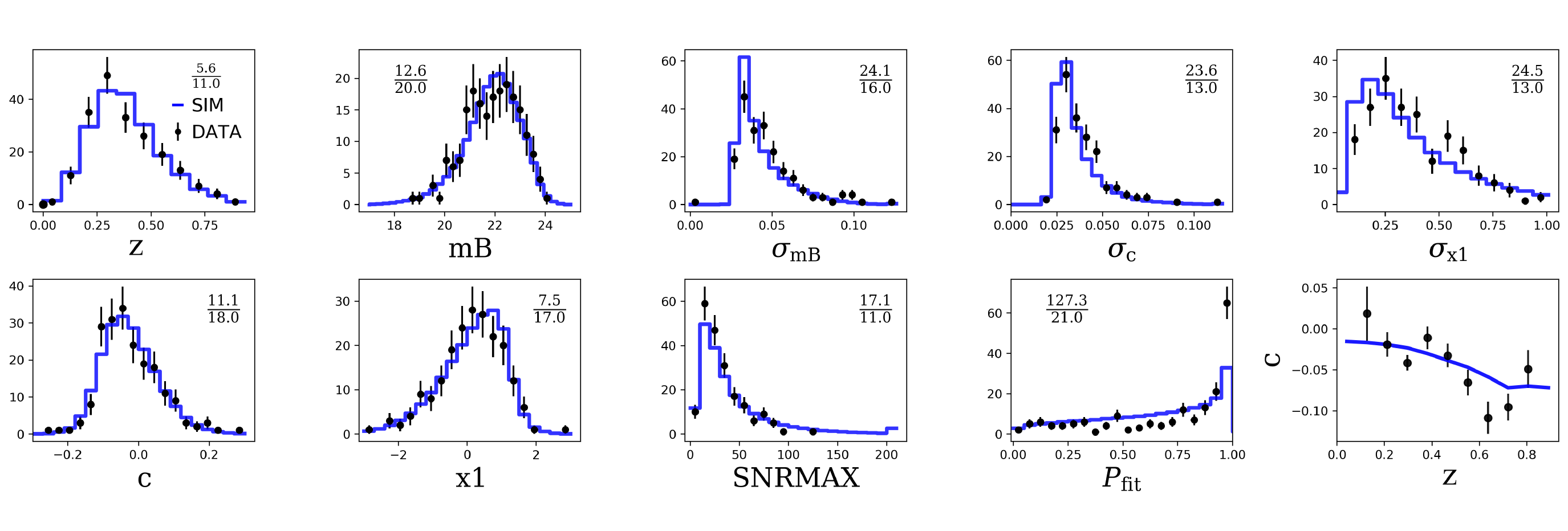}
  		\\
        
        \medskip
        
      \textbf{Low-$z$ Sample Data and Simulations}\par\medskip
  \includegraphics[width=0.98\textwidth]{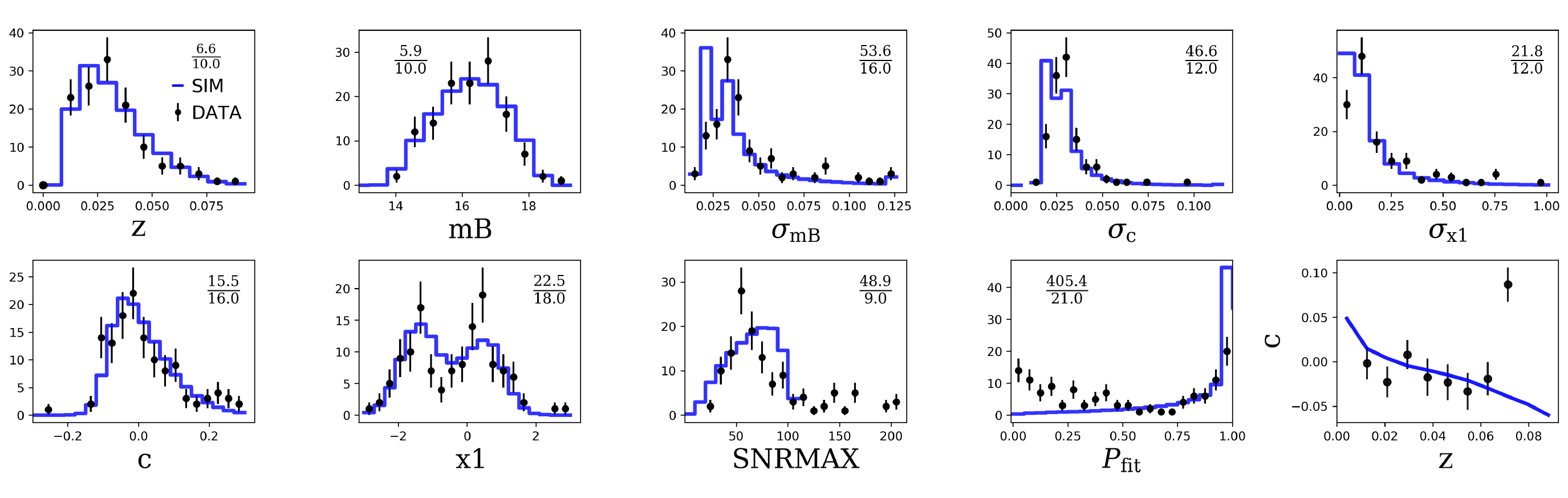}
\caption{Top: DES subset (black points) is compared to G10 simulations (blue histogram) that are used for bias correction. The simulations have $\sim$600,000 SNe for each subset but the histograms are scaled to match the size of the \SAMPLENAME~dataset. The distributions shown are: redshift in the CMB reference frame ($z$), the SALT2 $m_B$, uncertainty in $m_B$, stretch $x_1$, uncertainty in $x_1$, color $c$, the uncertainty in $c$, the maximum SNR of the light curve, the light curve fit probability ($P_{\rm fit}$), and lastly $c$ as a function of redshift. Bottom: Same as top but for the external low-$z$ sample. The fractions shown in each panel are $\chi^2/$ndof.}
\label{fig:overlay}
\end{centering}
\end{figure*}

S18 placed a $P_{\rm fit}>0.001$ cut on the low-redshift sample. While decreasing the fit probability cut to agree with Pantheon gained us 20 SNe~Ia, those additional SNe~Ia come in a region of parameter space that is poorly modeled by our simulations (see $P_{\rm fit}$ panel of Figure~\ref{fig:overlay}). Additionally, we find that applying a more conservative cut of $P_{\rm fit}>0.01$ to both the DES and low-$z$ subsets resulted in similar statistical constraints on distance. The distribution of low-$z$ sample light curve parameters after quality cuts is shown in the bottom half of Figure~\ref{fig:overlay}. 

In the second to last row of Table \ref{Tab:cuts} (`Valid BiasCor'), a few SNe are lost due to their SN properties falling within a region of parameter space for which the simulation does not have a bias prediction. Bias corrections are discussed in detail in Section \ref{biascor}. 

Each SN cosmology analysis that has utilized the historical CfA and \mbox{CSP-1} \lowz~samples has dealt with the fact that their Hubble diagram residuals have non-Gaussian tails that are discarded from the cosmological fit. In the last row of Table \ref{Tab:cuts} (`Chauvenets criterion'), we place a final set of cuts before running cosmological parameter fits.  This is the same cut on Hubble diagram residuals that was made in S18 of $3.5\sigma$.

\begin{deluxetable}{clll}
\centering
\tablecolumns{4}
\tablewidth{18pc}
\tablecaption{\# SN After Iteratively Applied Cuts}
\tablehead {
\colhead {}                &
\colhead {DES-SN}   &
\colhead {Low-$z$}  &
\colhead {Total SN}  \\
\colhead {Requirement}                &
\colhead {\# [Cut]}   &
\colhead {\# [Cut]}  &
\colhead {\# [Cut]} } 
\startdata

Initial 	& 251\footnote{Discovered by \texttt{DiffImg} and spectroscopically confirmed (\citealt{D'Andrea18}).}  & 333\footnote{CfA3, CfA4, and CSP-1 samples.}  &  542 \\

\hline \\

$z>0.01$ & 251 [0] & 261 [72] & 512 [72] \\
Fit~Convergence & 244 [7] & 257 [4] & 501 [11] \\
$S/N > 5$ in 2 bands & 239 [5] & 250 [7] & 439 [12]   \\
$T_{\rm rest}>10$, $T_{\rm rest}<0$& 230 [9] & 248 [2] & 481 [11]   \\

$E(B-V)_{\rm MW} < 0.25$ & 230 [0] & 243 [5] & 473 [5]   \\

$-0.3 < c < 0.3$ & 224 [6] & 170 [73] & 394 [79] \\
$-3 < x_1 < 3$ & 221 [3] & 150 [20] & 371 [23] \\
$\sigma_{x_1} < 1$ & 211 [10] & 150 [0] & 361 [10] \\

$P_{\rm fit} > 0.01$ & 208 [3] & 127 [23] & 335 [26]   \\
Valid BiasCor & 207 [1] & 125 [2] & 332 [3]   \\
Chauvenet’s criterion & 207 [0] & 122 [3] & 329 [3]  \\

\\
\hline \\
Cosmo. Sample 	& 207 & 122 &  329 \\

\enddata
\label{Tab:cuts}
\vspace{.18in}
\end{deluxetable}

\subsection{Host-galaxy Stellar Masses}
\label{hostmass}

Previous analyses of large SN~Ia samples have found a correlation between standardized SN luminosities and host-galaxy properties (\citealt{Gallagher08}; \citealt{Kelly10}; \citealt{Lampeitl10}; \citealt{Sullivan10}, low-$z$: \citealt{Childress13} and \citealt{Pan14}, JLA: B14, PS1: S18). Here we focus on the stellar mass ($\mathcal{M}_{\rm stellar}$) ratio of the host galaxy
\begin{equation}
\mathcal{R}_{\rm host} = \log_{10}(\mathcal{M}_{\rm stellar}/\mathcal{M}_{\sun}),
\end{equation}
as this quantity has been used in SN-cosmology analyses to correct standardized luminosities since \cite{Conley11}.  

Using catalogs from Science Verification DECam images \citep{SVcat},
the directional light radius method \citep{Sullivan06,Gupta16}
is used to associate a host galaxy with each SN~Ia.
The stellar masses of the DES-SN host galaxies are derived from 
fitting SEDs to $griz$ broadband fluxes with {\tt ZPEG} \citep{ZPEG},
where the SEDs are generated with Projet d'Etude des GAlaxies par Synthese Evolutive (\texttt{PEGASE}: \citealt{PEGASE}).

We define $\dmuHost$ to be a distance modulus correction, often referred to as an SN mag-step correction, between SNe with $\mathcal{R}_{\rm host}<\mathcal{R}_{\rm step}$ and SNe with $\mathcal{R}_{\rm host}>\mathcal{R}_{\rm step}$:
\begin{equation}
\label{Eq:hm}
\dmuHost = \gamma \times [1+e^{(\mathcal{R}_{\rm host}-\mathcal{R}_{\rm step})/0.01}]^{-1}-\frac{\gamma}{2},
\end{equation}
where $\mathcal{R}_{\rm step} = 10$. Here, the magnitude of $\dmuHost$ is determined by fitting for $\gamma$ where $\dmuHost$ is between [$+\gamma/2$, $-\gamma/2$], with a rapid transition near $\mathcal{R}_{\rm host}=10$. We find that because we have characterized $\dmuHost$ as a step function, its dependence on host mass uncertainties is weak, and therefore uncertainties are not accounted for in this calculation. Additionally, because S18 found little dependence between $\mathcal{R}_{\rm step}$ and cosmological parameters, we fix the location in our cosmology fit. While SN~Ia host-galaxy properties may change with redshift, we could allow for $\gamma$ to have a redshift dependence, and this possibility is discussed in Section \ref{results}. 

\begin{figure}
\includegraphics[width=0.49\textwidth]{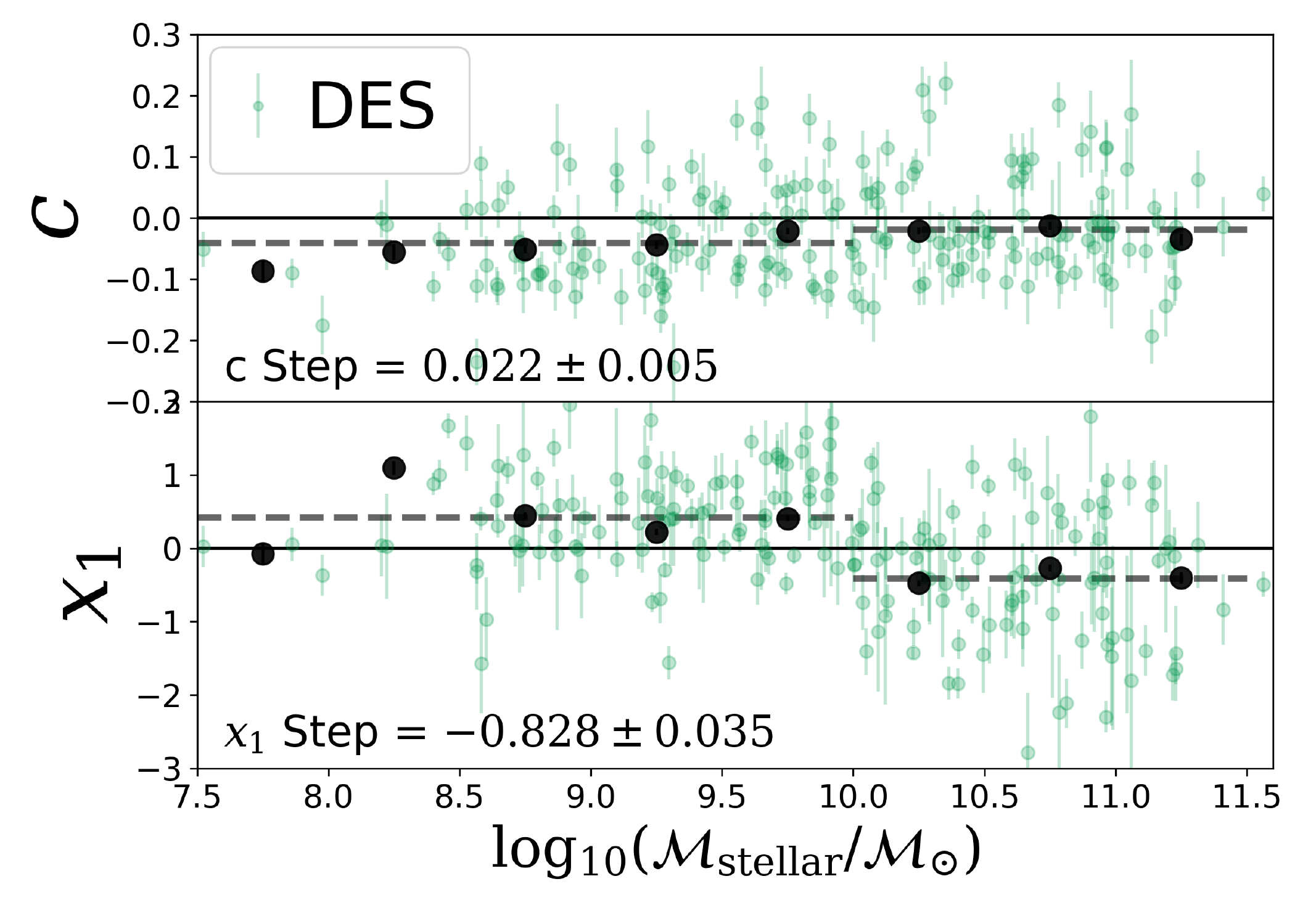}
\caption{Relations of color, stretch with host-galaxy stellar mass for the DES SN~Ia subset before bias corrections have been applied. Steps across $\log_{10}(\mathcal{M}_{\rm stellar}/\mathcal{M}_{\rm sun})$ are shown in the dashed lines. Binned data points are also shown.}
\label{fig:hostmassbefore}
\end{figure}

For galaxies that \texttt{ZPEG} was not able to determine a host mass, we first confirm that the hosts are faint and have not been mis-identified, and then we assign them to the low-mass bin. For the DES subset, there are 116 host galaxies with $\mathcal{R}_{\rm host}< 10$ and 91 host galaxies with $\mathcal{R}_{\rm host} > 10$. In Figure~\ref{fig:hostmassbefore} we show the distributions of color and stretch as a function of $\mathcal{R}_{\rm host}$. Correlations between SN~Ia light curve parameters and $\mathcal{R}_{\rm host}$ have been reported in previous analyses (B14, S18) and are characterized as an average difference (step) for events with $\mathcal{R}_{\rm host}< 10$ and $\mathcal{R}_{\rm host}> 10$. As shown in Figure~\ref{fig:hostmassbefore}, we find steps in stretch, $\Delta x_1=-0.828\pm0.035$, and color, $\Delta c=0.022\pm0.005$. These correlations are significantly larger than what was observed by S18 ($\Delta x_1=-0.210\pm0.041$ and $\Delta c=0.012\pm0.004$). While selection effects may play a role in this difference, a comprehensive study is left to a future work.

\subsection{Simulations}
\label{simulations}

Here we discuss the use of fakes so that our simulations incorporate the subtleties of the photometric pipeline that cannot be computed from first principles.  In addition, we describe here the simulations that are used for determining bias corrections. Because 11 different types of simulations are used throughout the analysis and validation, we refer to Table \ref{tab:simlist}, which lists key attributes for each.

\begin{deluxetable*}{clcccl}
\tablecolumns{6}
\tablewidth{.9\textwidth}
\tablecaption{Simulations Used in DES-SN3YR}
\tablehead {
\colhead {}                &
\colhead {Description}                &
\colhead {Samples}                &
\colhead {Scatter Model}                &
\colhead {Size}  &
\colhead {Used In}
}

\startdata
\\
&\textbf{$\mu$ bias}\tablenotemark{a} \\

1&Fiducial & DES+low$z$ & G10 \& C11 & $\sim$1,300,000 SNe& \S\ \ref{simlight curves} \& \ref{cosmology}, Figure \ref{fig:overlay}\\
2&Spec. Eff. Syst. & DES+low$z$ & G10 \& C11 & $\sim$1,300,000 SNe& \S\ \ref{cosmology} \& \ref{sys:speceff}, Figure \ref{fig:speceff}\\
3&$\mu$-bias Cosmo. & DES+low$z$ & G10 \& C11 & $\sim$1,300,000 SNe& \S\ \ref{cosmology} \& \ref{sys:biascor}, Figure \ref{fig:sysmudif}\\
4&5\% Flux Err. & DES+low$z$ & G10 \& C11 & $\sim$1,300,000 SNe& \S\ \ref{cosmology} \& \ref{sys:phot}, Figure \ref{fig:sysmudif}\\
5&$c$, $x_1$ Parent & DES+low$z$ & G10 \& C11 & $\sim$1,300,000 SNe& \S\ \ref{simlight curves}, \ref{cosmology} \& \ref{sys:parent}, Figure \ref{fig:sysmudif}\\
6& Two $\sigma_{\rm int}$ & DES+low$z$ & G10 \& C11 & $\sim$1,300,000 SNe& \S\ \ref{sys:intrinsicscatter}\\

\\
\\
&\textbf{Validation}\tablenotemark{b} \\
7&Fake Sample\tablenotemark{c} & DES & N/A & 100,000 SNe& \S\ \ref{fakes}\\
8&Fake $\mu$ bias& DES & N/A & $\sim$700,000 SNe& \S\ \ref{fakes}, Figure \ref{fig:fakesims}\\
9&Stat & DES+low$z$ & G10 & 200xDES-SN3YR& \S\ \ref{analysisbiases}\\
10&Zero point\tablenotemark{d} & DES+low$z$ & G10 & 200xDES-SN3YR& \S\ \ref{analysisbiases}, Figure \ref{fig:validation2d}\\
11&Scatter Model & DES+low$z$ & G10 \& C11 & 100xDES-SN3YR& \S\ \ref{analysisbiases}\\

\\
\enddata
\tablenotetext{a}{Simulations used to compute distance bias ($\mu$-bias) corrections (Section \ref{simulations}). }
\tablenotetext{b}{Simulations used in the validation of the analysis (Section \ref{validation}).}
\tablenotetext{c}{Intrinsic scatter set to zero. The simulated fluxes are inserted into DECam images as point sources.}
\tablenotetext{d}{For each band and each sample, a random zero point offset is chosen from Gaussian PDF with $\sigma=0.02$~mag.\\}
\label{tab:simlist}
\vspace{.18in}
\end{deluxetable*}

\subsubsection{Fakes overlaid on images}
Ideally, a large sample of fakes would be used for characterizing cosmological distance biases. However, our sample of 10,000 fakes that have been processed with \SMP\ is insufficient for multiple reasons. First, 10,000 fakes is more than an order of magnitude smaller than what is needed for the bias-correction sample used in the BBC method. Second, \SMP\ (or other similar methods) is far too computationally intensive for the large number of systematic iterations that are needed to test against varying SN properties and assumptions. These tests include multiple iterations of bias corrections, with varying properties, parent populations, and assumptions. 
For the many analysis iterations that are needed, it is vital to have a rapid method for obtaining simulated catalog photometry that approximates \SMP. Using the sample of fakes processed by \SMP, we tune our catalog simulations to replicate \SMP\ flux uncertainties. As shown in Figure~2 and Eq.~13 of K18, the SN flux uncertainties of the simulated SNe are scaled ($\hat{S}_{\rm sim}$) as a function of host-galaxy surface brightness by the ratio between the observed scatter in the fakes relative to the `observed' scatter in the simulation. As a result we obtain simulations of DES-SN with the same distribution of photometric uncertainties found in our real dataset and that can be used for rapid analysis iterations.

\subsubsection{Simulated light curves for bias corrections}
\label{simlight curves}
We use catalog-level simulations of large samples of SNe~Ia to model the expected biases in measured distances that follow from the known selection effects and our light curve analysis. The simulations of the DES-SN and low-redshift samples used for this analysis follow the description of K18. For individual events, distance biases can reach 0.4~mag as shown in Figure~9 of K18, and it is therefore imperative to have accurate simulations in order to predict biases. The simulation utilizes \texttt{SNANA} and, as detailed in Figure~1 of K18, consists of 3 main steps: 1) generating a SN source for each epoch (Source model), 2) applying instrumental noise (Noise model), and 3) simulating DES-SN observing and selection (Trigger model). Here we discuss each of these steps briefly along with specific choices made for this analysis

\textbf{Source model:} Our simulations first generate rest frame SN~Ia SEDs with the SALT2 model from B14. The model includes SN Ia parent populations of color and stretch, intrinsic luminosity variations, and cosmological effects.

For the DES subset, we test the parent distributions of $c$ and $x_1$ found in Table 1 of \cite{scolnickessler16} (hereafter SK16) and find that the High-z row, representative of the populations of all recent high-z surveys combined (SDSS, SNLS, PS1), results in the best agreement in the observed distributions of light curve parameters when comparing to our DES dataset. 

For the \lowz\ subset we follow S18. We do not re-derive $x_1$ and $c$ parent populations after removal of the CfA1 and CfA2 samples, which compose less than $16\%$ of the low-redshift sample, because population parameters have little dependence on selection efficiencies.

A model of SN brightness variations, called `intrinsic scatter,'
is needed to account for the observed Hubble residual scatter
that exceeds expectations from measurement uncertainties.
Most cosmology-fitting likelihoods characterize the excess
Hubble scatter with an additional $\sigma_{\rm int}$ term added in 
quadrature to the measured distance uncertainty. From an
astrophysical perspective, this $\sigma_{\rm int}$ term is equivalent 
to an intrinsic scatter model described by a Gaussian profile
where each event undergoes a coherent fluctuation that
is 100\% correlated among all phases and wavelengths. 
Many previous analyses, however, have demonstrated that this 
simple coherent model does not adequately describe intrinsic scatter. 
Following K13, we simulate intrinsic scatter with two different 
intrinsic scatter models in order to investigate the sensitivity 
to bias corrections and to the $\sigma_{\rm int}$ approximation in the 
cosmology-fitting likelihood.

Our intrinsic scatter models include a combination of coherent 
(Gaussian $\sigma_{\rm int}$) variations, and wavelength-dependent SALT2 SED 
variations that introduce scatter in the generated SN~Ia colors.
From K13 the first model, ``G10,'' is based on \cite{Guy10} and describes
$\sim 70$\% of the excess Hubble scatter from coherent variations,
and the remaining scatter from wavelength-dependent variations.
The second model, ``C11,'' is based on \cite{Chotard11} and describes
$\sim 30$\% of the excess Hubble scatter from coherent variations,
and the remaining scatter from wavelength-dependent variations.

Cosmological effects are applied, which include redshifting, dimming, lensing, peculiar velocity, and Milky Way extinction. The simulations used for bias corrections are performed in \LCDM~($w=-1.0$, $\Omega_M=0.3$, $\Omega_k=0.0$). We integrate the redshifted SED with the DECam filter transmission functions to obtain true top-of-atmosphere DECam magnitudes.

\textbf{Noise model:} We simulate the DES-SN cadence and observing conditions (PSF, sky noise, zero point) using the catalog of DES-SN images. A sample of simulated SNe are drawn from 10,000 random sky locations over the DES-SN observing fields and for each epoch, the observing conditions are taken from the corresponding DES-SN image. For simulations of more than 10,000 events, sky locations are repeated. We assign a host-galaxy surface brightness and determine photometric uncertainties from PSF, sky, and zero point. A photometric uncertainty scaling as a function of $m_{\rm SB}$ (Sec Sec 5 of K18) is then applied. The final product of the noise model is a set of DECam fluxes and flux uncertainties.

\textbf{Trigger model:} The last step is to apply the DES-SN detection criteria and spectroscopic selection. We require two detections on separate nights within 30 days. The spectroscopic selection function for the DES subset ($E_{\rm spec}$) is determined as a function of peak $i$ band magnitude (Section 6.1 of K18).

The \lowz\ subset trigger model, which is detailed in Section 6 of K18, is based on the procedure developed in B14, S14, and S18, which assume that the \lowz~subset is magnitude limited. Separate spectroscopic selection functions are determined for each of the \lowz~surveys (CfA3, CfA4 and \mbox{CSP-1}). With the assumption of a magnitude limited sample, we are able to obtain good agreement between simulations and data for the distribution of observed redshifts as shown in Figure~\ref{fig:overlay}.  However, since it is unclear how selection was done for the low-redshift surveys and that it involved a targeted search of galaxies, we simulate as a systematic uncertainty the assumption that the \lowz~subset is in fact volume-limited. The determination of the \lowz~efficiency function and the implementation of the volume-limited assumption in simulations is discussed in detail in K18. 

For a volume-limited \lowz~subset, redshift evolution of color and stretch are interpreted as astrophysical effects rather than manifestations of Malmquist bias. 
This allows for the combination of the volume-limited assumption and the uncertainty in parent populations of color and stretch to be analyzed with a single simulation. The parent populations used for the simulations of the \lowz~subset are documented in Table \ref{Tab:parent}.

\subsubsection{Data-Simulation Comparisons}
\label{comparisons}
We discuss here the method for evaluating the quality of our simulations. To characterize the level of agreement between data and simulated distributions, we define the $\chi^2_{\rm p}$ between the simulation and data for each population parameter (p) from the comparison of a binned light curve fit parameter distribution of the data and the normalized binned distribution of the high statistics simulation as follows:

\begin{equation}
\label{chisqp}
\chi^2_{\rm p} = \sum_i (N^{\rm data}_i - R \times N^{\rm sim}_i)^2/N_i^{\rm data},
\end{equation}
\begin{equation*}
R = \sum N_i^{\rm data}/\sum N_i^{\rm sim},
\end{equation*}
for parameter bins $i$ and simulation normalization $R$. The simulations have sufficiently high statistics that we ignore statistical fluctuations in the simulations and only use the Poisson uncertainties in the dataset.  

The agreement between simulations and our DES-SN dataset is shown by comparing the distributions of light curve fit parameters and uncertainties, redshift, and maximum S/N among all epochs (\texttt{SNRMAX}) in Figure~\ref{fig:overlay}. For each subplot in Figure~\ref{fig:overlay} we report $[\chi^2_{\rm p}]$/$[$dof$]$. Although only the simulations using the G10 scatter model are shown, the distributions using C11 simulations are indistinguishable by eye. 

We find good agreement between the data and simulations for many of the observed parameters, but most notably in redshift (Figure~\ref{fig:overlay}). In simulating the DES subset, there was no explicit tuning of the redshift distribution. This gives us confidence in our models used to generate the simulations.

It is important to note that we obtain relatively poor agreement between the DES subset and simulations for the light curve fit probability ($P_{\rm fit}$) distribution. However, because the agreement for the  \texttt{SNRMAX} distribution is good, it is possible that more subtle modeling of photometric uncertainties is needed or that there is variation in the SN population that is not captured by a SALT2 model. Agreement between data and simulations for the \lowz~subset for \texttt{SNRMAX} and $P_{\rm fit}$ is worse than for the DES subset. This suggests the need for significant improvements in flux uncertainty modeling. In Section \ref{discussion:simulating} we discuss the need for improvements to simulations of SNe~Ia datasets.

\begin{deluxetable*}{lcccccc}
\tablecolumns{7}
\tablewidth{0.99\textwidth}
\tablecaption{Parent Populations Parameters table}
\tablehead {
\colhead {Description}& 
\colhead {Scatter}& 
\colhead { $c_{\rm peak}~~(\sigma_+,~\sigma_-)$}               &
\colhead { $\frac{dc}{dz}$}               &
\colhead {$x_{1 ~{\rm peak1}}~~(\sigma_+,~\sigma_-)$} &
\colhead {$x_{1 ~{\rm peak2}}~~(\sigma_+,~\sigma_-)$  }      &        
\colhead { $\frac{dx_1}{dz}$}               \\
\colhead {}& 
\colhead {Model}& 
\colhead { }               &
\colhead { }               &
\colhead { }               &
\colhead { }               &
\colhead {  }             
}
\startdata

DES Nominal & G10 & $-0.054~~(0.043,~0.101)$ &0& $0.973~~(1.472,~0.222)$& $0.000~~(0.000,~0.000)$&0\\
DES Systematic & G10 & $-0.065~~(0.044,~0.120)$ &0& $0.964~~(1.232,~0.282)$& $0.000~~(0.000,~0.000)$&0\\
DES Nominal & C11 & $-0.100~~(0.003,~0.120)$ &0& $0.964~~(1.467,~0.235)$& $0.000~~(0.000,~0.000)$ &0\\
DES Systematic &  C11 & $-0.112~~(0.003,~0.144)$&0 & $0.974~~(1.236,~0.283)$& $0.000~~(0.000,~0.000)$ &0\\
\\

\hline \\

Low-$z$ Nominal & G10 & $-0.055~~(0.023,~0.150)$  &-1& $0.550~~(1.000,~0.450)$& $-1.500~~(0.500,~0.500)$ &25\\
Low-$z$ Vol. Lim. & G10 & $-0.055~~(0.018,~0.150)$&-1& $0.200~~(1.000,~0.450)$& $-2.100~~(0.500,~0.500) $&25\\

Low-$z$ Nominal & C11 & $-0.069~~(0.000,~0.150)$&-1 & $0.550~~(1.000,~0.450)$&$-1.500~~(0.500,~0.500)$ &25\\
Low-$z$ Vol. Lim. &  C11 & $-0.047~~(0.000,~0.110) $&-1& $0.200~~(1.000,~0.050)$& $-2.100~~(0.500,~0.500)$ &25\\

\enddata
\tablecomments{Parent population parameters of color ($c$) and stretch ($x_1$) used in \texttt{SNANA} simulations for bias corrections. The low-$z$ $x_1$ distributions are modeled as two Gaussians with two peaks shown in the table. \\ \\}
\label{Tab:parent}
\bigskip
\end{deluxetable*}

\subsection{Cosmology}
\label{cosmology}
Here we discuss the analysis steps taken to extract cosmological distances, fit for nuisance parameters, and correct for expected biases. Additionally, we discuss the production of statistical and systematic distance covariance matrices. Finally, we discuss the cosmological parameter fitting process.

\subsubsection{BBC}
\label{bbc}
\label{biascor}

\newcommand{\LHBBC}{{\cal L}_{\rm BBC}}

\newcommand{\chisqBBC}{\chi^2_{\rm BBC}}

\newcommand{\zindex}{\cal Z}
\newcommand{\Dmuz}{\Delta_{\mu,\zindex}}
\newcommand{\muzAvg}{\langle\mu\rangle_{\zindex}}
\newcommand{\zAvg}{\langle z\rangle_{\zindex}}

\newcommand{\sigmui}{\sigma_{\mu,i}}
\newcommand{\sigmu}{\sigma_{\mu}}
\newcommand{\mumodeli}{\mu_{{\rm model},i}}
\newcommand{\mumodelAvg}{\langle\mumodeli\rangle}
\newcommand{\mumodelzAvg}{\langle\mumodeli\rangle_{\zindex}}
\newcommand{\NZBIN}{20}

We use the ``BEAMS with Bias Corrections (BBC)'' fitting method \citep[KS17]{kesslerscolnic17} to convert the light curve fit parameters ($m_B$, $x_1$, $c$) 
into bias-corrected distance modulus values in \NZBIN\ discrete redshift bins, 
and to determine nuisance parameters ($\alpha$, $\beta$, $\gamma$). 
This BBC fit uses a modified version of the Tripp formula (\citealt{Tripp98}) where the measured distance modulus 
($\mu$) of each SN is determined by
\begin{equation}
\label{Eq:tripp}
  \mu = m_B - M + \alpha x_1 - \beta c + \dmuHost + \dmuBias~.
\end{equation}
$\alpha$ and $\beta$ are the correlation coefficients of $x_1$ and $c$ with luminosity,
respectively, and $M$ is the absolute magnitude of a fiducial SN~Ia with $x_1=0$ and $c=0$. 
Following \cite{Conley11}, we include $\dmuHost$ (Eq.~\ref{Eq:hm}) which depends on $\gamma$. 
The bias correction, $\dmuBias$, is determined from large simulations (K18) and is computed
from a 5-dimensional grid of $\{ z, x_1, c, \alpha, \beta \}$.

The BBC likelihood ($\LHBBC$) is described in detail in Eq.~6 of KS17.
For the \SAMPLENAME\ sample of spectroscopically classified events, 
we set the core collapse SN probability to zero and $\LHBBC$ simplifies to
\begin{eqnarray}
\label{eq:chisqBBC}
  & & -2\ln(\LHBBC) \equiv  \chisqBBC =  \nonumber \\
  & & \sum_i \left[ (\mu_i - \mumodeli - \Dmuz)^2/\sigmui^2  + 2\ln(\sigmui) \right]~, 
\end{eqnarray}
where the $i$-summation is over SN~Ia events,  $\mu_i$ is the distance modulus of the $i^{th}$ SN (Eq.~\ref{Eq:tripp}),
$\mumodeli$ is the distance modulus computed from redshift $z_i$ 
and an arbitrary set of reference cosmology parameters ($\Omega_M=0.3,~\Omega_\Lambda=0.7,~w=-1$),
and $\Dmuz$ is the fitted distance offset in redshift bin-index $\zindex$
determined from $z_i$.
To obtain similar distance constraints in each $\zindex$ bin,
the redshift bin size is proportional to $(1+z)^{n}$ with $n=6$, and we use \NZBIN\ $\zindex$ bins.

Dropping the $i$ index in Eq.~\ref{eq:chisqBBC}, 
the distance uncertainty of each SN is
\begin{multline}
  \label{sigmastat}
   \sigma_\mu^2 = C_{m_B,m_B} + \alpha^2 C_{x_1,x_1} + \beta^2 C_{c,c} + \\ 
      2\alpha C_{m_B,x_1} - 2\beta C_{m_B,c} - 2\alpha\beta C_{x_1,c} + \\ 
 \sigma^2_{\rm vpec} + \sigma^2_z + \sigma^2_{\rm lens} + \sigint^2~,
\end{multline}
where $C$ is the fitted covariance matrix from the light curve fit,
$\sigma_{\rm vpec}$ is from the peculiar velocity correction, 
$\sigma_{z}$ is from the redshift uncertainty, 
$\sigma_{\rm lens}$ is from weak gravitational lensing, and
$\sigint$ is determined such that the reduced $\chisqBBC$ is 1.
Prior to BBC, $\chi^2$-based analyses had ignored the $2\ln(\sigmu)$ term of Eq.~\ref{eq:chisqBBC} because 
it resulted in large biases (e.g., Appendix B in \citealt{Conley11}). 
However, KS17 found that including the $\dmuBias$ term removes the previously found biases,
and that including the $2\ln(\sigmu)$ is essential within the BBC framework.

To fit for cosmological parameters in \S~\ref{cosmofit},
the redshift-binned Hubble diagram is defined from the BBC fit as
\begin{eqnarray}
    \zAvg   & = &  {\rm INVERSE}(\mumodelzAvg) \\
    \label{eq:zz}
    \muzAvg & = & \Dmuz + \mumodelzAvg~,
    \label{eq:muz}
\end{eqnarray}
where INVERSE is a numerical function which computes redshift from the distance modulus,
and $\mumodelzAvg$ is the weighted-average $\mumodeli$,
\begin{equation}
    \mumodelzAvg = \left[\sum_{z_i \in \zindex} \mumodeli/\sigmui^2 \right]  \bigg/
        \left[ \sum_{z_i \in \zindex} \sigmui^{-2} \right]
\end{equation}
where the summations are over the subset of \SAMPLENAME\ events in redshift bin $\zindex$.

$\LHBBC$ has 3 types of approximations.  The first is the characterization of intrinsic scatter with a single $\sigint$ term in $\mathcal{L}_{\rm BBC}$,
which does not correspond to either of the scatter models. The second approximation in the $\chi^2$ likelihood is the implicit assumption of symmetric Gaussian uncertainties 
on the bias-corrected SALT2 fitted parameters (\citealt{March11}).
The final type of approximation is in the modeling for bias corrections, which are determined from simulations that
include approximations resulting from limited precision in the:
SALT2 model, 
color and stretch populations, 
intrinsic scatter model (G10 and C11), 
estimation of \SMP\ flux uncertainties, and 
choice of cosmology parameters.

The first two approximations are not included as systematic uncertainties because 
KS17 performed extensive testing on nearly one million simulated SNe~Ia to demonstrate 
that the resulting $w$ bias is below 0.01.
In addition, we perform our own \SAMPLENAME\ validation tests for both bias and uncertainty in \S~\ref{validation}.
Lastly, the third set of approximations in simulated bias corrections are included as systematic uncertainties.

\begin{figure*}
\begin{centering}
\includegraphics[width=0.8\textwidth]{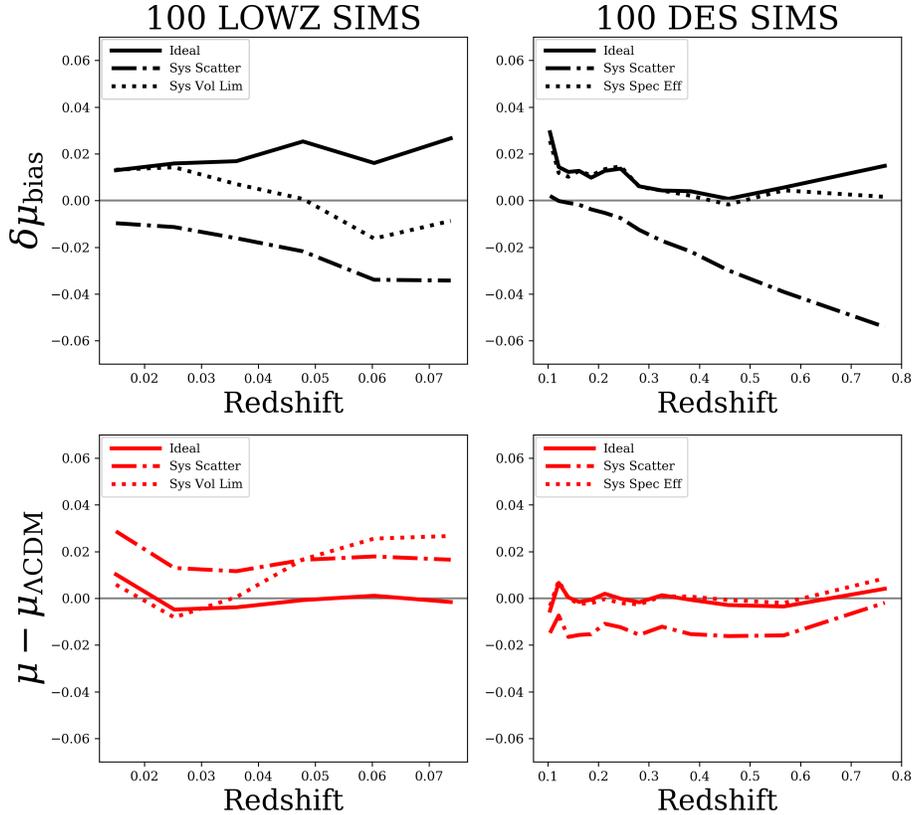}
\caption{\textbf{Top:} bias correction vs. redshift average over 100 \SAMPLENAME\ simulated samples (left: low-$z$, right: DES-SN). `Ideal' corrections have the same scatter model and same selection function in both the simulated data and simulated bias corrections. `Sys scatter' has C11 model for data and G10 model for bias corrections. `Sys Vol Lim' (left) and `Sys Spec Eff' (right) bias corrections are computed using the volume-limited \lowz~subset and the systematic variation on the spectroscopic efficiency function respectively (short dashed lines) . \textbf{Bottom:} Hubble diagram residuals after bias corrections are applied. Residuals are consistent with zero for the Ideal bias corrections.}
\label{fig:speceff}
\end{centering}
\vspace{.1in}
\end{figure*}

Here we illustrate the BBC method using 100 realizations of DES-SN3YR for both the G10 and C11 scatter models. The top panels of Figure~\ref{fig:speceff} show the calculated $\dmuBias$ as a function of redshift. In the bottom panels of Figure~\ref{fig:speceff}, we show the BBC-fitted distance residuals after bias corrections have been applied. For our `Ideal' analysis (solid lines), the bias corrections have the same scatter model and same selection function as the simulated data, and the BBC-fitted distance residuals are consistent with zero. While the average $\mu$-bias correction differs by up to 0.08~mag when the wrong model of intrinsic scatter is used for bias corrections (`Sys Scatter'), the BBC-fitted distance residuals differ by no more than $\sim 0.02$~mag. The reduced effect on
distance biases is caused by the different $\beta$ values from
the BBC fit.

In summary, $\chisqBBC$ (Eq.~\ref{eq:chisqBBC}) is minimized to determine 24 parameters: 
a distance modulus in each of the \NZBIN\ redshift bins (2 of which have no events), 
3 nuisance parameters ($\alpha$, $\beta$, $\gamma$), and 
the intrinsic scatter term ($\sigint$).
The ensemble of \NZBIN\ $[\zAvg$,$\muzAvg]$ pairs is the redshift-binned
Hubble diagram used to fit for cosmological parameters in \S~\ref{cosmofit}.

\subsubsection{Covariance Matrix}

Following \cite{Conley11}, we compute a systematic covariance matrix $C_{\rm stat+syst}$, accounting for both statistical and systematic uncertainties. However instead of a $N\times N$ matrix where $N$ is the number of SNe, here $N$ is the number of redshift bins. $C_{\rm stat}$ is a diagonal matrix whose $\mathcal{Z}^{th}$ entry is the BBC-fitted $\mu$-uncertainty in the $\mathcal{Z}^{th}$ redshift bin. The statistical uncertainties from the binned distance estimates form the diagonal matrix $C_{\rm stat}$, and $C_{\rm syst}$ is computed from all the systematic uncertainties summarized in Section \ref{systematics}.

Using BBC fitted distances, for each source of systematic uncertainty (`SYS') we define distances relative to our nominal analysis (`NOM') as follows:
\begin{equation}
\label{eq:deltamu}
\Delta\langle\mu_{\rm SYS}\rangle_{\zindex} \equiv \langle\mu_{\rm SYS}\rangle_{\zindex}- \langle\mu_{\rm NOM}\rangle_{\zindex} ,
\end{equation}
for redshift bins $\mathcal{Z}$. For each source of systematic uncertainty (`SYS'), we compute $\langle\mu_{\rm SYS}\rangle_{\zindex}$ by varying that source and re-computing bias corrected distances. Groupings of systematic variations are outlined in Table \ref{Tab:inputsys}, and there are a total of 74 individual systematic uncertainty contributions that are evaluated.

We build our redshift-binned 20$\times$20 systematic covariance matrix $C_{\rm syst}$ for all sources (${\rm SYS}_k$),
\begin{equation}
\label{eq:csys}
C_{{\zindex}_i{\zindex}_j,{\rm syst}} = \sum_{k=1}^{K=74} \frac{\partial \Delta\langle\mu_{\rm SYS}\rangle_{{\zindex}_i}}{\partial {\rm SYS}_k} ~ \frac{\partial \Delta\langle\mu_{\rm SYS}\rangle_{{\zindex}_j}}{\partial {\rm SYS}_k} ~ \sigma_k^2,
\end{equation}
which denotes the covariance between the $\mathcal{Z}_i^{th}$ and $\mathcal{Z}_j^{th}$ redshift bin summed over the $K$ different sources of systematic uncertainty ($K=74$) with magnitude $\sigma_k$.

The binned covariances and distances are provided in machine readable format in Appendix \ref{dr}. At the link in Appendix \ref{dr} there is also an un-binned version where the corrections to individual SNe~Ia are computed on a 2D 40-bin interpolation grid to create a covariance matrix for the full SN dataset.

The covariance matrix used to constrain cosmological models is defined as
\begin{equation}
\label{eq:cstatplussyst}
C_{{\rm stat+syst}} = C_{\rm stat} + C_{{\rm syst}}
\end{equation}
where $C_{\rm stat}$ is the diagonal matrix of $\sigma_{\mu}^2$ binned in redshift and where the indices $\mathcal{Z}_i,\mathcal{Z}_j$ have been dropped for convenience.

\subsubsection{Fit for Cosmological Parameters}
\label{cosmofit}
Constraining cosmological parameters with SN data using $\chi^2$ was first adopted by \cite{riess98} and again by \cite{Astier06}. The systematic covariance treatment was improved upon by \cite{Conley11}. Here we follow closely the formalism of S18.

Cosmological parameters are constrained by minimizing a $\chi^2$ likelihood.

\begin{equation}
\label{likelihood}
\chi^2_{\Delta} = \vec{D}^T~C_{\rm stat+syst}^{-1}~\vec{D}
\end{equation}
\begin{equation*}
D_{\zindex} = \muzAvg - \langle\mu_{{\rm model}}\rangle_\mathcal{Z}
\end{equation*}
where $\vec{D}$ is the vector of 20 distances binned in redshift with each element defined by $D_{\zindex}$. In our case $\langle\mu_{{\rm model}}\rangle_\mathcal{Z} = +5\log(d_L/10{\rm pc})$ where for a flat \wCDM~model
\begin{equation}
d_L(z) = (1+z)c\int_0^{z}\frac{dz^\prime}{H(z^\prime)},
\end{equation}
where for simplicity $z \equiv \zAvg$ (Eq.~\ref{eq:zz}) and with
\begin{equation}
H(z^\prime) = {H_0}\ \sqrt[]{\Omega_M(1+z^\prime)^3+\Omega_{\Lambda}(1+z^\prime)^{3(1+w)}},
\end{equation}
where $d_L(z)$ is calculated at each step of the cosmological fitting process and where flatness is assumed in the fits to determine the systematic error budget.

In our analysis we consider two intrinsic scatter models in simulated bias corrections, G10 and C11 (Section \ref{simlight curves}), to span the range of possibilities in current data
samples. We assign equal probability to each model and compute $\vec{D}$ and $C_{\rm stat+syst}$ twice, once for G10 and once for C11. We average the binned distance estimates and covariance matrices for each of the models for intrinsic scatter as follows:

\begin{equation}
\label{eq:sysscatterdata}
\vec{D} = \frac{\vec{D}^{\rm G10} + \vec{D}^{\rm C11}}{2},
\end{equation}

\begin{equation}
\label{eq:sysscatter}
C_{\rm stat+syst} = \frac{C^{\rm G10}_{\rm stat+syst} + C^{\rm C11}_{\rm stat+syst}}{2},
\end{equation}
where the superscripts `G10' and `C11' indicate bias corrections assuming that specific model of intrinsic scatter. The covariances, $C^{\rm G10}_{\rm stat+syst}$ and $C^{\rm C11}_{\rm stat+syst}$, each include the covariance to the other model of intrinsic scatter with scaling $\sigma_k=0.5$ following Eq.~\ref{eq:csys}. The average in Eq.~\ref{eq:sysscatterdata} results in a set of cosmological distances that are roughly half way between that of a G10 only assumption and that of a C11 only assumption, where the systematic uncertainty is half the difference instead of the entire difference. Implicit in this characterization of our distances is that the true intrinsic scatter model lies between that of G10 and C11 with 68\% confidence.

The fitting of cosmological parameters is done with CosmoMC (\citealt{cosmomc}) which is available online and described in Appendix \ref{publiccodes}. We fit the flat \wCDM\ model above to our \SAMPLENAME\ dataset and we combine with Planck 2016 priors. The best fit parameters and further extensions to \LCDM\ are given in the companion key paper (\KEYPAPERalt). In Section \ref{validation} we validate our analysis and uncertainties and in Section \ref{bhm} we discuss ongoing development of a more complex likelihood using
a Bayesian hierarchical modeling framework.

\begin{deluxetable*}{cll}
\tablecolumns{3}
\tablecaption{Sources of Uncertainty}
\tablehead {
\colhead {Size\tablenotemark{a}}              &
\colhead {Description}  &
\colhead{Reference}
}
\startdata
&\textbf{SN Photometry}\\
1 mmag & From astrometry & \cite{bernsteinInstrumentResponse}  \\
1 mmag & Non-linearity of the CCD. & \cite{bernsteinInstrumentResponse} \\
1-2 mmag & Photometric zero pointing. & B18-SMP   \\
3 mmag & Photometric bias determined by fakes. & B18-SMP \\
\\
&\textbf{Calibration}\\
$6/\sqrt[]{3}$ mmag & DECam $\sigma_{\rm uniformity}$ & \cite{burke17}\\
0.6 nm & DECam filter curves uncertainty. & \cite{dr1} \\
$\left[ -2,-2,-1,5\right]$ mmag & Modeling of C26202 implemented as coherent shift $\left[ g,r,i,z\right]$ & Figure \ref{fig:calibration} \\
\\
5mmag/700 nm & HST Calspec spectrum modeling uncertainty & \cite{Bohlin14}\\
1/3 No SuperCal & SuperCal process & S18, \cite{supercal}\\
Following S18 & Low-$z$ samples photometric calibration.&S18, CfA3-4, \mbox{CSP-1}\\
Following S18 & Low-$z$ samples filter curve measurement.&S18, CfA3-4, \mbox{CSP-1}\\
\\
Following B14& SALT2 light curve model calibration. & B14\\
\\

&\textbf{Bias Corrections (Astrophysical)}\\
Table \ref{Tab:parent} & $c$, $x_1$ Parent populations resulting in $\Delta \chi^2 = 2.3$ & \S\ \ref{sys:parent} \\
1/2 (G10 $-$ C11) & Model of intrinsic scatter variations & \S\ \ref{sys:intrinsicscatter} \\
Two $\sigma_{\rm int}$ & Separate fit $\sigma_{\rm int}$ for each subset & \S\ \ref{sys:intrinsicscatter} \\
0.05 in $w$&\tablenotemark{$\dagger$}Cosmology in which the bias correction sample is simulated. & \S\ \ref{sys:biascor} \\
4\% Scaling & MW Extinction maps & \S\ \ref{sys:mw}, \cite{sfd2} \\
\\

&\textbf{Bias Corrections (Survey)}\\
3.5$\sigma$ $\rightarrow$ 3$\sigma$ outlier cut & \tablenotemark{$\dagger$}Low-$z$ Hubble diagram outlier cut. & \S\ \ref{sys:outlier}\\
1$\sigma_{\rm stat}$ Fluctuation & Spectroscopic selection function statistical fluctuations. & \S\ \ref{sys:speceff}, Figure \ref{fig:speceff}\\
Low-$z$ Selection & Low-$z$ subset magnitude $\rightarrow$ volume limited survey. & \S\ \ref{sys:parent}\\
5\% $\sigma_{\rm phot}$ Underestimation & \tablenotemark{$\dagger$}Incorrect SN photometric uncertainties. & \S\ \ref{sys:phot} \\
\\

&\textbf{Redshifts}\\
$4\times10^{-5}$ in z&\tablenotemark{$\dagger$}Coherent $z$-shift. & \S\ \ref{sys:z}, \cite{Calcino17} \\
$0.9 \times \beta_{\rm bias}$ & Peculiar velocity modeling & \S\ \ref{sys:z}, \cite{Zhang17} \\

\enddata
\tablenotetext{a}{Size adopted for each source of systematic uncertainty.}
\tablenotetext{$\dagger$}{Sources of systematic uncertainty that have not been included in previous analyses.\\}
\label{Tab:inputsys}
\vspace{.1in}
\end{deluxetable*}

\begin{figure}
\includegraphics[width=0.49\textwidth]{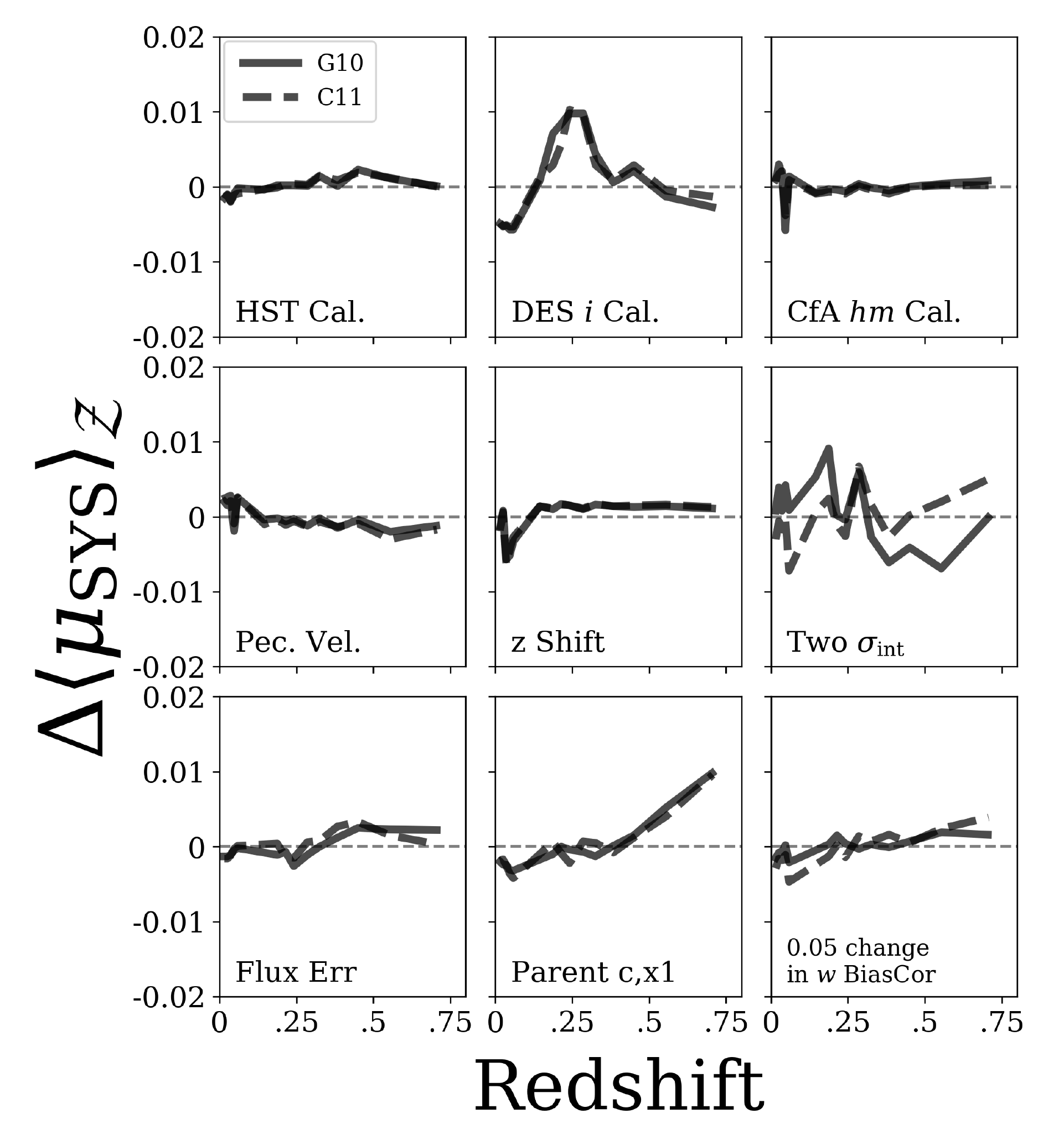}
\caption{Residuals to the nominal cosmological analysis for the \SAMPLENAME~dataset. Distance residuals are calculated for several sources of systematic uncertainty and using bias correction simulations of each model of intrinsic scatter (G10 and C11).}
\label{fig:sysmudif}
\vspace{.1in}
\end{figure}

\subsubsection{Blinding the Analysis}
\label{blinding}

We blind our analysis in two ways simultaneously as there are a number of steps in the analysis in which one could infer changes to cosmological parameters. First, we blind the binned distances output by the BBC fit. Additionally, to prevent accidental viewing of results, the cosmological parameter constraints were perturbed with unknown offsets.

The cosmological parameters were blinded until preliminary results were presented at the 231st meeting of the American Astronomical Society in 
January \citeyear{aas231}. After un-blinding we 
restored the blinding procedure and made the following changes.
First, we fixed the DECam filter transmissions after realizing that
atmospheric absorption had been mistakenly ignored.
Next,
we re-tuned simulations of SMP photometric errors and improved our 
host-galaxy library. 
Finally, we included several additional sources of systematic uncertainty: a global shift in our redshifts, two additional 
calibration systematics (`1/3 No SuperCal' and `SuperCal Coherent Shift'), 
and a systematic uncertainty for the use of two $\sigint$. 

We unblinded again during the internal review process;
w increased by 0.024 and the the total uncertainty increased by 4\% (0.057 to 0.059).

\section{Treatment of Systematic Uncertainties}
\label{systematics}
Here we summarize the treatment and value of each systematic uncertainty from the analysis steps in Section \ref{analysis} in order to create $C_{\rm sys}$ from Eq.~\ref{eq:csys}. A summary table of the systematics used is provided in Table \ref{Tab:inputsys}. In Figure~\ref{fig:sysmudif} we compare the $\Delta\langle\mu_{\rm SYS}\rangle_\mathcal{Z}$ for several systematics, which allows us to visualize the change in distances for some of the major sources of systematic uncertainty. Systematics which have a large change in distance between low and high redshift (i.e.~Parent $c$, $x_1$) are the largest contributors to the total cosmological parameter error budget which is discussed in Section \ref{results}. 

\subsection{Calibration}
\label{sys:cal}
There are several systematic uncertainties related to calibration which include but are not limited to the uncertainty from the photometry (as discussed in B18-SMP), the calibration to the AB system, and the calibration uniformity across the 10 observing fields. The uncertainty in calibration uniformity across the sky is defined as $\sigma_{\rm syst} = \sigma_{\rm uniformity}/\sqrt[]{N}$ where $N$=3 is the number of DES-SN field groups overlapping PS1 (see C,S,X in Section \ref{dataset}), and where we adopt $\sigma_{\rm uniformity}=~6$ mmag from \cite{burke17}.
Within a field group (e.g., C=C1+C2+C3), we do not count each field (for $N$) because the calibration uniformity over ~1 degree scales is expected to be better than the uniformity over the large separations between field groups.

Uniformity uncertainty due to the location of C26202 is already accounted for here because C26202 is located in one of our SN fields that overlap with PS1.
For DES, we combine the photometric uncertainty, uniformity uncertainty, and statistical uncertainty in the AB calibration and propagate a single uncertainty in the photometric zero point per band. A final uncertainty is propagated independently by band such that there is a separate entry in $C_{\rm syst}$ for each band.

To evaluate the agreement of the absolute calibration of the DES-SN fields with the absolute calibration that is used for the \lowz~sample as described in SuperCal, we utilize the overlap of DES stars with those of PS1 which have also been calibrated following SuperCal. We compute $\chi^2_{\rm cal}$ from the difference in absolute calibration, $\Delta M_{{\rm SuperCal}_i-{\rm DES}_i}$, between PS1-SuperCal (red) and DES (grey dashed) shown in (Figure~\ref{fig:calibration}) as follows
\begin{equation}
\chi^2_{\rm cal} = \sum_{i}^{N_{\rm filter}} \frac{\langle \Delta M_{{\rm SuperCal}_i-{\rm DES}_i} \rangle^2}{\sigma_{{\rm SuperCal}}^2+\sigma^2_{{\rm syst}}}.
\end{equation}
where $\Delta M_{{\rm SuperCal}-{\rm DES}_i}$ are the offsets to synthetic magnitudes in each filter (red line of Figure~\ref{fig:calibration}) relative to the DES calibrated to C26202, $\sigma_{\rm SuperCal}$ is the uncertainty from \cite{supercal} of [$3,3,2,4$] mmag in [$g,r,i,z$] bands, and $\sigma_{{\rm syst}}$ is the uncertainty in the uniformity of the fields used for comparison between PS1 and DES ($6/\sqrt[]{3}$ mmag). We find that $\chi^2_{\rm cal}=1.5$ for 4 degrees of freedom, indicating that the DES calibration to C26202 is consistent with SuperCal.

In order to account for the possibility that the C26202 brightness measured by HST is biased due to incorrect modeling of the C26202 spectrum, we include a coherent shift in the absolute calibration of DES amongst all bands simultaneously to SuperCal as an additional uncertainty. That is, we shift our DES-SN magnitudes to an absolute system where the vertical red lines in Fig \ref{fig:calibration} are defined as zero. Thus, we obtain a new set of SN distances using this new calibration definition and the $\Delta\langle\mu_{\rm SYS}\rangle_\mathcal{Z}$ for this choice are propagated in our covariance matrix $C_{\rm syst}$.

The uncertainty in the calibration of the \lowz~sample is adopted from SuperCal. Additionally, as was done in S18, we adopt an overall uncertainty associated with the SuperCal itself which \cite{supercal} characterized as 1/3 the size of the impact on distances if SuperCal was not applied.

A number of calibration systematics are propagated separately from the absolute and relative calibration treatment above. Uncertainty in the DECam filter transmission functions propagate to uncertainties in absolute calibration because FGCM utilizes these transmission functions to predict the flux of C26202. A 0.5nm wavelength uncertainty arises in the determination of the filter transmission function due to  the precision on wavelength in the measurement. Additionally, there is a 0.3nm effect arising from illumination lamps on the flat field screen that should be, but are not exactly, on the same optical axis. These two wavelength uncertainties are added in quadrature for a total of 0.6nm. 

We also include the uncertainty in modeling the spectrum of C26202, which is 5mmag over 700nm. Lastly, we have not retrained the SALT2 model, and therefore we use the same SALT2 calibration uncertainty as in B14.

We do not include a systematic uncertainty from chromatic corrections, since we have already included FGCM uncertainties which are based on applying these corrections. Furthermore, \cite{Lasker18} find that if chromatic corrections are not applied, the change in fit $w$ is 0.005. This change in $w$ is consistent with the statistical uncertainty associated with this correction, and it is well below the systematic uncertainty from our analysis.

\subsection{Intrinsic Scatter Model}
\label{sys:intrinsicscatter}
One of the largest systematic uncertainties results from the modeling of intrinsic scatter in the simulations used to predict bias corrections. We include two intrinsic scatter models, G10 and C11, and assign equal probability to each model. Because of the parallel treatment of the scatter models (Eqs.~\ref{eq:sysscatterdata}~\&~\ref{eq:sysscatter}), we end up with two sets of nuisance parameters. From here on in this paper, unless otherwise noted, results and nuisance parameters are stated in the context of the G10 model. 

As will be shown in Sec \ref{sec:nuissiancesigmaint}, the $\sigint$ values show 
$>3\sigma$ tension when determined separately for the 
\lowz\ and DES subsets, and this tension persists
for both intrinsic scatter models.
In addition, our DES-SN $\sigint$ value is the smallest of 
any rolling SN search, suggesting that it is a fluctuation. 
To account for the possibility that this $\sigint$ difference 
is real, we include a systematic uncertainty based on an analysis 
using two $\sigint$ values, and compare to the nominal analysis 
that assumes a single $\sigint$ value.
For the ``Two $\sigint$'' analysis, we scale the spectral flux
variations from the intrinsic scatter model (G10 or C11) so that 
analyzing the simulation results in the same $\sigint$ values as 
for the \lowz\ and DES-SN data subsets. These scaled scatter models
are used to generate bias correction simulations, and the BBC fit
is modified to constrain the ratio, \mbox{$\sigint$(\lowz)/$\sigint$(DES-SN)},
to match that of the data.  
To summarize, there are two uncertainties related to the unknown 
source of intrinsic scatter. First is the relative contribution
of coherent vs. wavelength dependent scatter (G10 vs. C11). 
Second is the overall amplitude difference in scatter between
the \lowz\ and DES subsets.

\subsection{Color and Stretch Parent Populations}
\label{sys:parent}

In order to estimate the uncertainty in parent color and stretch distributions, we vary the mean and width of each parent population in the simulation until we achieve $>1\sigma$ deviations between the observed and simulated distributions. We alter the systematic parent populations of color and stretch in order to increase the $\Delta \chi_p^2$, as defined in Eq.~\ref{chisqp}, by $\sim 2.3$ , following Table 39.2 of \cite{pdg}. The population parameter values used for the nominal and systematic simulations are shown in Table \ref{Tab:parent}. 
The dependence between observed populations and the spectroscopic efficiency function is sufficiently weak to justify solving for each independently. The differences in assumed parent populations manifest themselves in different bias corrections to the dataset and are visualized in the lower-central panel of Figure~\ref{fig:sysmudif}.

For the uncertainty in the parent populations of color and stretch for the \lowz~subset, this is encompassed in the volume-limited case. In this case, redshift evolution of color and stretch are interpreted as astrophysical effects rather than manifestations of Malmquist bias (Section \ref{simlight curves}). A different set of parent population parameters are determined for the volume-limited case and are shown in Table \ref{Tab:parent}. 

\subsection{Spectroscopic Selection}
\label{sys:speceff}

We generate 200 realizations of the DES subset with only statistical fluctuations. We run our $E_{\rm spec}$ fitting procedure on each realization and find that biases in recovering the input $E_{\rm spec}$ are limited to 7\% ($E_{\rm fit}$/$E_{\rm input}$-1) across the range $19<i_{\rm peak}<24$ whilst $1\sigma$ statistical fluctuations are up to 25\% at $23^{rd}$~mag. Because neither the simulation nor BBC fit
 take into account the statistical uncertainty in the $E_{\rm spec}$, we adopt the $1\sigma$ statistical fluctuation and propagate it as a systematic uncertainty.

We do not include a spectroscopic efficiency systematic for the \lowz~subset. Instead, the \lowz~subset is assumed to be magnitude limited and the systematic uncertainty for simulating this sample is to model it as volume-limited (see Table 3 and \S\ 6.2 of K18).

\subsection{Cosmology Assumption in Bias Corrections}
\label{sys:biascor}

We include the systematic uncertainty from our choice to simulate selection biases with a fixed set of $w$CDM parameters ($\Omega_M$=0.3, $\Omega_\Lambda$=0.7, $w$=-1).  Here we redetermine the distance bias after changing the reference cosmological model in our simulations to $w_{\rm ref}=w_{\rm bestfit}-0.05$, a change that matches the statistical precision of our measurements.  The difference in distance biases for these two reference cosmology values is illustrated in Figure~\ref{fig:sysmudif} and is less than 2 mmag across the entire redshift range. 

\subsection{Redshifts}
\label{sys:z}

We include two systematic uncertainties for our treatment of redshifts.  The first is from our modeling of the peculiar velocities, and following \cite{Zhang17} we modify the light-to-matter bias parameter ($\beta_{\rm bias}$) by 10\% and remeasure the redshift corrections.  The second is a coherent shift in each redshift of $4\times 10^{-5}$ as conservatively constrained in \cite{Calcino17}.

\subsection{Low-$z$ Hubble Residual Outliers}
\label{sys:outlier}

We include the systematic uncertainty associated with Hubble residual outlier rejection of SNe~Ia in the \lowz~subset. S18 placed a 3.5$\sigma$ cut on their sample. For our dataset of 329 SNe~Ia, Chauvenets criterion corresponds to a 3$\sigma$ cut. We investigate the systematic effect of applying both $3.5\sigma$ and $3\sigma$ cuts on Hubble diagram residuals to the \lowz~subset. Because this cut depends on the best fit cosmological model, it is discussed later in Section \ref{budget}.

\subsection{Photometry}
\label{sys:phot}

For the \SMP\ pipeline, there is an additional systematic uncertainty beyond the 0.3\% biases mentioned in Section \ref{photometry}. Our \SMP\ pipeline performs stellar position fits independently on each night, but uses a globally-fitted position of the SN across all nights (B18-SMP). Fitting for stellar positions each night independently accounts for the proper motion of the stars, but B18-SMP find this difference in the treatment of the stars and SNe can cause a $\sim$2~mmag bias towards brighter fluxes. This small additional systematic uncertainty is added in quadrature to the calibration uncertainty.

We also test for the underestimation of photometric uncertainties. B18-SMP showed that after using fakes to correct the flux uncertainties as a function of host-galaxy local surface brightness, SN flux uncertainties are accurate to within 5\%. We therefore consider a systematic underestimation of uncertainties of 5\%. 

\subsection{Milky Way Extinction}
\label{sys:mw}

Lastly, we account for Milky Way extinction using maps from \cite{schlegel98}, with a scale of 0.86 based on \cite{sfd1}, and the Milky Way (MW) reddening law from \cite{Fitzpatrick99}.  We adopt a global 4\% uncertainty of $E(B-V)_{\rm MW}$ based on the fact that \cite{sfd2}, in a re-analysis of \cite{sfd1},  derive smaller values of reddening by $ 4\%$, despite using a very similar SDSS footprint.

\section{Results}
\label{results}

\begin{figure}
\includegraphics[width=0.49\textwidth]{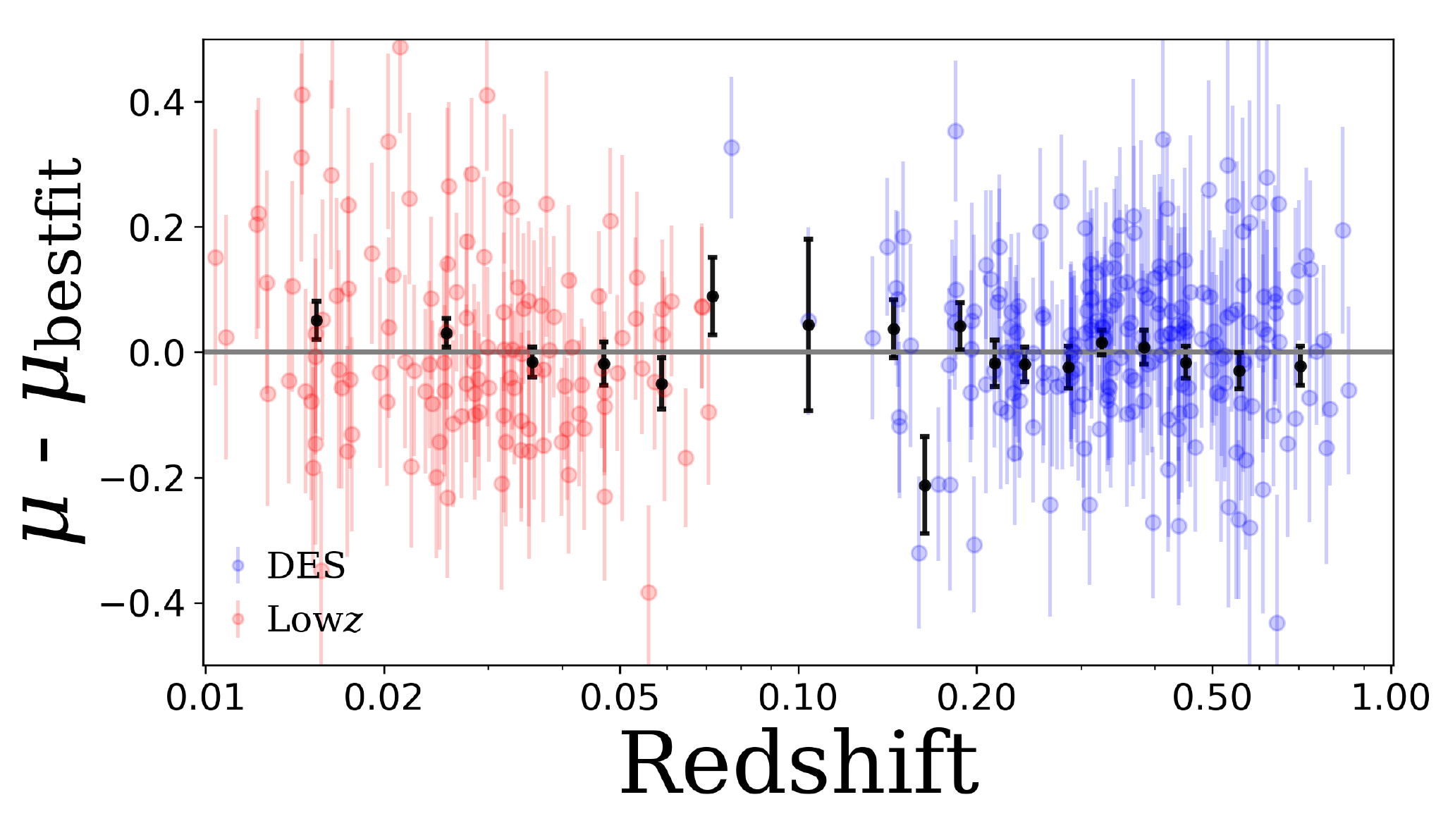}
\caption{Residuals in distance to the best fit flat $w$CDM cosmology as a function of redshift. Blue: DES subset. Red: Low-$z$ subset. Black: Binned distances used for cosmological fits. BBC fitted distances shown are averaged assuming each model of intrinsic scatter (G10 and C11).}
\label{fig:wwobiascor}
\vspace{.1in}
\end{figure}

\begin{figure}
\begin{centering}
\includegraphics[width=0.49\textwidth]{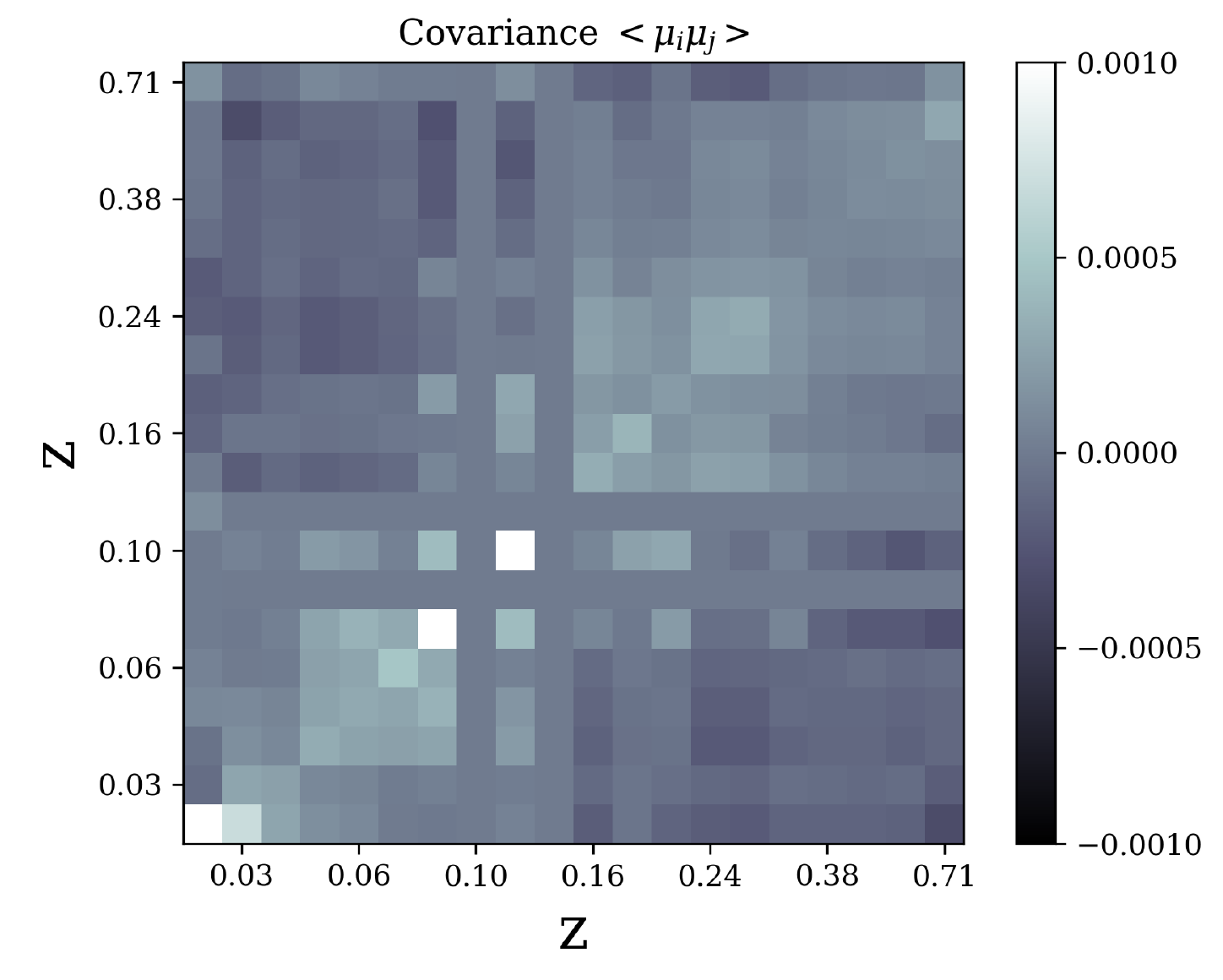}
\caption{Distance covariance matrix in redshift bins without statistical uncertainties on the diagonal. \\}
\label{fig:covmat}
\end{centering}
\end{figure}

We perform a cosmological fit to our redshift-binned and bias-corrected Hubble diagram. The distances obtained in this analysis are shown as binned residuals to the best fit cosmology in Figure~\ref{fig:wwobiascor} after bias corrections have been applied. 
The covariance matrix used in our cosmological fits with each of the systematics components ($C^{\rm Cosmo}_{\rm syst}$) is shown in Figure~\ref{fig:covmat}. In this section we report the fit values for the nuisance parameters in Eq.~\ref{Eq:tripp} and the systematic error budget on cosmological parameters. Several of our results require further discussion which can be found in Section \ref{discussion}. We refer the reader to \KEYPAPER~ for the unblinded best fit constraints of cosmological parameters.

\begin{deluxetable*}{rcccc}
\tablecolumns{5}
\tablewidth{35pc}
\tablecaption{Nuisance Parameters from BBC Fit}
\tablehead {
\colhead {Parameter} &
\colhead {Description} &
\colhead { G10}               &
\colhead { C11  }     &           
\colhead { AVG\tablenotemark{$\dagger$}  }                
}
\startdata

$\alpha$ & DES-SN3YR & 0.146 $\pm$  0.009 &  0.147 $\pm$  0.009 & 0.147 $\pm$  0.009\\
$\alpha$ & DES subset & 0.151 $\pm$  0.012 &  0.152 $\pm$  0.012 & 0.152 $\pm$  0.012\\
$\alpha$ & Low-$z$ subset &0.145 $\pm$  0.014 &  0.144 $\pm$  0.014 & 0.145 $\pm$  0.014\\
$\alpha$ & PS1 &  0.167 $\pm$ 0.012 & 0.167 $\pm$ 0.012 & 0.167 $\pm$ 0.012 \\
$\alpha$ & SNLS & 0.139 $\pm$ 0.013 & 0.139 $\pm$ 0.013 & 0.139 $\pm$ 0.013\\

\\

$\beta$ & DES-SN3YR & 3.03 $\pm$  0.11 &  3.58 $\pm$  0.14 & 3.30 $\pm$ 0.13\\
$\beta$ & DES subset & 3.02 $\pm$  0.13 & 3.56 $\pm$  0.17 & 3.29 $\pm$ 0.15 \\
$\beta$ & Low-$z$ subset & 3.06 $\pm$  0.19 & 3.61 $\pm$  0.24 & 3.34 $\pm$ 0.15\\
$\beta$ & PS1 &  3.02 $\pm$ 0.12 & 3.51 $\pm$ 0.16 & 3.26 $\pm$ 0.14\\
$\beta$ & SNLS &  3.01 $\pm$ 0.14 & 3.59 $\pm$ 0.17 & 3.30 $\pm$ 0.16\\
\\

$\gamma$ & DES-SN3YR & 0.025 $\pm$  0.018 & 0.016 $\pm$  0.018 & 0.021 $\pm$  0.018\\
$\gamma$ & DES subset& $\ghrstep$ $\pm$ $\ghrsteperr$& $\chrstep$ $\pm$ $\chrsteperr$ & 0.007 $\pm$ 0.018\\
$\gamma$ & Low-$z$ subset & 0.070 $\pm$  0.038 & 0.043 $\pm$  0.038 & 0.057 $\pm$ 0.038\\
$\gamma$ & PS1 &  0.039 $\pm$ 0.016 & 0.041 $\pm$ 0.016 & 0.040 $\pm$ 0.016 \\
$\gamma$ & SNLS &  0.045 $\pm$ 0.020 & 0.037 $\pm$ 0.020 & 0.041 $\pm$ 0.020\\

\\
\tablenotemark{*}$\sigma_{\rm int}$ & DES-SN3YR & $0.094 \pm 0.008$ &  $0.117 \pm 0.008$ & $0.106 \pm 0.008$\\
\tablenotemark{*}$\sigma_{\rm int}$  & DES subset& $\desgsigint  \pm 0.007$ & $\descsigint \pm 0.008$ & $0.077 \pm 0.008$\\
\tablenotemark{*}$\sigma_{\rm int}$  & Low-$z$ subset & $0.120 \pm 0.015$ & $0.144 \pm 0.015$ &  $0.132 \pm 0.015$ \\
$\sigma_{\rm int}$ & PS1 &  0.08 & 0.10 & 0.09 \\
$\sigma_{\rm int}$ & SNLS &  0.09 & 0.10 & 0.10\\

\enddata
\tablecomments{Nuisance parameters and uncertainties for the \SAMPLENAME~and the DES and low-$z$ subsets with comparisons to other datasets. The values for PS1 and SNLS are taken from S18 which does not report uncertainties on $\sigma_{\rm int}$.}
\tablenotetext{$\dagger$}{AVG is presented here solely for comparison purposes and is not used in the analysis.}
\tablenotetext{*}{ $\sigma_{\rm int}$ uncertainty is the RMS from 100 simulated realizations of the dataset. \\}

\label{Tab:nuisance}
\end{deluxetable*}

\begin{deluxetable}{cccc}
\tablecolumns{4}
\tablecaption{Comparison of $\sigma_{\rm tot}$}
\tablehead {
\colhead {} &
\colhead { $\sigma_{\rm tot}$(G10)}               &
\colhead {$\sigma_{\rm tot}$(C11) }         \\ 
\colhead {Dataset} &
\colhead {{\scriptsize 5D [1D]}}               &
\colhead {{\scriptsize 5D [1D]  }}                
}
\startdata

 DES subset & 0.129 [0.156] &  0.128 [0.156]\\
 Low-$z$ subset & 0.156 [0.158] &  0.157 [0.158]\\
 DES-SN3YR & 0.142 [0.155] &  0.141 [0.155]\\
 PS1 & 0.14 [0.16] &  0.14 [N/A]\\
 SNLS & 0.14 [0.18] &  0.14 [N/A]\\

\enddata
\tablecomments{Comparison of RMS of Hubble diagram residuals ($\sigma_{\rm tot}$) for the subsets of SNe. Comparisons between performing 5D and 1D bias corrections are also shown. The values for PS1 and SNLS are taken from S18. \\}

\label{Tab:sigmatot}
\end{deluxetable}

\begin{deluxetable}{crr}
\tablecolumns{3}
\tablewidth{15pc}
\tablecaption{Systematic variations for $\gamma_{\rm DES}$}
\tablehead {
\colhead {Variation} &
\colhead { $\gamma$ [mag]  } &
\colhead {\# SNe~Ia}
}
\startdata

Nominal & $0.009 \pm 0.018$ & 207\\
$c>0$ & $0.069 \pm 0.039$ & 70 \\
$c<0$ & $-0.005 \pm 0.020$ & 137\\
$x_1>0$ & $0.018 \pm 0.025$ & 119 \\
$x_1<0$ & $-0.013 \pm 0.029$ & 88\\
no $z$ band &  $0.000 \pm  0.021$ & 202 \\
1D BiasCorr. &  $0.041 \pm 0.021$ & 207\\
\texttt{DiffImg} Photometry &  $0.001 \pm  0.020$ & 207\\
$\mathcal{M}_{\rm stellar} \neq$ null &  $0.010 \pm  0.020$ & 207\\
$\mathcal{R}_{\rm step} = 10.1$ & $0.021 \pm 0.019$ & 207\\
10 $z$-bins & $0.015 \pm 0.018$ & 207\\
Le Phare & $0.008 \pm 0.020$ & 207\\

\enddata
\tablecomments{Changes in $\gamma$ for the DES subset after perturbations to analysis. Parameter values are shown for the G10 model of intrinsic scatter only. \\}
\label{Tab:perturbations}
\end{deluxetable}

\subsection{Nuisance Parameters}
\label{sec:nuisance}

The BBC fit output includes 4 nuisance parameters: $\alpha$, $\beta$, $\sigma_{\rm int}$, and $\gamma$. The values for these parameters are summarized in Table \ref{Tab:nuisance} along with a comparison with those of the PS1 and SNLS samples from S18. Here we describe the values found, their consistency with those of previous samples, as well as various perturbations to our analysis and the affect on the recovered nuisance parameters.

\subsubsection{ $\alpha$, $\beta$}
\label{alphabeta}
A comparison of $\alpha$ and $\beta$, the standardization coefficients of stretch and color, with those of the PS1 and SNLS samples are shown in Table \ref{Tab:nuisance}. We find that $\alpha$ and $\beta$ are in agreement with various surveys. We test for $\alpha$ or $\beta$ dependence with redshift:
\begin{equation}
\alpha = \alpha_0 + z \times \alpha_1,\ \beta = \beta_0 + z \times \beta_1,
\end{equation}
and we find that $\alpha_1$ and $\beta_1$ are consistent with zero, with the possible exception of $\beta_1$ in our G10 analysis which we detect at $-1.9\sigma$. However in our C11 analysis we detect $\beta_1$ at $-0.5\sigma$ and thus we consider the evolution in the G10 case to be a statistical fluctuation. 

\subsubsection{$\sigma_{\rm int}$ and $\sigma_{\rm tot}$}
\label{sec:nuissiancesigmaint}
The nominal analysis assumes a single value for the amount of intrinsic scatter needed to bring $\chi^2/$dof$=1$ ($\sigma_{\rm int}$). We perform the nominal analysis twice, once for each model of intrinsic scatter (G10 and C11) and the values of $\sigma_{\rm int}$ are found to be 0.094 and 0.117 respectively (Table \ref{Tab:nuisance}). These are in agreement with the values found by previous analyses (PS1, SNLS). However, we also examine the $\sigma_{\rm int}$ for each subset in our analysis.
For the DES subset we find $\sigma_{\rm int} = \desgsigint$~mag for G10 and $\descsigint$~mag for C11, which are the smallest observed values of any rolling supernova survey to date using the SALT2 framework. For the low-$z$ subset we find $\sigma_{\rm int} = 0.12$~mag for G10 and $0.14$~mag for C11. In analyzing 100 simulated statistical realizations of \SAMPLENAME, we find that the $RMS(\sigma_{\rm int})$ for the DES subset is 0.007 and for the low-$z$ subset it is 0.015. Thus, the $\sigma_{\rm int}$ values for DES-SN and low-$z$ subsets differ by more than $3\sigma$. In Section \ref{budget} we discuss the change in fit $w$ if two $\sigma_{\rm int}$ are used in our analysis ($\sigma_{\rm int}^{{\rm low-}z}$ and $\sigma_{\rm int}^{\rm DES}$). 

In Table \ref{Tab:sigmatot} we show the total scatter about the Hubble diagram, $\sigma_{\rm tot}$, for the subsets in this analysis and we compare with other surveys. We find the lowest observed value of $\sigma_{\rm tot}$, \desgsigtot~mag. We also confirm that the 5D bias corrections performed in BBC provide improved Hubble residual scatter over 1D corrections. 1D corrections in this analysis are only used as crosschecks to previous analyses such as B14.

\begin{figure}
\includegraphics[width=0.49\textwidth]{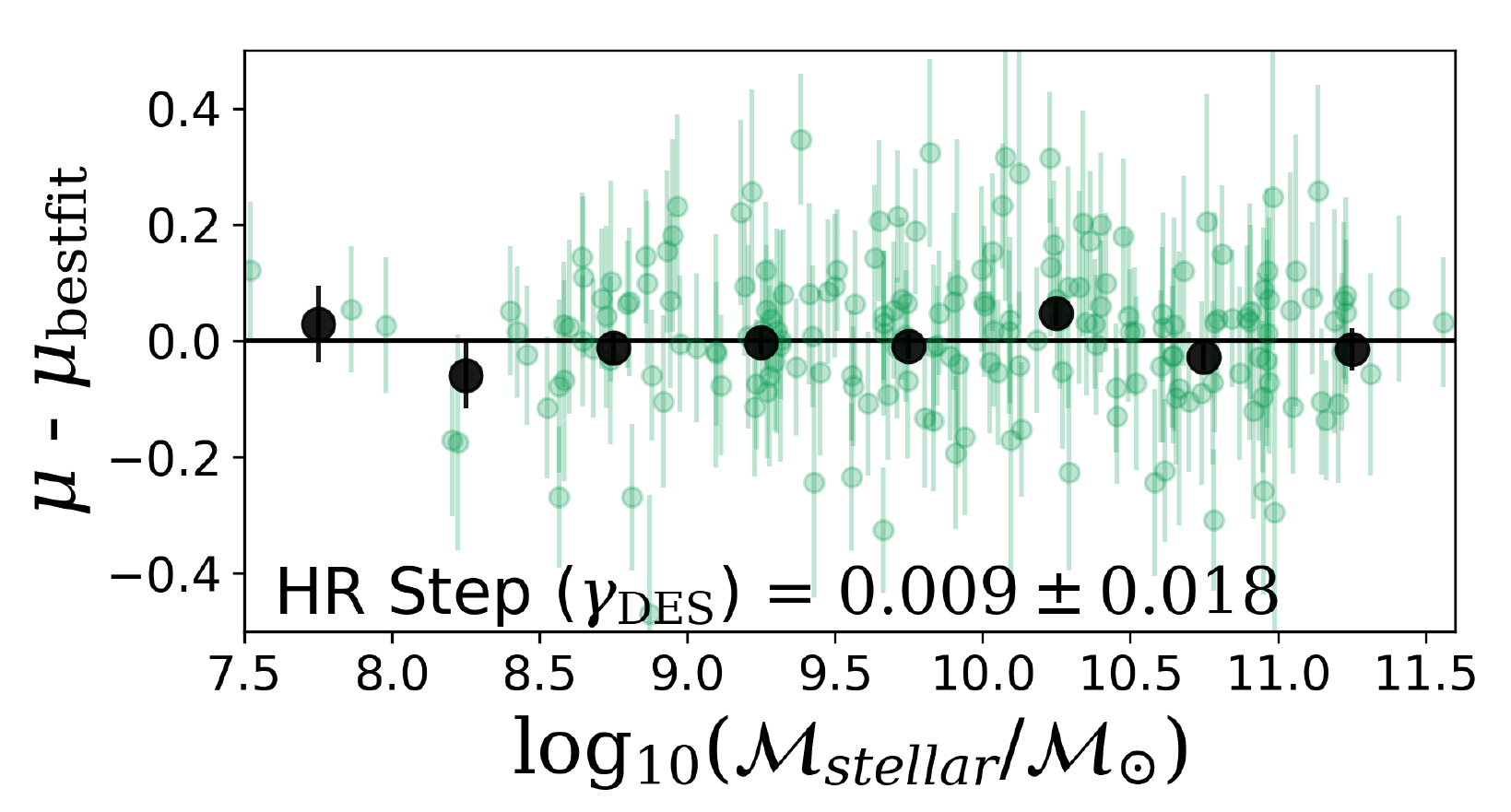}
\caption{Residuals in distance to the best fit cosmology as a function of $\log_{10}(\mathcal{M}_{\rm stellar}/\mathcal{M}_{\odot})$ for the DES subset only. Correlation between residuals and mass is characterized as a step function at $10^{10}\mathcal{M}_{\odot}$ however we find no clear trend in the DES-SN data.}
\label{fig:after}
\vspace{.1in}
\end{figure}

\subsubsection{Host-galaxy Stellar Mass Step $\gamma$}

Somewhat surprisingly, we find little correlation between host-galaxy stellar mass and Hubble diagram residuals  ($\gamma = 0.025 \pm 0.018$) for \SAMPLENAME. This is driven by the fact that for the DES subset alone, we find no evidence of a correlation ($\gamma_{\rm DES} = \ghrstep \pm \ghrsteperr$ mag).  For the low-$z$ subset we find $\gamma_{{\rm low-}z} = 0.070 \pm 0.038$ mag, which is consistent with previously seen results. The Hubble diagram residual vs. host mass relation for the DES subset are plotted in Figure~\ref{fig:after}. The DES subset value is \gammasigma$\sigma$ smaller than $\gamma_{\rm Pantheon}$ found in S18 which used the same BBC fitting method. As a crosscheck, we have obtained a second set of host-galaxy stellar mass estimates using a different set of SED templates (\citealt{Bruzual03}) and fit the $griz$ magnitudes with Le Phare (\citealt{lephare}) spectral fitting code. With the separate set of mass estimates, we find the $\gamma_{\rm DES}$ value is still consistent with zero (Table \ref{Tab:perturbations}).

Another crosscheck is to replace the 5D bias correction in the
   BBC fit with a 1D correction that depends only on redshift,
   which is similar to the JLA (B14) analysis. We find that using 1D bias correction in $z$, analogous to that of the JLA (B14) analysis, results in a larger $\gamma_{\rm DES}$ of $0.041 \pm 0.021$ mag. This is in agreement with S18 who find that the 5D bias correction reduces scatter about the Hubble diagram and reduces $\gamma_{\rm DES}$ by $\sim0.02$~mag compared to using the 1D bias correction from B14.  This will be studied in a forthcoming DES-SN analysis (Smith et al. 2018 in prep) of simulations that include correlations between the SN properties and the host in simulations. We note that using 5D bias corrections, S18 find significant values for $\gamma$ for each of their subsets of SNe and that the value found here for the \lowz\ sample is in agreement with S18.

To examine potential systematics in measuring $\gamma_{\rm DES}$, Table \ref{Tab:perturbations} shows several 
variations in our BBC fitting procedure. As DECam has better $z$-band sensitivity compared to previous surveys, we ran our analysis without $z$ band and found a consistent $\gamma_{\rm DES}$ ($0.007 \pm  0.023$ mag) with a slightly larger uncertainty.

Additionally, because color and stretch are both correlated with host-galaxy stellar mass (Figure~\ref{fig:hostmassbefore}), we investigate the effect of various cuts to our dataset on $\gamma_{\rm DES}$. Splitting the DES subset into two sub-samples of color, we find that $c>0$ results and $c<0$ differ by $1.6\sigma$. When performing the analogous test in stretch, we find $x_1>0$ and $x_1<0$ differ by 1$\sigma$. Statistically these measurements are self-consistent. As a precautionary check that the small $\gamma_{\rm DES}$ value is not an artifact of our \SMP\ pipeline, we perform a BBC fit with the \texttt{DiffImg} photometry and find that $\gamma_{\rm DES}$ remains consistent with zero.  

Since we have included host-galaxies whose mass could not be determined (S/N too low), and assigned them to the $\mathcal{R}_{\rm host}< 10$ bin, we perform the BBC fit with these events excluded (`$\mathcal{M}_{\rm stellar} \neq$ null') and still find $\gamma_{\rm DES}$ consistent with zero. We also test using 10 redshift bins instead of 20 and again the recovered value for $\gamma_{\rm DES}$ is consistent with zero.

We perform a separate check for redshift evolution of $\gamma$ parametrized as 
\begin{equation}
\gamma = \gamma_0 + \gamma_1 \times z.
\end{equation}
We find $\gamma_1$ is consistent with zero for the DES subset ($-0.11 \pm  0.10$ mag). 

Finally, because specific star formation rate (sSFR) is known to correlate with host galaxy stellar mass (\citealt{rigault15}; \citeyear{Rigault18}), we explicitly check for a sSFR step with Hubble residuals in the same fashion as Eq.~\ref{Eq:hm} and find $0.037\pm0.025$ mag.

\begin{deluxetable*}{l|ccc}
\tablecolumns{4}
\tablewidth{28pc}
\tablecaption{$w$ Uncertainty Contributions for $w$CDM model\tablenotemark{a} }
\tablehead {
\colhead {Description\tablenotemark{b}} &
\colhead { $\sigma_{w}^\prime$}               	&
\colhead {$\sigma_{w}^\prime/\sigma^{\rm stat}_{w}$}   &      
\colhead { $w$ shift}               
}
\startdata
		 Total Stat ($\sigma^{\rm stat}_{w}$)	 &   \statwerr\	&   1.00& 0.000 \\
         Total Syst\tablenotemark{c} ($\sigma^{\rm total\ syst}_{w}$)& \systwerr\	&	1.00&	-0.006  \\
\hline \\
   {\bf [Photometry \& Calibration]} & {\bf [\sigmacal]}& {\bf [0.50]}& {\bf [-0.005]}  \\
     Low-$z$ 		 & 0.014	&	0.33& -0.003 \\
     DES			& 0.010	&	0.33&0.001 \\
     SALT2 model 	 & 0.009	&	0.21& -0.003 \\
     HST Calspec     &0.007& 0.17&   0.001\\
     1/3 No SuperCal     &0.005& 0.12&   -0.001\\
     SuperCal Coherent Shift\tablenotemark{d}	 &0.005&	0.12&	-0.001\\
         & & \\
    {\bf [$\mu$-Bias Corrections: Survey]} & {\bf  [\werrSystBiasSurvey]} & {\bf [0.55]}&{\bf [-0.001]} \\
   \tablenotemark{$\dagger$}Low-$z$ 3$\sigma$ Cut	& 0.016	&	0.38&0.005 \\
   Low-$z$ Volume Limited				& 0.010	&	0.24&0.009 \\
   Spectroscopic Efficiency				& 0.007	&	0.17&0.001 \\
   \tablenotemark{$\dagger$}Flux Err Modeling	& 0.001	&	0.02&-0.001 \\
        & & \\
   {\bf [$\mu$-Bias Corrections: Astrophysical]}  & {\bf [\werrSystBiasAstro]} & {\bf [0.62]}& {\bf [-0.003]} \\
   Intrinsic Scatter Model					& 0.014	&	0.33&-0.001 \\
   $c$, $x_1$ Parent Population				& 0.014	&	0.33&0.000 \\
   \tablenotemark{$\dagger$}Two $\sigma_{\rm int}$	& 0.014	&	0.33&-0.005 \\
   MW Extinction							& 0.005	&	0.12&-0.001 \\
   \tablenotemark{$\dagger$}$w,\Omega_M$ for bias corr.	& 0.006	&	0.14&0.001 \\
     & & \\

 {\bf [Redshift]} & {\bf [\werrSystRedshift]} & {\bf [0.29]} & {\bf [0.003]} \\ 
   \tablenotemark{$\dagger$}$z+0.00004$	& 0.006	& 0.14&-0.001 \\
   Peculiar Velocity		    	& 0.007	& 0.17&0.004 \\
\enddata
\tablenotetext{a}{The sample is DES-SN3YR (DES-SN + low-$z$ sample) plus CMB prior.}
\tablenotetext{b}{Items in {\bf [bold]} are sub-group uncertainty sums.}
\tablenotetext{c}{The quadrature sum of all systematic uncertainties does not equal $\systwerr$ because of redshift-dependent correlations when using the full covariance matrix.}
\tablenotetext{d}{Uncertainty is also included in Photometry \& Calibration: DES.\\}
\tablenotetext{$\dagger$}{Uncertainty was not included in previous analyses.}
\label{Tab:finalbudget}
\vspace{.1in}
\end{deluxetable*}

\subsection{Systematic Error Budget}
\label{budget}

The uncertainties on $w$ are presented in Table \ref{Tab:finalbudget} for fits to a flat $w$CDM model with Planck 2016 CMB priors. The systematic uncertainties shown in Table \ref{Tab:finalbudget} are defined as 
\begin{equation}
\label{wsyst}
\sigma^{\prime}_{w} = \sqrt{(\sigma_w^{\rm stat+syst})^2-(\sigma_w^{\rm stat})^2} 
\end{equation}
where $\sigma_w^{\rm stat+syst}$ is the uncertainty when only a specific systematic uncertainty (or group of uncertainties) is applied such that $\sigma^{\prime}_{w}$ is the contribution to the total uncertainty from the specific systematic alone.
Small shifts in $w$ are expected when including systematic uncertainties due to different inverse-variance weights as a function of redshift from the BBC fit. We characterize this effect in Table \ref{Tab:finalbudget} by including
\begin{equation}
\label{wshift}
w{\rm -shift} = w_{\rm stat+syst} - w_{\rm stat}, 
\end{equation}
which is the difference between including and excluding systematic uncertainties. Additionally, we show the contribution to the uncertainty budget for each systematic grouping in column $\sigma_{w}^{\rm syst}$, and the ratio of systematic uncertainty to statistical uncertainty ($\sigma^{\rm syst}_{w}$/$\sigma^{\rm stat}_{w}$). We note that simply summing errors in quadrature
from Table \ref{Tab:finalbudget} will not result in the uncertainty for `ALL' because of redshift-dependent correlations among the systematic effects.

We find that the statistical and systematic uncertainties on $w$ for the \SAMPLENAME~dataset are  
\begin{equation*}
\sigma^{\rm stat}_{w} = \statwerr, 
\end{equation*}
\begin{equation*}
\sigma^{\rm total\ syst}_{w} = \systwerr, 
\end{equation*}
where $\sigma^{\rm total\ syst}_{w}$ is the $w$ uncertainty from all systematics and excluding statistical uncertainties. This indicates that our result is equally limited by systematic and statistical uncertainties. In Section \ref{discussion} we discuss how additional data may aide in the reduction of systematic uncertainties. 

In Table \ref{Tab:finalbudget}, we break down the independent contributions to the $w$-error budget. We also group the systematic uncertainties into four main categories and find that nearly equal contributions to the total uncertainty from the largest three groupings: 1) photometry and calibration, 2) astrophysical bias corrections, and 3) survey bias corrections, all of which are associated with estimation of distances. The final and smallest grouping, 4) describes the systematic uncertainties associated with the redshifts in our analysis.

\textbf{Photometry and Calibration:} Because the low-redshift samples are calibrated to the PS1 absolute magnitude system and because the DES subset has been calibrated to a single CalSpec standard star, we have included an additional calibration uncertainty. We assume coherent offsets to SuperCal to be our systematic uncertainty for the potential incorrect modeling of the single CalSpec standard. We find that this results in an uncertainty on $w$ of 0.005. This uncertainty is included in the `DES' calibration uncertainty.

\textbf{Astrophysical $\mu$-Bias Corrections:}
As mentioned in Section \ref{sys:intrinsicscatter}, we run the entire analysis pipeline separately for G10 and C11 models of intrinsic scatter. 
The contribution to the error budget from intrinsic scatter model alone is found to be $\sigma_w=0.014$ 
While we derive separate parent populations associated with each intrinsic scatter model, we also assess the systematic uncertainty in these parent populations.  This systematic (`$c$, $x_1$ Parent Pop') is as large as that due to the intrinsic scatter model itself.

Our nominal analysis assumes that all SNe~Ia samples have the same amount of intrinsic variation. However, upon examining the $\sigma_{\rm int}$ of the DES subset, we find that it is in tension with the value found for the \lowz~subset. We therefore implement another set of BiasCor simulations with separate $\sigma_{\rm int}$ for each subset and we re-derive distances allowing for two separate $\sigma_{\rm int}$ in the nuisance parameter fitting stage of SALT2mu. This introduces a systematic uncertainty of 0.014 in $w$.

\textbf{Survey $\mu$-Bias Corrections:} For our nominal analysis we have followed the treatment in S18 and placed a cut on the Hubble residuals at $3.5\sigma$ from the best fit cosmological model. This cut results in a loss of 3 \lowz~SNe~Ia. In addition, we test a 3$\sigma$ cut that results in an additional 2 SNe~Ia cut from the \lowz~subset. No SNe~Ia from the DES subset are lost to outlier cuts. The size of the systematic uncertainty in the outlier cut is $\sigma_w=\lowzcutsys$. The uncertainty arising from statistical fluctuations in the determination of the spectroscopic selection efficiency is 0.007.

\textbf{Redshifts:} we have included two sources of systematic uncertainty associated with the redshifts used in this cosmological analysis. We find that while both the uncertainty in the peculiar velocities and a systematic redshift measurement offset must be accounted for, their contribution to the $w$-uncertainty budget is not yet comparable to that of distance uncertainties.

\textbf{New:} We have included several sources of systematics that have not been included in previous analyses. These are the redshift uncertainty, an uncertainty on the reported photometric errors, a change in the reference cosmology for simulations, outlier cuts to the \lowz~subset, and separate $\sigma_{\rm int}$'s for each subset. The outlier cut is the largest single source of uncertainty in our analysis and the separate treatment of $\sigma_{\rm int}$ is tied for the second largest. When all of these new systematic uncertainties are combined, we find $\sigma_w$=\werrSystNew, which is comparable to other systematic uncertainty groupings found in in Table \ref{Tab:finalbudget}.

\section{Validation of Analysis}

Here we describe our validation of the analysis using two separate sets of simulations. The first is based on 10,000 fake SNe~Ia light curves overlaid on images, and processed with \SMP, light curve fitting, BBC and CosmoMC. The second test uses a much larger catalog-level simulation from K18, and is processed as if they were a catalog produced by \SMP. While these validation tests could have revealed problems leading to additional systematic uncertainties, no such issues were identified and therefore no additional uncertainties are included. Nonetheless, the validation  tests were essential tools in developing the analysis framework and they provide added confidence in the final analysis. Since these validation tests are not sensitive to errors in calibration, nor to assumptions about SN properties, we caution their interpretation.

\label{validation}
\subsection{Fake SNe~Ia Overlaid Onto Images}
\label{fakes}

For the DES subset we simulate a sample of fake SNe~Ia light curves and insert light curve fluxes onto DES-SN images at locations near galaxy centers.  B18-\SMP\ use these fake transients to 1) measure biases associated with \SMP, 2) assess the accuracy of \SMP\ uncertainties and subsequently adjust errors in both data and simulations, and 3) optimize the photometric pipeline outlier rejection.  Here, we take this fake analysis one step further and perform a cosmology analysis resulting in a measurement of $w$. The benefit is that we can investigate potential biases that are not correctly modeled in early stages of the analysis (i.e.~the search pipeline), which could propagate to uncorrected biases in distances and fit cosmological parameters. While previous analyses (e.g., \citealt{Astier06}, B14) used fake transients to test their photometry pipelines, our test is the first to validate the cosmology analysis with fakes.  

\begin{figure}
\begin{centering}
\includegraphics[width=0.47\textwidth]{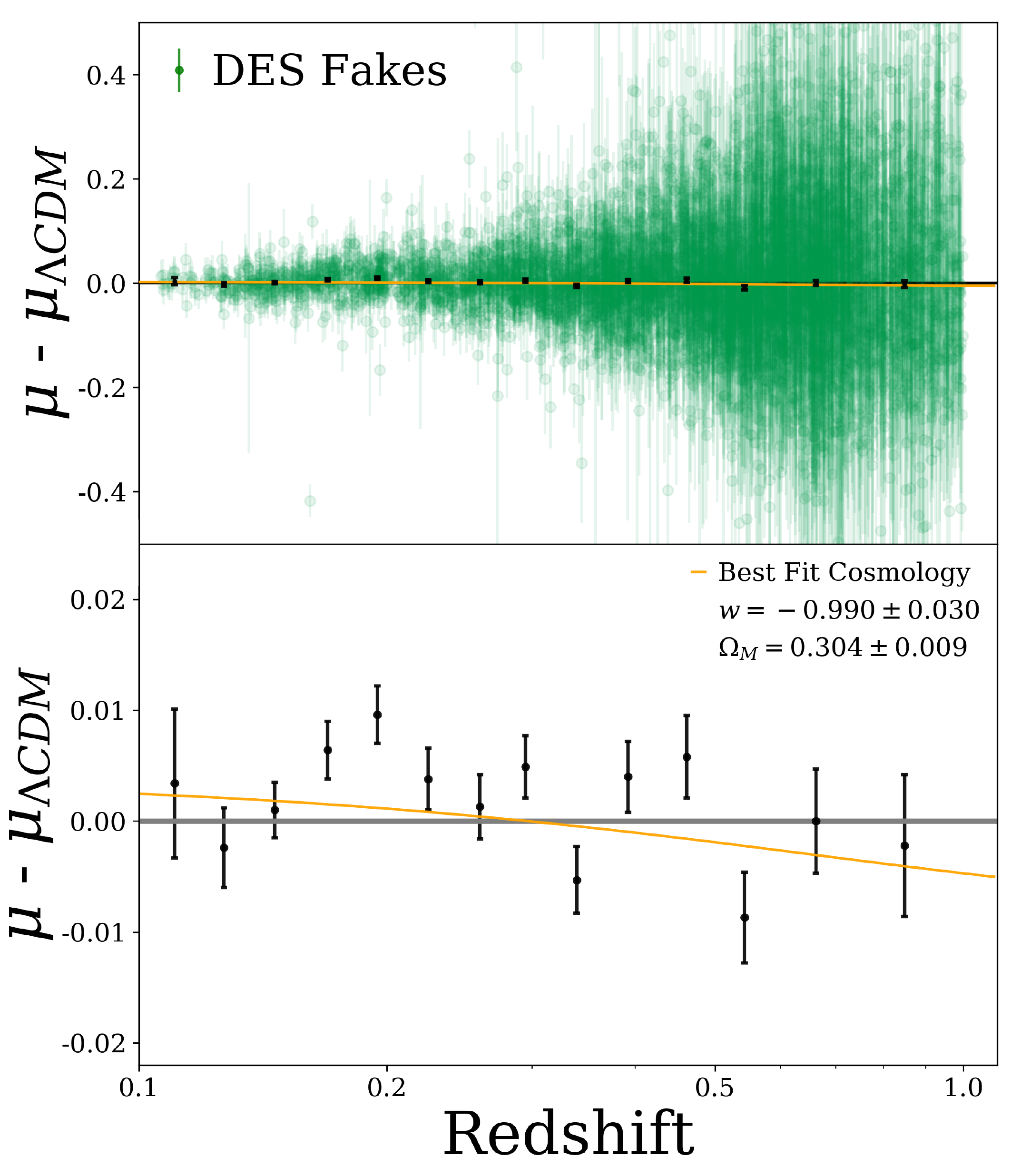}
\caption{Hubble residuals from 6586 fake SNe using the same analysis 
procedure as for the DES-SN3YR sample, except the CMB prior
is replaced with a Gaussian $\Omega_M$ prior.
\textbf{Upper:} zoomed out showing BBC bins and individual SNe on same y-scale as Fig \ref{fig:wwobiascor}.
\textbf{Lower:} zoomed in to show BBC-binned residuals more clearly. 
Black horizontal line corresponds to the flat $\Lambda$CDM model ($\Omega_M$=0.3) 
used to generate the fakes. Orange line is the best fit 
$w$CDM model, and best fit $w$ and $\Omega_M$ are shown on the lower panel.}
\label{fig:fakemu}
\vspace{.1in}
\end{centering}
\end{figure}

A sample of 10,000 fake SNe~Ia light curves are discovered by \texttt{DiffImg}, processed by \SMP, bias corrected with BBC, and run through our cosmological parameter fits with CosmoMC in the same fashion as the real dataset. A detailed description of the analysis of the 6,586 fakes that pass quality cuts is found in Appendix \ref{fakesimappendix}. The agreement between the BiasCor sample used to model the fakes dataset, and the fakes processed through our analysis pipelines is shown in Figure~\ref{fig:fakesims}, which is analogous to Figure~\ref{fig:overlay} for the real data. We analyze the fakes with BBC (Section \ref{bbc}) to produce a redshift-binned Hubble diagram and the BBC distances residuals to the input \LCDM~distances are shown in Figure~\ref{fig:fakemu} as a function of redshift. Cosmological fits to the fake SNe~Ia are not performed with Planck 2016 CMB priors because the underlying cosmology of Planck is unknown and therefore we cannot check for cosmological parameter biases. Instead, we perform $w$CDM fits on the binned distances with a prior on $\Omega_M \sim \mathcal{N}(0.3,0.01)$. The $\chi^2/$dof in Figure~\ref{fig:fakemu} is 2.5, however the amount of additional distance uncertainty per SN required to bring $\chi^2/$dof to unity is 4 mmag, which is much smaller than the intrinsic scatter in the DES-SN subset ($\sigma_{\rm int}^{\rm DES}=0.070$~mag).  Finally, we find $w=-0.990\pm0.030$ (yellow) which is consistent with the \LCDM~cosmology in which the fake SNe~Ia were generated. Since the $w$ bias from analyzing the fakes is consistent with zero, we do not assign a systematic uncertainty from this test.

\begin{figure}
\includegraphics[width=0.49\textwidth]{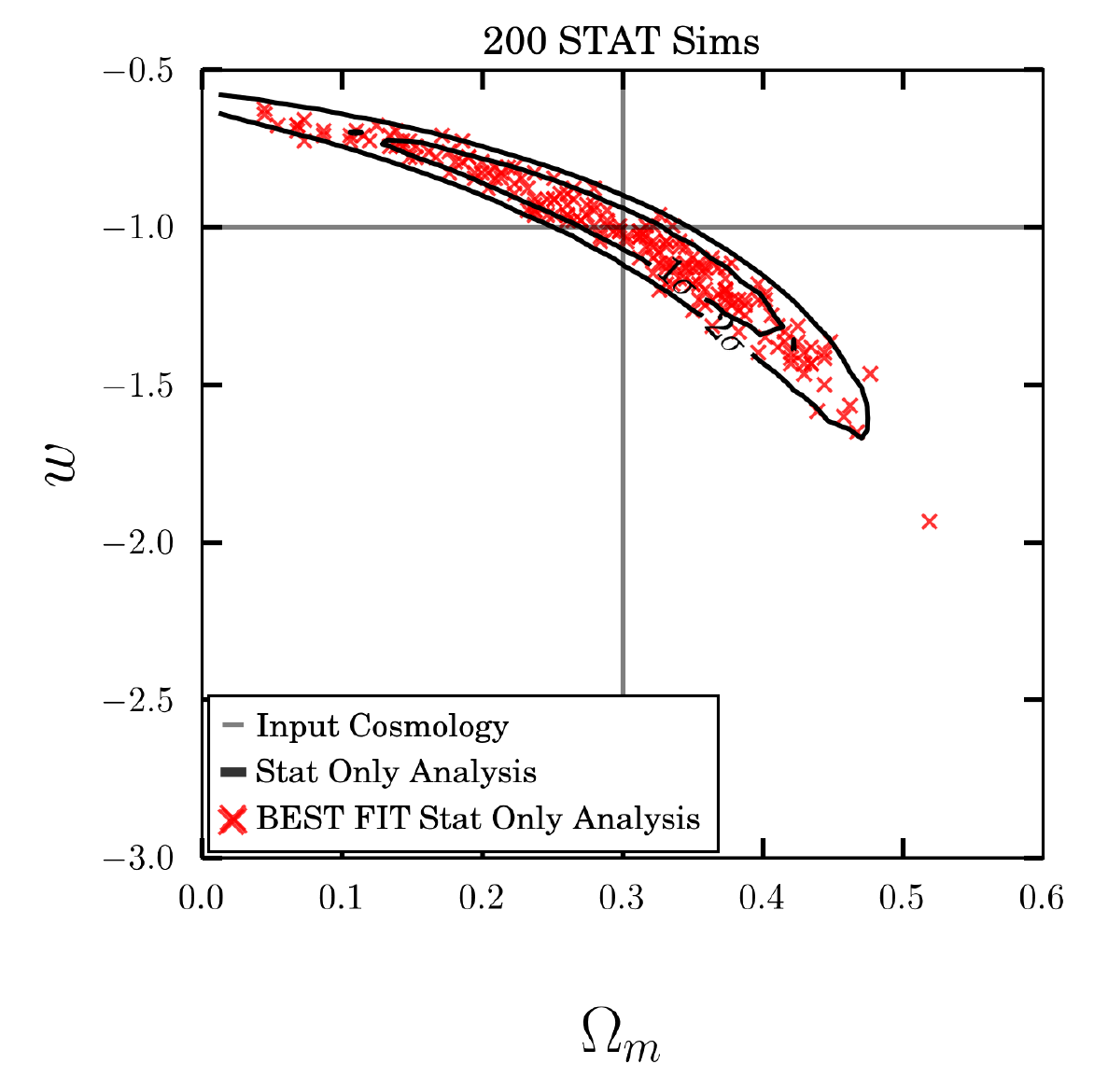}
\includegraphics[width=0.49\textwidth]{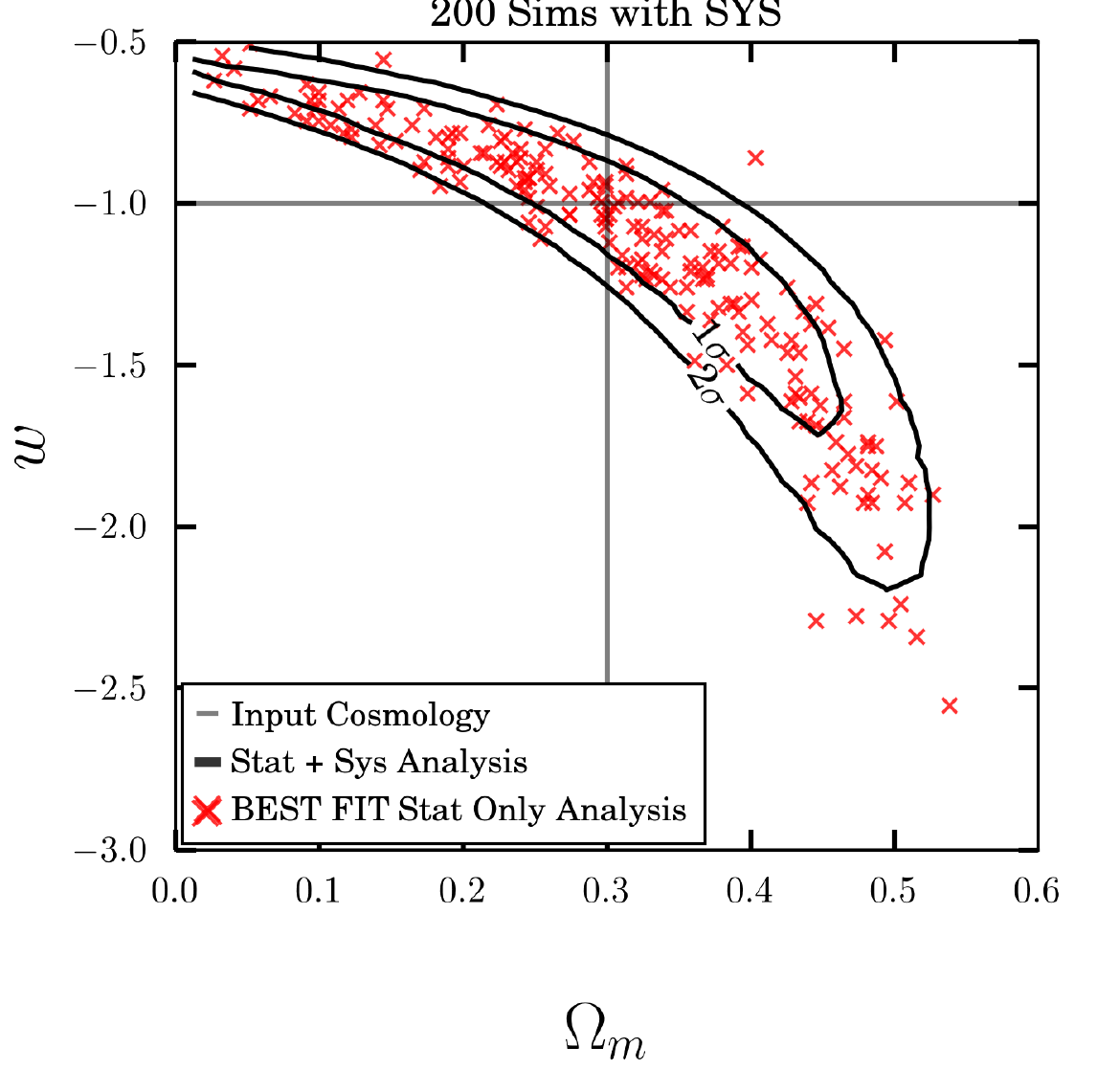}
\caption{\textbf{Top}: 200 simulated \SAMPLENAME~datasets with statistical only fluctuations. Best fits (red) and average posterior (black curve) are shown.  \textbf{Bottom}: 200 simulated \SAMPLENAME~datasets with input calibration systematic of 0.02~mag per filter. The best fit cosmological parameters for each of the 200 simulations from a BBC+CosmoMC analysis using ($C=C_{\rm stat}$) are shown in red. The average posterior from fits to the 200 simulations using $C=C_{\rm stat}+C_{\rm cal}$ is shown in black. All simulations are generated in the same input cosmology shown in the grey cross-hairs. All fits have a tophat prior on $\Omega_M \in  [0,1]$.}
\label{fig:validation2d}
\vspace{.1in}
\end{figure}

\subsection{Large Catalog-Level \texttt{SNANA} Simulations}
\label{analysisbiases}
\label{validationwithsim}

In contrast to the analysis with fakes, we perform our analysis on \texttt{SNANA} simulations that include systematic variations in both the DES-SN and low-$z$ samples. These simulations are used to check that our recovered cosmological parameters and their uncertainties are determined accurately.  

We begin by generating 200 data samples of comparable size to the \SAMPLENAME, each with independent statistical fluctuations, and with no systematic variations. Here we simulate and analyze using the G10 model only.
Each sample is processed with light curve fitting, BBC and CosmoMC using an $\Omega_M$ prior of $\mathcal{N}(0.3,0.01)$. We find an average $w$ bias consistent with zero (\fakestatbias) as shown in row 1 column 1 (r1,c1) of Table \ref{Tab:validation}. We also validate the size of our reported uncertainties. We compare the RMS of the 200 fitted $w$ values with the average reported $w$ uncertainty, defined as:
\begin{equation}
\label{eq:rsigma}
R_\sigma(w) = \langle\sigma_{w}\rangle/{\rm RMS}(w_{\rm statonly})
\end{equation}

We find that $R_\sigma(w)$ = 1.06 as shown in (r1,c3) of Table \ref{Tab:validation}, indicating that the average reported errors are in agreement with the RMS of fitted $w$ values
($R_{\sigma}(w)=1$ for perfect agreement).
In the top panel of Figure~\ref{fig:validation2d}, we combine the cosmological parameter posteriors of each of the 200 BBC fits by adding the $\chi^2$ contours in order to achieve an ``average" contour for the 200 realizations with size corresponding to the typical statistical uncertainty. We also show the best fit parameters for each of the 200 statistical realizations, calculated from each of the individual posterior peaks, and find that 135 (68\%) of the 200 best fits lie within the 1$\sigma$ contour.

In order to assess the treatment of multiple independent systematics, we run simulations with systematic biases in the zero point. For each band in each of 200 simulated G10 samples, we perturb the calibration with a randomly selected zero point shift from a Gaussian distribution with $\sigma$ = 0.02. This perturbation is for each sample, not each event, and is artificially inflated compared to our data calibration uncertainties ($\sim 6$ mmag) in an effort to improve the sensitivity of this test. Upon analyzing these simulations with BBC+CosmoMC in which we only account for the zero point uncertainty in the covariance matrix, we find again that the $w$ bias is consistent with zero (\fakezpbias) as shown in (row 2, column 1) of Table \ref{Tab:validation}.

In order to demonstrate the effect of the simulations with zero point systematics (bottom of Figure~\ref{fig:validation2d}), we show best fit parameters from stat-only analyses of each of the 
200 simulations with perturbed calibrations (red points).
Here we show the average CosmoMC contour using the stat+syst covariance matrix that accounts for zero point systematic uncertainty (black contour). 
We find that 139 (70\%) of the 200 best fits (stat-only) fall within the averaged one sigma contour (stat+syst), consistent with a 1$\sigma$ interpretation of the contour. This is also shown in  (row 2, column 4) of Table \ref{Tab:validation} where we demonstrate that after combining with the $\Omega_M$-prior, the RMS in fit $w$ from analyses with $C=C_{\rm stat}$ agrees with the average output uncertainties on $w$ from analyses where $C=C_{\rm stat} + C_{\rm syst}$: $R_\sigma(w)$ = 1.03. 

\newcommand{\aBiasG}{-0.0008\pm 0.0009} 
\newcommand{\bBiasG}{-0.024\pm 0.012}   
\newcommand{\RaVALG}{0.91} 
\newcommand{\RbVALG}{1.16}  

\newcommand{\aBiasC}{-0.0003\pm 0.0008} 
\newcommand{\bBiasC}{-0.022\pm 0.016}   
\newcommand{\RaVALC}{1.05} 
\newcommand{\RbVALC}{0.98} 

\newcommand{\Ra}{R_\sigma(\alpha)}
\newcommand{\Rb}{R_\sigma(\beta)}

\begin{deluxetable}{cllll}
\tablecolumns{5}
\tablecaption{ Bias and Uncertainty Precision for $\alpha$ and $\beta$ }
\tablehead {
\colhead { model\tablenotemark{a} }        &
\colhead { $\alpha$ bias } &
\colhead { $\beta$ bias } &
\colhead { $\Ra$ } &
\colhead { $\Rb$ }
}
\startdata
  G10   & $\aBiasG$ & $\bBiasG$ & $\RaVALG$ & $\RbVALG$ \\
  C11   & $\aBiasC$ & $\bBiasC$ & $\RaVALC$ & $\RbVALC$ \\
\enddata
\tablenotetext{a}{ Intrinsic scatter model used in simulated samples for data and bias corrections.}
\label{Tab:validate_ab}
\end{deluxetable}

In order to validate the treatment of the intrinsic scatter model systematic, we generate 100 realizations of \SAMPLENAME\ using both G10 and C11. When analyzing all 200 results from the 100 G10 simulations and 100 C11 simulations together using the averaged distances and covariances of Equations~\ref{eq:sysscatterdata}~\&~\ref{eq:sysscatter}, we find no biases (\fakescatbias) in recovered cosmological parameters as shown in (r3,c1) of Table \ref{Tab:validation}. We perform the same test on the output uncertainties described above  for the scatter model systematic and we find $R_\sigma$ = 1.00 (shown in r3,c3 of Table \ref{Tab:validation}). However, because our set of distances used to compute cosmological parameters is averaged between the best fit distances of each scatter model, we expect subtle biases when evaluating simulations created with a single model of intrinsic scatter. In analyzing only the 100 G10 realizations combined with $\Omega_M \sim \mathcal{N}(0.3,0.01)$, we find a $w$ bias of $-0.03$, and for the 100 C11 realizations a $w$ bias of $+0.03$. We note that combining SNe with the prior on $\Omega_M$ is weaker than combining SNe and \cite{Planck16_cosmo} CMB constraints by roughly 50\%. The $w$ shift for each scatter model when combining with CMB becomes $\pm0.014$, which is in agreement with the systematic uncertainty derived from Eq.~\ref{eq:csys} and shown in Table \ref{Tab:finalbudget}.

We check the recovery of the BBC fitted parameters for $\alpha$ and $\beta$
in Table~\ref{Tab:validate_ab}. 
Analogous to $R_{\sigma}(w)$ for $w$-uncertainties (Eq.~\ref{eq:rsigma}), we define $\Ra$ and $\Rb$ for statistical-only fits of $\alpha$ and $\beta$ respectively.
For both intrinsic scatter models, the $\alpha$ bias is consistent with zero.
For $\beta$, there is a hint of bias at the sub-percent level.
The uncertainties and RMS ($\Ra,\Rb$) agree at the 10\% level for G10, and at the few percent level for C11.

Finally, we generate large simulations of \SAMPLENAME\ with two separate values of $\sigma_{\rm int}$ for each subset to examine the biases in our analysis. We analyze with BiasCor simulations generated with two separate values of $\sigma_{\rm int}$ and find that $\langle w \rangle = -1.002 \pm  0.008$ after combining with $\Omega_M \sim \mathcal{N}(0.3,0.01)$. The lack of bias when accounting for the two $\sigint$ in BiasCor simulations ensures that our treatment of this systematic has been implemented correctly. We also analyze the same realizations our Nominal BiasCor, which use a single value for $\sigint$, and find $\langle w \rangle = -1.036 \pm 0.008$. The observed bias in $w$ when analyzing with our nominal analysis justifies the inclusion an additional systematic uncertainty. We note again that combining SNe with the prior on $\Omega_M$ is weaker than combining with CMB by roughly 50\% and thus the associated systematic uncertainty reported in Table \ref{Tab:finalbudget} is smaller.

\begin{deluxetable*}{c|ccccc}
\tablecolumns{6}
\tablewidth{.9\textwidth}
\tablecaption{Summary of Validation Results from Simulations}
\tablehead {
\colhead { Column}                &
\colhead { 1}                &
\colhead {2}                &
\colhead {3} &
\colhead {4} &
\colhead {} \\ \hline \\
\colhead { Row}                &
\colhead { $\bar{w}+1$\tablenotemark{a}}                &
\colhead {RMS($w_{\rm statonly}$) }                &
\colhead {$\langle\sigma_{w}\rangle$} &
\colhead {$R_\sigma$\tablenotemark{b}} &
\colhead {Description} 
}
\startdata
1&\fakestatbias&	0.047&	0.050&		1.06& Statistical  \\
2&\fakezpbias& 0.098&  0.101&          1.03 & ZP Systematic\tablenotemark{c}  \\
3&\fakescatbias&    0.076&    0.076&    1.00 & Intrinsic Scatter Model\tablenotemark{d}  \\
\enddata
\tablecomments{200 ``DES Like" realizations with and without input sources of systematic uncertainty. All simulations are fit with an $\Omega_M = 0.3 \pm .01$ prior.} 
\tablenotetext{a}{Inverse variance weighted average.}
\tablenotetext{b}{$R_\sigma$, defined in Eq.~\ref{eq:rsigma}. }
\tablenotetext{c}{ZP Systematic corresponds to a zero point magnitude offset drawn from a Gaussian distribution of width 0.02~mag for each band independently in each of the 200 simulations.} 
\tablenotetext{d}{Intrinsic scatter model systematic corresponds to 200 simulations, 100 with each model of intrinsic scatter (G10 and C11).\\}
\label{Tab:validation}
\vspace{.1in}
\end{deluxetable*}

\section{Development of Bayesian Model Fitting}
\label{bhm}

\newcommand{\BHMname}{\textit{Steve}}

One of the predominant issues in supernova cosmology is that color and stretch uncertainties
are assumed to be Gaussian and symmetric. This assumption is not valid when the uncertainties
are comparable to the intrinsic width of the underlying population.

This issue has historically been addressed in two different ways. 
The first method, used by BBC, determines the true populations of stretch and color (SK16)
and in a separate step determines bias corrections with simulations.
The second method is to construct a model in which the true underlying values for color and
stretch are parametrized \citep{March11}. Bayesian Hierarchical Models (BHM) have been developed that both utilize
bias-correct observables \citep{BAHAMAS} and incorporate selection effects
directly into the model \citep{UNITY}. Here we summarize a new method
called \BHMname\ (H18: \citealt{Hinton18}),
which makes use of detailed \SNANA\ simulations to describe the
selection efficiency as part of the likelihood. 
In addition, \BHMname\ does not make assumptions about the underlying 
intrinsic scatter model, and it uses a parameterized treatment 
of systematic uncertainties. Although this method is still under 
development, here we illustrate progress by describing its performance 
on simulated validation samples and the DES-SN3YR sample.

H18 validate \BHMname\ on the same set of 200 DES-SN3YR 
simulations as described in Section 6. 
For the sample generated with the G10 model there is no $w$ bias,
while for the sample generated with C11 there is a bias of 0.05.
When evaluating all 200 validation simulations (G10 and C11 combined), 
\BHMname\ results in an average $w$ bias of $+0.03$ and an average $w$ difference 
($\Delta w$) between \BHMname\ and the nominal method (BBC+CosmoMC) is $+0.04$.
The corresponding RMS on $\Delta w$ is $0.06$, where this additional scatter
comes from the inclusion of fitted parameters in \BHMname\ that are fixed in the BBC fit. For example, \BHMname\ allows for redshift dependent populations, that are not in the BBC fit because we find no evidence for such a dependence (Section \ref{alphabeta}). The extra parameters also result in a larger $w$ uncertainty for \BHMname\ in comparison to BBC. 

To predict $\Delta w$ for the \SAMPLENAME\ sample, we take the mean $\Delta w$
from the validation sims. For the RMS, however, the validation
sims are fit with a Gaussian $\Omega_M$ prior, $\mathcal{N}(0.3, 0.01)$, which is less
stringent than the CMB prior used to fit the data. Fitting with 
both priors shows that the validation uncertainties are over-estimated
by a factor of 1.7, and therefore for \SAMPLENAME\  we expect
${\rm RMS}(\Delta w) = 0.04$. On the \SAMPLENAME\ dataset, we find a $w$-difference of $0.07$, which is consistent with our simulated prediction of $0.04 \pm 0.04$.

\begin{deluxetable}{clll}
\tablecolumns{4}
\tablecaption{Comparison of \BHMname\ and BBC \\ Nuisance Parameters for \SAMPLENAME}
\tablehead {
\colhead {} &
\colhead { \BHMname\ } &
\colhead { BBC(G10)} &
\colhead { BBC(C11)} 
}
\startdata

  $\alpha$     & $0.166\substack{+0.008 \\ -0.015}$ &    $0.146+ \pm 0.009$ & $0.147 \pm 0.009$\\ \\
  $\beta$      & $3.54\substack{+0.12 \\ -0.20}$       &    $3.03 \pm 0.11$ & $3.58 \pm 0.14$\\ \\
  $\gamma$      & $0.029\substack{+0.020 \\ -0.028}$   &    $0.025 \pm 0.018$ & $0.016 \pm 0.018$\\ \\
  $\sigma_{\rm int}$(low-$z$) &$0.197\substack{+0.018 \\ -0.017}$ & $0.120 \pm 0.015$ & $0.144 \pm 0.015$\\ \\
  $\sigma_{\rm int}$(DES-SN) & $0.034\substack{+0.030 \\ -0.016}$  & $0.066 \pm 0.006$ &  $0.087 \pm 0.006$\\

\enddata
\tablecomments{BBC(G10) and BBC(C11) refer to intrinsic scatter model used to compute bias corrections.}
\label{Tab:bhmvbbc}
\end{deluxetable}

%
%
\newcommand{\aBHM}{$\alpha_{\BHMname}$}
\newcommand{\aBBC}{$\alpha_{\rm BBC}$}
\newcommand{\bBHM}{$\beta_{\BHMname}$}
\newcommand{\bBBC}{$\beta_{\rm BBC}$}
\newcommand{\gBHM}{$\gamma_{\BHMname}$}
\newcommand{\gBBC}{$\gamma_{\rm BBC}$}

The fitted nuisance parameters from \BHMname\ are compared to those
from the BBC method in Table \ref{Tab:bhmvbbc}. The \aBHM\ value is about 0.02 higher than \aBBC\ 
and \bBHM\ is consistent with \bBBC\ using the C11 
intrinsic scatter model in the bias-correction simulation.
$\gamma$ for both methods is consistent with zero, although
\gBHM\ is more consistent with \gBBC\ using the G10 model.
Both methods show that the intrinsic scatter term ($\sigma_{\rm int}$) 
is significantly different between the \lowz\ and DES subsets,
although the $\sigma_{\rm int}$ agreement between the two methods 
is marginal.

\section{Discussion}
\label{discussion}

\subsection{Comparison with Other Samples}

\begin{figure}
\includegraphics[width=0.48\textwidth]{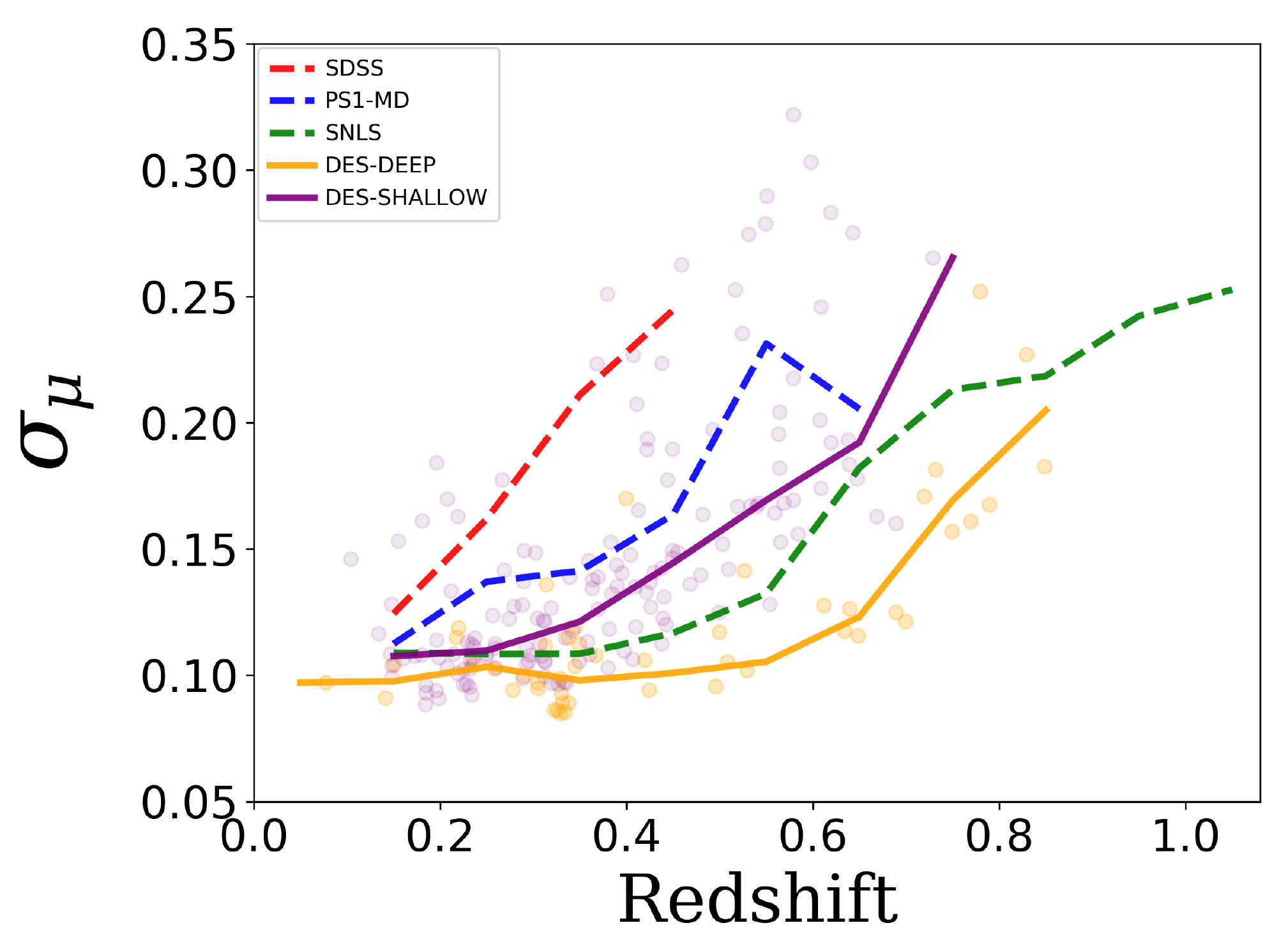}
\caption{Distance modulus uncertainty vs. redshift for DES, PS1, SNLS, and SDSS. Distance modulus measurement uncertainties reported by each survey are combined with the $\sigma_{\rm int}$ from this work (DES) and from S18 (PS1,SNLS,SDSS). The colored dots are the individual SNe~Ia from the DES Shallow (purple) and Deep (yellow) fields. The solid (DES) and dashed (other) lines are the binned medians of the respective distributions. \\ }
\label{fig:muerr_raw}
\vspace{.1in}
\end{figure}

For the nominal analysis using BBC+CosmoMC, statistical and systematic uncertainties on $w$ from \totsn~ \SAMPLENAME~SNe~Ia are \statwerr~(stat) and \systwerr~(syst). Previous surveys have also found that their statistical and systematic uncertainties are roughly equal. S18 analyzed the PS1 plus low-$z$ subset of the Pantheon sample, and these 451 events result in a statistical and systematic uncertainties on $w$ of 0.046 (stat) and 0.043 (syst). Additionally, in the Joint Light Curve Analysis (B14) they report an uncertainty on $w$ of 0.057 (stat+syst) using 740 SNLS+SDSS+low-$z$+HST SNe~Ia. The \SAMPLENAME~result is a notable improvement in constraining power on $w$ for the given sample size (\totsn~SNe~Ia), despite the consideration of new sources of systematic uncertainty. Much of this improvement is due to the quality of the DECam CCDs which include higher sensitivity to redder wavelengths (\citealt{holland03}, \citealt{diehl08}) resulting in improved distance constraints for the most distant supernovae. A comparison of distance uncertainties is shown in Figure~\ref{fig:muerr_raw} using the measurement uncertainties from each respective survey combined with the $\sigma_{int}$ for each survey that was derived in S18. We find that the DES-SN deep field SNe~Ia have smaller uncertainties in distance modulus than SNLS, and the DES-SN shallow field SNe~Ia have smaller uncertainties than PS1 but larger than SNLS.

\subsection{Prospects for Improving Systematic Uncertainties}

There are several prospects for future reduction of systematic uncertainties, the largest of which is due to calibration. Multiple improvements are in development for the calibration of the DES photometric system. In this work we used a single HST Calspec standard in one of the SN fields to link our photometric magnitudes to the AB system. In the last two seasons of the survey, we measured $ugrizY$ photometry for two other CalSpec standards (DA White Dwarfs) that are within the DES footprint. We have identified a large number of hot DA White Dwarfs ($\sim100)$ which are faint enough to avoid saturation in our nominal 90 sec exposures, but bright enough to collect ground-based spectra of suitable quality for analysis. In addition to absolute calibration, there are also prospects to reduce the uncertainty due to internal calibration. A publication dedicated to detailing bandpass measurement corrections, stellar catalog improvements, and code improvements is forthcoming.

The next largest source of systematic uncertainty is from the model of intrinsic scatter, with $\sigma_w$ = 0.014.  Our 
\lowz\ subset is redder than the DES-SN and other high-$z$ populations because it was part of a targeted selection of host-galaxies. The different color population of the 
\lowz~subset results in increased sensitivity to the change in scatter model. Additionally, we find that our dataset is more sensitive to the intrinsic scatter model uncertainty than S18. This is because the \lowz~sample is a larger fraction of our cosmological sample (\SAMPLENAME) in comparison to S18. The two intrinsic scatter models are nearly 8 years old and there are currently more than $\sim$1300 SNe~Ia from Pantheon + DES-SN that could potentially test the validity of either G10 or C11. We leave this study for a future analysis.

Equally as large as the intrinsic scatter systematic uncertainty is the $w$ uncertainty in the parent populations of color and stretch. There is room for improvement here in two respects: in our analysis methods and in the external dataset used.

First, in Section \ref{sys:parent} of our analysis, we employed a similar ad-hoc procedure as S18 to characterize the uncertainty in the 6 parameters describing the parent populations of color and stretch based on estimates from \cite{scolnickessler16}. A more rigorous method of accounting for these parameter uncertainties and covariances in the BBC method is needed for future analyses.

Second, there is room for improvement from combining with \lowz\ datasets with selection effects that are less severe and better understood. The Foundation supernova survey (\citealt{Foundation}; hereafter F18) has the potential to reduce this uncertainty for the low-$z$ sample. Foundation measures light curves for SNe~Ia discovered by other rolling surveys (ASA-SN, ATLAS, etc...) and as a result, obtains a sample with less galaxy-selection bias than the current low-$z$ sample. The Foundation \lowz~survey on the Pan-STARRS telescope has released 225 \lowz~SNe~Ia in DR1 and they are still collecting data with the goal of obtaining up to 800 $griz$ light curves with high quality calibration. They find that the median color ($c=-0.035$) and stretch  ($x_1=0.160$) of the Foundation SNe~Ia DR1 sample are much closer to that of the high-z surveys  (i.e.~DES: $c=-0.037$ , $x_1=0.115$) compared to the medians of the distributions of the current \lowz~sample (i.e.~CfA,\mbox{CSP-1}: $c=-0.021$ , $x_1=0.048$).

The Foundation \lowz~survey may also provide insight into the distribution of residuals to the Hubble diagram at low redshift. In the \SAMPLENAME~analysis we find a significant source of systematic uncertainty ($\sigma_w = \lowzcutsys$) associated with the outlier cut of the \lowz~subset due to non-Gaussian tails in residuals to the best fit cosmological model. Additional statistics will better allow us to characterize the distribution of \lowz~SNe~Ia about the Hubble diagram. The non-Gaussian Hubble residuals could be related to data quality, galaxy selection effects, unknown astrophysical effects, or poor SN modeling which is more apparent at high S/N. In any case, the Foundation \lowz~sample will facilitate further study of this systematic. 

\subsection{Host Mass Hubble Residual Step and Intrinsic Scatter}

For DES-SN3YR, we find small values for both $\gamma$ and $\sigma_{\rm int}$. For the DES subset, $\gamma$ is consistent with zero, indicating no evidence of a correlation between the Hubble residuals to our best fit cosmology and host-galaxy stellar mass. A significant correlation has been seen to varying degrees in previous analyses (\citealt{Sullivan10}, \citealt{Kelly10}, \citealt{Lampeitl10}, \citealt{gupta11}, \citealt{dandrea11}, \citealt{smith12}, \citealt{wolf}, \citealt{Rigault13}) and S18 recalculated these quantities within the BBC framework and recovered non-zero steps of size: SDSS ($0.057 \pm 0.015$ mag), Pan-STARRS ($0.039 \pm 0.016$ mag), SNLS ($0.045 \pm 0.020$ mag), and low-$z$ ($0.076 \pm 0.030$ mag). In an upcoming work, we plan to simulate the correlations between color and host-galaxy stellar mass, and the host-galaxy stellar mass Hubble residual step itself. However, because we recover a non-zero $\gamma$ value for the low-$z$ sample as seen in previous analyses, we suspect that the null correlation found for the DES subset may be the result of selecting a different population of SNe or host galaxies, but not the result of our analysis techniques.

For future surveys such as LSST and WFIRST, as well as for low redshift studies of SNe~Ia for precision $H_0$ measurements, it will be important to improve analysis techniques and study selection effects on the host-galaxy stellar mass correlation, especially if this effect evolves with redshift (\citealt{Childress14}). Although, in \SAMPLENAME, we did not find evidence of evolution of $\gamma$ as a function of redshift.

\begin{figure}
\begin{centering}
  \includegraphics[width=0.48\textwidth]{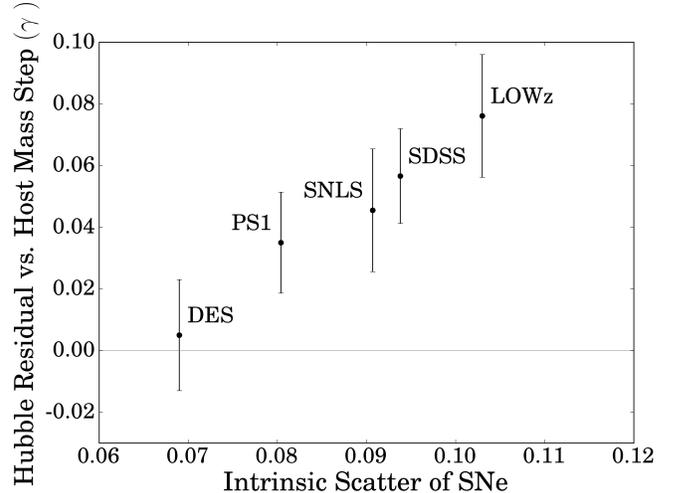}

\caption{Hubble residual step size in mags ($\gamma$) as a function of the intrinsic scatter ($\sigma_{\rm int}$) of SNe~Ia samples. The largest rolling surveys (DES, PS1, SNLS, SDSS) are shown in addition to the targeted \lowz~subset used in this analysis. Values for the non-DES points are taken from \cite{scolnic18} and all are calculated using 5D bias corrections using BBC fit for consistency. \\}
\label{fig:scolniceffect}
\end{centering}
\end{figure}

Future SN cosmology analyses will also be faced with the decision whether to include two $\sigint$. We have found that the $\sigint$ values of the \lowz~and DES subsets are incompatible. In this work our nominal analysis assumes a single value for $\sigint$ for historical reasons, however we find that the systematic associated with this choice is one of the largest sources of uncertainty in our analysis. Interestingly, looking at recent SNe~Ia datasets all analyzed with the SALT2 model and BBC 5D formalism, we find a correlation between $\gamma$ and $\sigma_{\rm int}$ of the individual samples. Figure~\ref{fig:scolniceffect} shows this correlation for the DES and low-$z$ subsets as well as for the other surveys analyzed in S18. The incompatibility between the DES subset and the low-$z$ subset does not appear to be unique to the \lowz~data used in this analysis (CfA and \mbox{CSP-1}). \cite{Foundation} report in their initial data release an intrinsic scatter of 0.111. The $\sigma_{\rm int}$-$\gamma$ correlation could be a measurement artifact or $\sigma_{\rm int}$ could have astrophysical dependence. Future work will be focused on probing the possibility of a redshift dependent intrinsic scatter term, but will require the use of larger datasets. As uncertainty budgets shrink with new and larger SNe~Ia samples, it will become important for future analyses to better characterize this effect and model it in simulations.

\subsection{Simulating SNe~Ia Samples}
\label{discussion:simulating}
We have shown that there is still room for improvement in modeling the simulated $P_{\rm fit}$ distributions (Figure~\ref{fig:overlay}). We find that the agreement for the DES-SN sample is better than that of the low-$z$ sample, especially in the range $P_{\rm fit}<0.5$. This is in part due to the extensive care taken to accurately simulate the DES-SN sample as described in Sec 5.1.1 of K18, however it is unclear if the lesser agreement in the low-$z$ sample could be the result of unmodeled astrophysics. For the DES-SN sample, there is disagreement between the $P_{\rm fit}$ distributions of the simulations and the data in the highest bin ($P_{\rm fit}>0.95$). We also see a similar disagreement at the high end when comparing the simulated and fake SN distributions (Figure~\ref{fig:fakesims}). Since the same discrepancy is seen with the fakes, we rule out the possibility that this is entirely due to SN modeling. 

The $P_{\rm fit}$ agreement between simulations and data for the low-$z$ sample is poor at both low and high $P_{\rm fit}$ (Figure~\ref{fig:overlay}). This disagreement will hopefully be improved with the Foundation sample, which will facilitate more accurate simulations. In addition, our DES-SN sample has an additional 90,000 fake supernovae on which we can run \SMP\ and improve our modeling of flux uncertainties in the simulation.

\subsection{Improvements to the Validation}
\label{discussion:validationimprovements}
The validations described in Section \ref{validation}
are the most extensive for a
SN~Ia cosmology analysis pipeline to date. 
Using fakes we have validated from discovery on DECam images
to cosmological parameters,
and using catalog-level simulations
and we have validated the $w$ bias (\mbox{$<0.01$}) and treatment of systematic uncertainty.
Future work will expand the number of systematics in Table~\ref{Tab:validation}.  Additionally, because we utilize BBC, which uses an approximate $\chi^2$ likelihood assuming
symmetric Gaussian uncertainties, we will validate the BBC confidence region for binned distances, and this will eventually lead to comparing the cosmology likelihoods between the BBC+CosmoMC and Bayesian 
(\BHMname, Section \ref{bhm}) methods. In addition to comparing likelihoods between methods, ideally we would compare our
BBC+CosmoMC likelihood to a true likelihood such as from the Neyman construction (\citealt{pdg}).
However, such a comparison is computationally challenging.

\section{Conclusion}
\label{conclusion}

We have presented the analysis, cosmological parameter uncertainty budget, and validation of \SAMPLENAME~sample consisting of of 207 spectroscopically confirmed Type Ia Supernovae ($0.1 < z < 0.85$) discovered by DES-SN and an external sample of 122 low-redshift SNe~Ia after quality cuts ($0.01 < z < 0.1$).  The cosmology constraints are given in the DES-SN key paper (\KEYPAPERalt).  We find a total uncertainty $\sigma_w$ = 0.057 (stat+syst). The calibration of the various samples used is the largest source of systematic uncertainty. Additionally we find no correlation between host-galaxy stellar mass and Hubble residuals to the best fit cosmology.

Our validation using a population of fake SNe injected onto real images is the first such test for potential biases through the entire SNe~Ia discovery, photometry, and analysis pipelines. Resulting biases in distance are limited to 1\% and the fit value of $w$ is consistent with the cosmology in which the fakes were generated. Additionally, we discuss a rigorous method of validating the interpretation of the total uncertainty budget using hundreds of catalog-level simulations. We find that after accounting for sources of systematic uncertainty there are no significant biases in the cosmological parameter analysis pipeline and that the RMS($w$) and the average uncertainty agree to within 6\%.  The sample from DES used for this analysis is roughly $10\%$ of the full DES photometric sample, and treatment and validation of systematic $w$ uncertainties will become even more crucial with the larger sample.\\

\acknowledgments

This paper has gone through internal review by the DES collaboration. DB and MS were supported by DOE grant DE-FOA-0001358 and NSF grant AST-1517742. This research used resources of the National Energy Research Scientific Computing Center (NERSC), a DOE Office of Science User Facility supported by the Office of Science of the U.S. Department of Energy under Contract No. DE-AC02-05CH11231. We are grateful for the support of the University of Chicago Research Computing Center for assistance with the calculations carried out in this work. Part of this research was conducted by the Australian Research Council Centre of Excellence for All-sky Astrophysics (CAASTRO), through project number CE110001020.

Funding for the DES Projects has been provided by the U.S. Department of Energy, the U.S. National Science Foundation, the Ministry of Science and Education of Spain, 
the Science and Technology Facilities Council of the United Kingdom, the Higher Education Funding Council for England, the National Center for Supercomputing 
Applications at the University of Illinois at Urbana-Champaign, the Kavli Institute of Cosmological Physics at the University of Chicago, 
the Center for Cosmology and Astro-Particle Physics at the Ohio State University,
the Mitchell Institute for Fundamental Physics and Astronomy at Texas A\&M University, Financiadora de Estudos e Projetos, 
Funda{\c c}{\~a}o Carlos Chagas Filho de Amparo {\`a} Pesquisa do Estado do Rio de Janeiro, Conselho Nacional de Desenvolvimento Cient{\'i}fico e Tecnol{\'o}gico and 
the Minist{\'e}rio da Ci{\^e}ncia, Tecnologia e Inova{\c c}{\~a}o, the Deutsche Forschungsgemeinschaft and the Collaborating Institutions in the Dark Energy Survey. 

The Collaborating Institutions are Argonne National Laboratory, the University of California at Santa Cruz, the University of Cambridge, Centro de Investigaciones Energ{\'e}ticas, 
Medioambientales y Tecnol{\'o}gicas-Madrid, the University of Chicago, University College London, the DES-Brazil Consortium, the University of Edinburgh, 
the Eidgen{\"o}ssische Technische Hochschule (ETH) Z{\"u}rich, 
Fermi National Accelerator Laboratory, the University of Illinois at Urbana-Champaign, the Institut de Ci{\`e}ncies de l'Espai (IEEC/CSIC), 
the Institut de F{\'i}sica d'Altes Energies, Lawrence Berkeley National Laboratory, the Ludwig-Maximilians Universit{\"a}t M{\"u}nchen and the associated Excellence Cluster Universe, 
the University of Michigan, the National Optical Astronomy Observatory, the University of Nottingham, The Ohio State University, the University of Pennsylvania, the University of Portsmouth, 
SLAC National Accelerator Laboratory, Stanford University, the University of Sussex, Texas A\&M University, and the OzDES Membership Consortium.

Based in part on observations at Cerro Tololo Inter-American Observatory, National Optical Astronomy Observatory, which is operated by the Association of 
Universities for Research in Astronomy (AURA) under a cooperative agreement with the National Science Foundation.

This paper makes use of observations taken using the Anglo-Australian Telescope under programs ATAC A/2013B/12 and  NOAO 2013B-0317; the Gemini Observatory under programs NOAO 2013A-0373/GS-2013B-Q-45, NOAO 2015B-0197/GS-2015B-Q-7, and GS-2015B-Q-8; the Gran Telescopio Canarias under programs GTC77-13B, GTC70-14B, and GTC101-15B; the Keck Observatory under programs U063-2013B, U021-2014B, U048-2015B, U038-2016A; the Magellan Observatory under programs CN2015B-89; the MMT under 2014c-SAO-4, 2015a-SAO-12, 2015c-SAO-21; the South African Large Telescope under programs 2013-1-RSA\_OTH-023, 2013-2-RSA\_OTH-018, 2014-1-RSA\_OTH-016, 2014-2-SCI-070, 2015-1-SCI-063, and 2015-2-SCI-061; and the Very Large Telescope under programs ESO 093.A-0749(A), 094.A-0310(B), 095.A-0316(A), 096.A-0536(A), 095.D-0797(A).

The DES data management system is supported by the National Science Foundation under Grant Numbers AST-1138766 and AST-1536171.
The DES participants from Spanish institutions are partially supported by MINECO under grants AYA2015-71825, ESP2015-66861, FPA2015-68048, SEV-2016-0588, SEV-2016-0597, and MDM-2015-0509, 
some of which include ERDF funds from the European Union. IFAE is partially funded by the CERCA program of the Generalitat de Catalunya.
Research leading to these results has received funding from the European Research
Council under the European Union's Seventh Framework Program (FP7/2007-2013) including ERC grant agreements 240672, 291329, 306478 and 615929.
We  acknowledge support from the Australian Research Council Centre of Excellence for All-sky Astrophysics (CAASTRO), through project number CE110001020, and the Brazilian Instituto Nacional de Ci\^encia
e Tecnologia (INCT) e-Universe (CNPq grant 465376/2014-2).

This manuscript has been authored by Fermi Research Alliance, LLC under Contract No. DE-AC02-07CH11359 with the U.S. Department of Energy, Office of Science, Office of High Energy Physics. The United States Government retains and the publisher, by accepting the article for publication, acknowledges that the United States Government retains a non-exclusive, paid-up, irrevocable, world-wide license to publish or reproduce the published form of this manuscript, or allow others to do so, for United States Government purposes.

Based in part on data acquired through the Australian Astronomical Observatory, under program A/2013B/012. We acknowledge the traditional owners of the land on which the AAT stands, the Gamilaraay people, and pay our respects to elders past and present.

\bigskip

\appendix

\renewcommand{\thefigure}{D.\arabic{figure}}

\setcounter{figure}{0}

\renewcommand{\thetable}{C.\arabic{table}}
\setcounter{table}{0}

\section{A. Light curve Minimization Algorithms}
\label{minuit}
Light curve parameter minimization is performed with \texttt{SNANA}'s implementation of SALT2 (\citealt{SALT2}) based on CERNLIB’s \texttt{MINUIT} program (\citealt{MINUIT}) using \texttt{MINOS} minimization. There is an alternative minimization method, \texttt{MIGRAD}, however we found that it causes pathological errors for 2\% of our sample of SNe~Ia, resulting in incorrect weighting in the SALT2mu distance fitting process. \texttt{MINOS} was found to avoid the pathological color errors although it is 2.5x slower than \texttt{MIGRAD}. \texttt{MIGRAD}'s speed is useful for development and debugging, however for the final cosmological analysis we use \texttt{MINOS}.

There are additional fitting anomalies that occur for high-SNR events
for both \texttt{MIGRAD} and \texttt{MINOS}. These algorithms sometimes fall in false minima, and to avoid these anomalies we add 3\% of peak SN flux
to all flux uncertainties on the first of three fit iterations. 

\begin{deluxetable*}{c|ccccccc}
\tablecolumns{8}
\tablewidth{.95\textwidth}
\tablecaption{light curve Fit Parameters}
\tablehead {
\colhead { SN-ID }               &
\colhead {$z_{\rm CMB}$  }                &
\colhead {c  }                &
\colhead {$x_1$  }                &
\colhead {$m_B$  }                &
\colhead {$\log(\mathcal{M}_{\rm stellar}/\mathcal{M}_\sun)$ }                &
\colhead {$\mu$ } &
\colhead {$\mu_{\rm corr}$ }  }
\startdata

1248677 & 0.350 & -0.093 $\pm$ 0.022 & 1.01 $\pm$ 0.11 & 21.530 $\pm$ 0.027 & 9.845 $\pm$ 0.014 & 41.305 $\pm$ 0.103 & 0.014 $\pm$ 0.004 \\
1250017 & 0.182 & -0.096 $\pm$ 0.026 & 1.06 $\pm$ 0.16 & 20.038 $\pm$ 0.035 & 8.797 $\pm$ 0.038 & 39.827 $\pm$ 0.108 & 0.020 $\pm$ 0.006 \\
1253039 & 0.454 & -0.094 $\pm$ 0.026 & 0.29 $\pm$ 0.26 & 22.288 $\pm$ 0.030 & 10.795 $\pm$ 0.140 & 41.986 $\pm$ 0.119 & -0.018 $\pm$ 0.005 \\
1253101 & 0.460 & 0.027 $\pm$ 0.033 & 1.34 $\pm$ 0.36 & 22.412 $\pm$ 0.040 & 8.526 $\pm$ 0.194 & 41.873 $\pm$ 0.123 & -0.001 $\pm$ 0.008 \\
1253920 & 0.196 & -0.085 $\pm$ 0.027 & -0.78 $\pm$ 0.14 & 20.330 $\pm$ 0.033 & 9.234 $\pm$ 0.033 & 39.818 $\pm$ 0.110 & 0.007 $\pm$ 0.007 \\

\enddata
\tablecomments{SN-ID, redshift, light curve fit parameters, host-galaxy stellar mass, distance moduli, and distance bias corrections of DES-SNe after quality cuts using the G10 model of intrinsic scatter. A subset of SNe are shown here. The full version of this table can be found online following the link in Appendix \ref{dr} for both models of intrinsic scatter (G10 and C11) as well additional information including RA, DEC, fit parameter covariances, 5D bias corrections, and more.\\}
\label{lcfittable}
\vspace{.1in}
\end{deluxetable*}

\section{B. Public products used in the analysis}
\label{publiccodes}
PEGASE (\citealt{PEGASE}), Le Phare (\citealt{lephare}), \SMP\ (\citealt{Brout18-SMP}-SMP), AutoScan (\citealt{autoscan}), SALT2 models (\citealt{Guy10}, B14), \texttt{SNANA} (\citealt{Kessler09SNANA}, K18), CosmoMC (\citealt{cosmomc}), SNID (\citealt{SNID}), MARZ (\citealt{marz}), ZPEG (\citealt{ZPEG}),  Superfit (\citealt{superfit}).

\section{C. Data Release Products}
\label{dr}
DES-SN3YR binned and unbinned distances, measurement uncertainties and covariance are included at \urldr\ as well as the full Table \ref{lcfittable} in machine readable format.

\section{D. Analysis of the Fakes}
\label{fakesimappendix}

\newcommand{\SIMONE}{SIM1}
\newcommand{\SIMTWO}{SIM2}

Here we describe a few details about the cosmology analysis with fake SN light curve fluxes 
overlaid on DECam images. To avoid confusion between two sets of \SNANA\ simulations, 
we define \SIMONE\ for simulated fluxes overlaid on images, 
and \SIMTWO\ for the bias-correction simulation used in the BBC fitting stage.

For \SIMONE, SN~Ia lightcurve fluxes were generated in a LCDM cosmology over a redshift
range from 0.1 to 1.2. These fluxes and were inserted as point sources onto DECam images 
at galaxy locations chosen randomly with probability proportional to its surface brightness 
density. The generation of fake lightcurves and the procedure for image
overlays are described in detail in Section 2 of K18. \texttt{DiffImg} discovered 40\% of the 
100,000
fake SNe Ia lightcurves that were inserted on the 
DES-SN images and the \SMP\ pipeline was run on a representative subset of 
10,000
lightcurves. 
Analysis requirements and SALT2 lightcurve fitting resulted in a sample of
6586
fake SNe Ia that are fit with BBC and CosmoMC.

For the BBC fit we create a bias correction sample from \SNANA\ simulations (\SIMTWO).
The underlying SN~Ia light curve model is identical to that used in \SIMONE:
e.g., color \& stretch population, and no intrinsic scatter.
In the first season (Y1), there was a \SIMONE\ generation bug forcing the 
same galactic extinction ($E(B-V=0.043$) at all CCD image locations, 
and this same bug was intentionally preserved in \SIMTWO\ for Y1.
In contrast to the real data, $E_{\rm spec}=1$ for both \SIMONE\ and \SIMTWO.

\begin{figure*}[!b]
\begin{centering}
\includegraphics[width=0.95\textwidth]{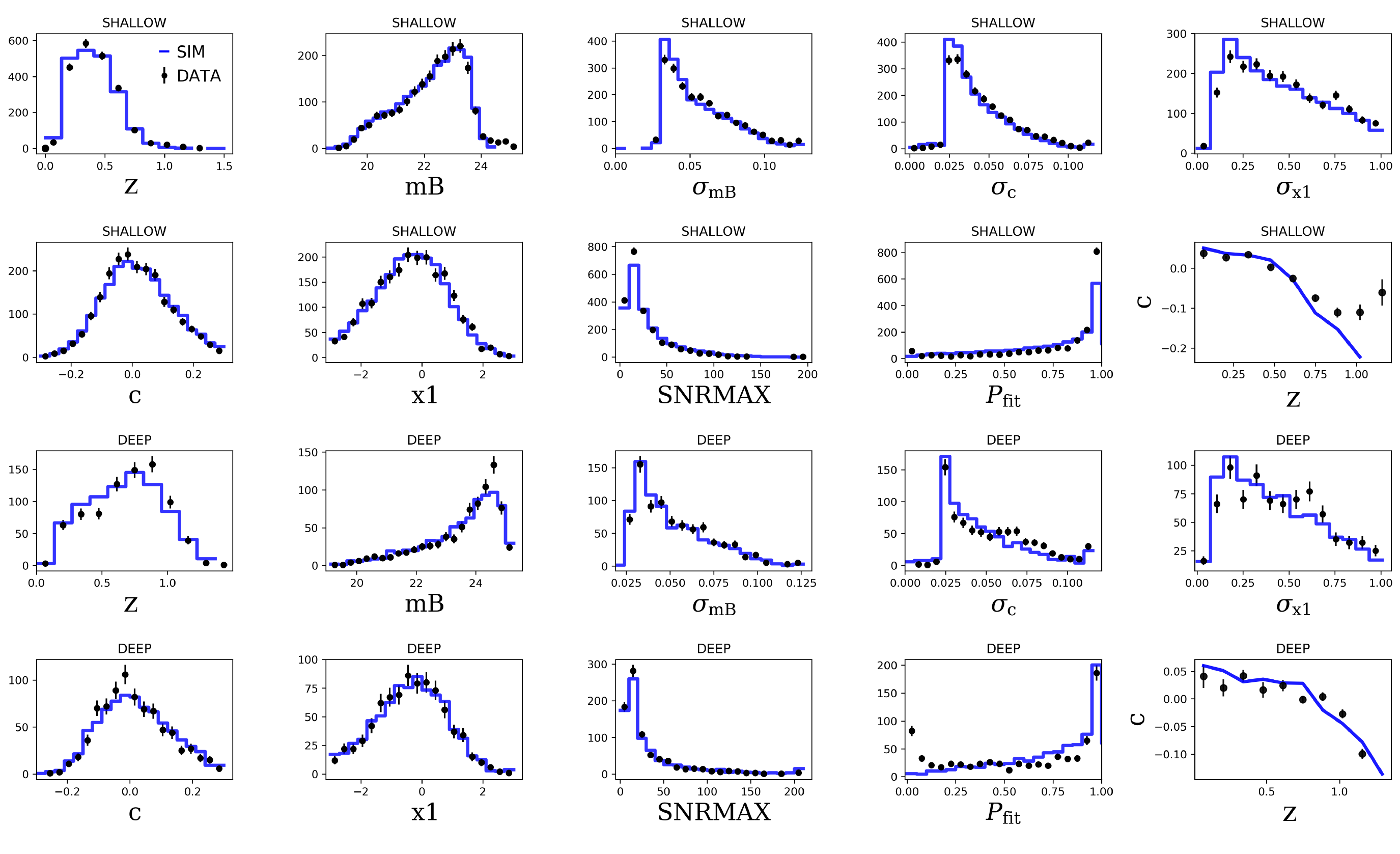}
\caption{Comparison of 6,586 fake supernova light curve fits with simulations used to compute biases in a fake cosmology analysis.}
\label{fig:fakesims}

\vspace{.1in}
\end{centering}
\end{figure*}

Finally, the \SIMTWO\ redshift distribution was tuned in each of the ten SN 
fields to match \SIMONE\ after cuts.
This field-dependent redshift tuning was needed because of the subtle way that
\SIMONE\ had selected real host-galaxies from the science verification (SV) catalog.
Although a single host-galaxy $z$ dependence was specified, the non-uniform depth 
of the SV galaxy catalog resulted in a different redshift distribution in each field. 
To illustrate this feature, consider an extreme example with just two fields 
(e.g., E1, E2). Next, suppose that the galaxy catalog for E1 only includes redshifts 
$z<0.5$ while for E2 we have $z>0.5$.
A simulation generating a flat galaxy redshift distribution over $0<z<1$ results
in non-overlapping (i.e., different) redshift distributions in E1 and E2. A comparison between the resulting bias correction simulation and the fakes is shown in Figure~\ref{fig:fakesims}.

\bibliographystyle{apj}
\bibliography{brout}

\end{document}